# Extended Ionized Fe Objects in the UWIFE Survey

Yesol Kim[1,2]★ , Bon-Chul Koo[1] , Tae-Soo Pyo[3,4] , Dirk Froebrich[5] , Woong-Seob Jeong[2,6] ,
Jae-Joon Lee[2] , Yong-Hyun Lee[1,7] , Ho-Gyu Lee[2] , Hyun-Jeong Kim[2,8] , Watson P. Varricatt[9]

[1] *Department of Physics and Astronomy, Seoul National University, Seoul 151-747, Korea*
[2] *Korea Astronomy and Space Science Institute, Daejeon 305-348, Korea*
[3] *Subaru Telescope, National Astronomical Observatory of Japan (NAOJ), National Institutes of Natural Sciences (NINS),*
 *650 North Aohoku Place, Hilo, HI 96720, USA*
[4] *School of Mathematical and Physical Science, SOKENDAI (The Graduate University for Advanced Studies), Hayama, Kanagawa 240-0193, Japan*
[5] *University of Kent, Canterbury CT2 7NH*
[6] *Korea University of Science and Technology (UST), 217 Gajeong-ro Yuseong-gu, Daejeon 34113, Republic of Korea*
[7] *Samsung SDS, Olympic-ro 35-gil 125, Seoul, Republic of Korea*
[8] *Department of Earth Science Education, Kongju National University, 56 Gongjudaehak-ro, Gongju-si, Chungcheongnam-do 314701, Republic of Korea*
[9] *UKIRT Observatory, Institute for Astronomy, University of Hawaii, 640 North Aohoku Place, University Park, Hilo, HI 96720, USA*

**ABSTRACT**
We explore systematically the shocked gas in the first Galactic quadrant of the Milky Way using the United Kingdom Infrared Telescope (UKIRT) Wide-field Infrared Survey for Fe$^+$ (UWIFE). The UWIFE survey is the first imaging survey of the Milky Way in the [Fe II] 1.644 $\mu$m emission line and covers the Galactic plane in the first Galactic quadrant ($7° < l < 62°; |b| \lesssim 1°.5$). We identify 204 extended ionized Fe objects (IFOs) using a combination of a manual and automatic search. Most of the IFOs are detected for the first time in the [Fe II] 1.644 $\mu$m line. We present a catalog of the measured sizes and fluxes of the IFOs and searched for their counterparts by performing positional cross-matching with known sources. We found that IFOs are associated with supernova remnants (25), young stellar objects (100), H II regions (33), planetary nebulae (17), and luminous blue variables (4). The statistical and morphological properties are discussed for each of these.

## 1 INTRODUCTION

Understanding the role of shocks is essential for comprehensively studying the ecology of the Milky Way, since they transfer mass and kinetic energy into the interstellar medium (ISM), provide heavy elements for future star formation by destroying dust grains, and regulate star formation. Shock waves are one of the principal mechanisms of the interaction between stars and the ISM in galaxies, thereby playing an important role in the evolution of the Galaxy. Among the most powerful shock-driving sources are outflows and jets from young stellar objects (YSOs), stellar winds from massive OB stars, and supernova (SN) explosions. To understand the physics of the interactions as well as the nature of the shock-driving sources, observations of emission lines from the shocks are essential.

The [Fe II] a$^4D_{7/2} \rightarrow$ a$^4F_{9/2}$ 1.644 $\mu$m transition results in one of the brightest emission lines in near-infrared (NIR). It originates from one of the 16 levels of Fe$^+$ that have a low excitation energy. Therefore they are easily excited in shocked gas, resulting in many lines, particularly in NIR. This emission line is thought to be bright in shock-excited gas; one suggested reason is that due to far-ultraviolet (FUV) radiation from the shock front, the Fe atom is in the form of Fe$^+$ over extended regions (McKee et al. 1984, Hollenbach et al. 1989, Oliva et al. 1989, Koo et al. 2016). In contrast, in photoionized regions, Fe atoms are predominantly at higher ionization states, except when the ionizing radiation is hard enough that it can penetrate further into the interstellar cloud (Koo et al. 2016). Therefore, [Fe II] emission lines from shocked gas are stronger than those from photoionized regions; for example, [Fe II] 1.257 $\mu$m / Pa $\beta$ is over 0.1 in supernova remnants (SNRs) compared to 0.01−0.03 in Orion (Koo & Lee 2015; Mouri et al. 2000). Furthermore, the Fe abundance can be enhanced by shocks owing to grain destruction, making the [Fe II] lines stronger (Koo 2014; Greenhouse et al. 1991; Mouri et al. 2000 and references therein). These characteristics of [Fe II] make its lines extremely useful for studying interstellar shocks (e.g., Dinerstein 1995; Nisini 2008). For example, the 1.644 $\mu$m emission line as a tracer of shocked atomic gas enables us to study shocked regions in jets/outflows of YSOs (Nisini et al. 2002; Caratti o Garatti et al. 2006; Takami et al. 2006; Pyo et al. 2006; Pyo et al. 2009; Oh et al. 2016), planetary nebulae (Welch et al. 1999; Smith et al. 2005), supernova remnants (Koo et al. 2007; Lee et al. 2009; Lee et al. 2013), and nebulae of luminous blue variables (Smith 2002). Since NIR [Fe II] lines suffer less extinction than widely used optical emission lines such as H$\alpha$, [S II], and [O III], the NIR lines can give us information on deeply embedded regions inaccessible by optical lines.

★ E-mail:yskim916@gmail.com

Lee et al. (2014) conducted an unbiased [Fe II] 1.644 $\mu$m narrow-band imaging survey, which is called the United Kingdom Infrared Telescope (UKIRT) Wide-field Infrared Survey for Fe$^+$ (UWIFE). The survey area ($7° < l < 62°; |b| \lesssim 1°.5$) is located in the first Galactic quadrant. This survey is the first unbiased, high-resolution [Fe II] survey of the Milky Way. It therefore enables us to discover more [Fe II]-emitting sources and conduct a statistically meaningful investigation of Galactic [Fe II] line sources. Alongside [Fe II]-emitting Galactic SNR to study similar to [Fe II] line objects in nearby galaxies, the survey is expected to systematically detect low-brightness [Fe II] line sources from other kinds of [Fe II] emitters. Therefore it enables us to assess the level of contribution of each [Fe II]-emitting population. Further spectroscopic studies of new [Fe II] sources found in UWIFE can be used to derive critical densities in the range of $\sim10^4$–$10^5$ cm$^{-3}$ and temperatures up to $10^4$ K (Pesenti et al. 2003), filling the gap in density between [S II] $\lambda6731\sim10^4$ cm$^{-3}$ and [O I] $\lambda6300\sim10^6$ cm$^{-3}$ (Osterbrock 1989). With other [Fe II] lines and emission lines such as [Fe II] 1.533 $\mu$m, density diagnostics of $\sim10^2$–$10^5$ cm$^{-3}$ can be measured and line ratio diagrams with other [Fe II] lines (Pesenti et al. 2003) can help us understand the new parameter range.

Shinn et al. (2014) systematically searched for outflows from ultra-compact H II regions (UCHIIs), inferred [Fe II] outflow mass-loss rates, and discussed the travel time of the [Fe II] outflows using the UWIFE data. The statistical [Fe II] line study of Galactic SNRs in UWIFE and the UKIRT Wide-field Infrared Survey for H$_2$ (UWISH2; Froebrich et al. 2011) survey revealed a detection rate of 24 per cent for both surveys and suggested a relatively higher coincidence with mixed-morphology and/or radio-bright SNRs (Lee et al. 2019).

A comprehensive catalog of UWIFE sources will give an opportunity to compare shocked [Fe II] line objects with other tracers in previous large-scale Galactic plane surveys. Particularly, the UWIFE survey area is fully covered with the complementary survey, UWISH2 (Froebrich et al. 2011), which was carried out using UKIRT and the Wide-Field Camera (WFCAM, Casali et al. 2007). The catalog of extended H$_2$-emitting sources identified in UWISH2 (Froebrich et al. 2015) will be useful for the comparison of shocked molecular gas with higher excitation atomic gas. Also, the Isaac Newton Telescope (INT) Photometric H$\alpha$ Survey of the Northern Galactic Plane (IPHAS; Drew et al. 2005) and the UWISH2 survey can provide a chance to compare different outflow/shock tracers. Surveys tracing continuum sources in embedded regions such as the UKIRT Infrared Deep Sky Survey (UKIDSS) Galactic Plane Survey (GPS; Lucas et al. 2008) in the near-infrared, the Galactic Legacy Infrared Mid-plane Survey Extraordinaire (GLIMPSE; Churchwell et al. 2009), the Multiband Imaging Photometer for *Spitzer* Galactic Plane Survey (MIPSGAL; Carey et al. 2009) in the mid-infrared (MIR), and the *Herschel* infrared Galactic Plane Survey (Hi-GAL) (Molinari et al. 2010) in the far-infrared (FIR) were published.

Furthermore, the source catalog of various kinds of objects, namely, the catalog of UCHIIs from the Co-Ordinated Radio 'N' Infrared Survey for High-mass star formation (CORNISH, Hoare et al. 2012) and the catalog of Extended Green Objects (EGO, Cyganowski et al. 2008) can be good candidates to compare with [Fe II] sources, as well as emission line source catalogs, viz., catalogs of H$\alpha$ emission-line sources from IPHAS (Witham et al. 2008), and Molecular Hydrogen emission-line Object (MHO, Davis et al. 2010). In accordance with these catalogs and aims, we designate [Fe II] 1.644 $\mu$m emission-line sources as ionized Fe objects (IFOs) and compile the first comprehensive catalog of Galactic extended IFOs. The catalog includes basic physical properties of IFOs, such as coordinates ($l$, $b$), size, position angle, and flux. Information about possible counterparts and their distance is also included.

The organization of the paper is as follows. In Section 2, we describe the data reduction, the source identification, the photometry of the detected sources, and the procedure for searching for counterparts or exciting sources of IFOs. In Section 3, we first present the catalog of IFOs. The catalog contains the sizes and fluxes of IFOs as well as their counterparts. The IFOs are classified by their counterpart types. We then explore the statistics of the physical properties and the distribution of IFOs. In Section 4, we discuss the results of the individual types of IFOs. In Section 5, we summarize our paper.

## 2 DATA AND SOURCE CATALOG

### 2.1 UWIFE Survey Data

We have used the UWIFE survey data to search for extended IFOs in the Galactic plane. The UWIFE survey was carried out using WFCAM at UKIRT in 2012 and 2013 (Lee et al. 2014). The [Fe II] narrow-band filter was used, having a central wavelength of 1.644 $\mu$m and an effective bandwidth of 0.026 $\mu$m. The WFCAM hosts four Rockwell Hawaii-II HgCdTe 2 k $\times$ 2 k arrays, each covering 13.65 arcmin $\times$ 13.65 arcmin in area at a pixel scale of 0.4 arcsec. Four pointings of the telescope covered a contiguous area of 0.75 deg$^2$ (designated as 'tile', following the WFCAM terminology). Each pointing was composed of a set of dithered and microstepped observations, fully sampling the point spread function in good seeing conditions (<0.8 arcsec). The total integration time per pixel was 720s. The final [Fe II] images have a nominal 5$\sigma$ detection limit of 18.7 mag for point sources, with a median seeing of 0.83 arcsec. For extended diffuse sources, the corresponding surface brightness limit is $8.1 \times 10^{-20}$ W m$^{-2}$ arcsec$^{-2}$.

Lee et al. (2014) also produced continuum-subtracted [Fe II] images (hereafter [Fe II]-H images) by using the $H$-band images from the GPS. The continuum subtraction was carried out in two steps, i.e., point-like continuum sources were first removed in both [Fe II] and $H$-band images, and then the point source removed $H$-band images were subtracted from the point source removed [Fe II] images to remove extended continuum sources. The details of the observation and data processing procedure can be found in Lee et al. (2014).

All [Fe II] and [Fe II]-H images from UWIFE are available at the UWIFE web page [1]. The images consist of 220 tiles, where a single tile is a square of 54' $\times$ 54' in equatorial coordinates. The tiles are arranged as 55 stripes of four consecutive tiles at constant declination along the Galactic plane, covering a region within the First Galactic Quadrant of $7° < l < 62°; |b| \lesssim 1°.5$ (see figure 1 of Lee et al. 2014). On the web page, the UWISH2 H$_2$ and GPS $JHK$-band images are also available.

### 2.2 Source Identification

In this study, we first aimed to identify IFOs in the continuum-subtracted images (hereafter, [Fe II]-H). We identified most of the IFOs through visual inspection and added several faint IFOs by mean of an automatic source identification, which uses the same algorithm as UWISH2 (Froebrich et al. 2015).

We focused on the extended sources in this study. Visual inspections were carried out twice for the whole survey area. We searched for all possible emission features and, for each feature, we defined an ellipse approximately surrounding the emitting area by eye. Then the central coordinates, radii, and P.A. of the ellipses are measured and listed in Table 1. All IFO candidates identified in the [Fe II]-H image were double-checked in both un-subtracted [Fe II] and GPS *H*-band images to confirm whether they were a real source or not. As the UWIFE and GPS observations were separated by several years, variable sources were seen as emission or absorption in the [Fe II]-H images. In particular, since artifacts with a negative digital number (DN) in GPS *H*-band resemble real sources in the [Fe II]-H image, we checked the position of all IFO candidates in the corresponding *H*-band data.

---

[1] http://gems0.kasi.re.kr/uwife/

In addition, there are various kinds of artifacts in the [Fe II]-H images. Representative artifacts are: the residuals of bright stars, ghosts, cross-talks, cross stripes after star subtraction, and the diffraction pattern of bright stars (see Appendix A.1 for examples). Residuals of high proper-motion stars were also left in the [Fe II]-H images. We also excluded the features hampered by the artifacts from bright stars. The sources that show diffuse structures in both [Fe II] and *H* which are significantly brighter in [Fe II] compared to the GPS *H*-band, or the sources with a low probability of being scattered emission from dust seen in the GPS *H*-band, were selected as real sources.

Using the [Fe II]-H images, we conducted an unbiased automatic detection with the code used for identifying MHOs in UWISH2 (Froebrich et al. 2015) to benefit from its objectiveness. We adjusted the code to fit the specifications of UWIFE data: (1) Remove small-scale features (residual of star subtraction), determine the large-scale background level from a 40 arcsec scale median filter, and calculate its noise value. (2) Draw contours at the $1\sigma$ level in ds9[2] and identify the isolated contours as 'regions'. The level was determined empirically to include faint emission of IFOs. The low ($1\sigma$) level produces contours around the remaining point sources and noise peaks, but those 'false' regions are removed by a minimum size limit in the next stage. (3) Remove contours that are too small ($<4$ arcsec$^2$) or near the image borders. (4) To avoid mistakenly identifying star residual as IFOs, remove contours smaller than 35 arcsec$^2$ if they are located within 3 arcsec to the Two Micron All-Sky Survey (2MASS) *H*-band stars brighter than 15 mag. This procedure was conducted for all UWIFE tiles except for tile 003, 080, and 196 due to the late release of the *H*-band data in the GPS survey.

All sources identified by the automatic detection were cross-checked by visual inspection. We first examined whether the identified source from the code is an image artifact or not. Appendix A.1 shows some examples of the artifacts, including residuals of detector cross-talk and diffraction patterns from saturated stars. These non-astronomical sources can be easily distinguished by comparing them in the [Fe II] and *H*-band images and were removed from the catalog. We also rejected point-like sources (e.g., high proper motion stars, variables, [Fe II]-emitting stars, etc.) that are not considered in this paper. Note that the visual identification treats a group of clumpy structures as a single object (e.g., shells of SNR). On the other hand, the code identifies the substructures separately. We fitted each automatically-identified IFO with an ellipse and derived the geometrical parameters of the semi-major axis, semi-minor axis, and position angle. This process added 14 IFOs, and the complete catalog is presented in Table 1, which also provides their coordinates, sizes, fluxes, and counterparts.

### 2.3 Photometry

We conducted photometry of the IFOs in the [Fe II]-H images. Since our targets have an extended structure, we adopted aperture photometry. In the [Fe II]-H data, artifacts often have higher digital counts than IFOs. Therefore, masking artifacts is a crucial process. We masked the identified artifacts to prevent large uncertainties in the aperture photometry. The residuals of point sources (stars) brighter than 14th magnitude in the *H*-band (based on the 2MASS point source catalog, Skrutskie et al. 2006) were also masked. The size of the masking area was 6 arcsec in diameter, which is large enough to cover general residual patterns. When instrumental artifacts such as electronic cross-talk or diffraction patterns intruded on the aperture, we manually masked them to prevent any contamination.

In order to derive the total flux ($F_{\rm tot}$) of the identified IFOs in a scientific unit (W m$^{-2}$), we used the following equation :

$$F_{\rm tot} = F_0 \cdot \left(\frac{DN}{t_{\rm exp}}\right) \cdot 10^{-0.4 \cdot m_{\rm zpt}}$$

$F_0$ is the in-band flux of Vega falling in the [Fe II] filter ($3.27 \times 10^{-11}$ W m$^{-2}$, Lee et al. 2019), whereas $t_{\rm exp}$ and $m_{\rm zpt}$ are the net exposure time (60 s) and the zero-point magnitude of each image, respectively ('EXPTIME' and 'MAGZPT' in the image header). DN is the total digital number falling in the aperture corrected for the sky background. This local background of each source was estimated from a sky annulus with an inner and outer radius of 1.2 and 1.5 times the aperture. We took the mode of the sky values to further avoid the effect of any possible artifacts. The uncertainty of the flux is estimated considering the photometric calibration error from the uncertainty of the zero-point magnitude of ~0.06 mag, which corresponds to ~6 per cent of the total flux (Lee et al. 2019). The contribution of Poisson noise from aperture photometry and sky subtraction is negligible. The former is less than one-fifth, and the latter is less than one-tenth of absolute calibration uncertainty.

### 2.4 Search for Associated Exciting Sources

We have searched for the possible driving source(s) of IFOs via positional cross-matching with previously known sources: SNRs, H II regions, compact and ultra-compact H II regions, luminous blue variables (LBVs), planetary nebulae (PNe), and YSOs. IFOs associated with these sources are classified as SNR-IFO, HII-IFO, CHII-IFO, LBV-IFO, PN-IFO, and YSO-IFO, respectively. The rest of the IFOs are classified as 'unknown-IFO'. In the following, we describe the processes and catalogs employed for the search for the individual exciting source types.

SNRs have complex and filamentary structures often with a large spatial extent. Thus, a careful identification and the separation of genuine SNR-origin from mere superposition was required. We first selected IFOs located within the boundary of known SNRs, using the central positions and sizes of SNRs in the Galactic SNR catalog of Green (2019). We then referred to the references in the catalog and also SIMBAD[3] for the multiwavelength morphology of SNRs for the confirmation of the association. If an IFO shows a coherent structure occupying a similar extent and/or its morphology implies a spatial correlation with the SNRs, we categorised it as an SNR-IFO. We also checked the area in SIMBAD for a possible superposition of unrelated, superposed sources such as PNe along the same line of sight. An IFO without noticeable morphological correlation but positionally coincident with evident SNR emission was categorised as an SNR-IFO (e.g., IFO 38).

---

[2] https://sites.google.com/cfa.harvard.edu/saoimageds9

[3] http://simbad.harvard.edu/simbad/

Diffuse H II regions also occupy a large spatial area and have complex morphology, so that a SIMBAD/VizieR query by IFO coordinate with an arcmin radius often returns various kinds of incidental sources such as sub-filaments of H II regions, jets/outflows from neighboring YSOs, and merely superposed sources along the line of sight. Therefore, keeping in mind that proximity alone does not necessarily guarantee a true correlation, a morphological correlation was also taken into account for identifying H II regions as a exciting source. If needed, a query with a larger angular scale was conducted to locate the diffuse H II region. We then compared the morphology of IFOs with that of H II regions obtained from high-resolution radio images (GPS; White et al. 2005, New-GPS; Helfand et al. 2006, and the H I, OH, recombination line survey of the Milky Way, THOR continuum; Beuther et al. 2016) and/or datasets from large-scale multi-wavelength studies (Fujita et al. 2021; Povich et al. 2009; Roshi et al. 2017). We also used small-scale surveys and targeted studies (see § 4.3). The IFOs with a positive correlation have been categorized as HII-IFOs. However, since the [Fe II] line emission from an H II region is inherently faint, morphological correlation with radio is occasionally hard to confirm. On the basis of this possibility, a few extended and faint IFOs have also been regarded as HII-IFOs although they do not have a clear morphological relationship with an H II region (see comments in § 4.3).

We further explored whether IFOs are associated with compact, ultra-compact, or hyper-compact H II regions (CHII, UCHII, and HCHIIs) by querying VizieR within an arcmin-scale radius. Two comprehensive lists of UCHII regions were selected for the VizieR positional matching: the CORNISH UCHII region catalog (Kalcheva et al. 2018), which is appropriate for the comparison with the UWIFE survey data in terms of comparable high-resolution ($1''.5$) and spatial coverage ($10° < l < 65°, |b| < 1°$), and the catalog presented by Bronfman et al. (1996) which is a large-scale compilation of Infrared Astronomical Satellite (*IRAS*) FIR color-selected UCHII regions with higher Galactic latitude coverage ($|b| < 2°$). The IFO positions were subsequently searched in SIMBAD to refer to targeted studies. We compared the [Fe II] line morphologies with available radio continuum images (see § 4.2). When the IFOs show morphological correlation with radio structures or delineate the boundary of radio structures, they are classified as CHII-IFOs. IFOs having counterparts supposedly earlier or at a lower-mass evolutionary stage of an UCHII region (e.g., hypercompact H II region, UC H II precursor, ultra-compact embedded cluster which was suggested as a lower-mass class of UCHII) are also included in this category (see § 4.2).

In order to identify IFOs associated with luminous blue variables (LBV-IFOs), the SIMBAD query was conducted with a radius criterion of 10 arcmin. But we noted that the list of LBVs and LBV candidates (hereafter cLBVs) has not been fully incorporated in SIMBAD, so we also used the catalog of (c)LBVs compiled by Nazé et al. (2012) which lists the coordinates of 68 (c)LBVs. As far as we know, this is the most comprehensive catalog of (c)LBVs. For example, Weis & Bomans (2020), in their review article of LBVs, presented a catalog, but it has a smaller number of (c)LBVs than Nazé et al. (2012), i.e., 47 versus 68. In the catalog of Nazé et al. (2012), twenty-two LBVs (including 19 candidates) are located inside the UWIFE area. There was also a [Fe II] survey of 9 LBVs by Smith (2002). Among the 9 LBVs, only one was located within the UWIFE survey area and it has been identified in our survey, too.

For IFOs associated with PNe (PN-IFOs), the SIMBAD query was used with a radius criterion of 10 arcmin. We additionally compared the morphology of IFOs with multiwavelength data from references in SIMBAD. In order to incorporate recently discovered PNe and PN candidates that have not been updated in SIMBAD, we made use of the following databases and catalogs. We used the Hong Kong/AAO/Strasbourg H$\alpha$ (HASH) planetary nebula database which lists multi-wavelength data of newly found ∼3500 PNe and PN candidates (Parker et al. 2016). The database includes three large-scale catalogs of Galactic PNe; the Strasbourg-ESO catalog, the catalog of Galactic Planetary Nebulae version 2000, and the Macquarie/AAO/Strasbourg H$\alpha$ (MASH) catalogs, together with 159 new PNe from the related IPHAS survey and ∼400 from the literature. A large number of unpublished, new PN candidates are accessible in this database, which are mostly (1) older, redder, and have lower surface brightness or (2) are more remote and small-scaled, faint PNe (Parker et al. 2016). When the counterpart is not a well-known source and is identified only in the HASH database, we checked the association using the references provided. There are PNe only detectable in NIR, so the recent study of PNe based on the UWISH2 data (Gledhill et al. 2018) was also checked for possible counterparts. This study complements faint or small-scale PNe previously undiscovered.

For the remaining IFOs, we made use of several large-scale catalogs of YSOs alongside catalogs for specific regions or targets. The large-scale survey of YSOs in four evolutionary stages (i.e., quiescent, YSO, protostellar, and massive star-forming stages, Urquhart et al. 2018) was used to find YSO-IFOs, keeping in mind the survey resolution (30 arcsec). The Infrared Array Camera (IRAC) red-source catalog was also used in the same manner (Robitaille et al. 2008) to locate YSO-candidate counterparts. When there was a positive match, we subsequently displayed their positions on the [Fe II]-H images with H$_2$ contours of UWISH2 data to confirm their association. H$_2$ images are useful since H$_2$ emission is usually more easily excited, forming a series of knots between an IFO and the YSO that drives a H$_2$ outflow. When the positional match and morphological information could not pinpoint an obvious YSO counterpart, we listed up to two YSOs. Also used are small-scale survey catalogs to benefit from a deeper searches for YSOs. Kim et al. (2015) conducted a detailed survey of YSO candidates in the infrared dark cloud (IRDC) G53.2 region and investigated their evolutionary stages. Povich & Whitney (2010) investigated the M17 region where we have identified many IFOs, and the study provided the evolutionary stages of the YSOs. Ragan et al. (2009) covered multiple IRDC regions in the UWIFE survey area and a YSO Class with MIR color and distance information was provided. Other small-scale catalogs of YSOs available in Vizier were also used when available (§ 4.1).

Since Herbig–Haro (HH) objects are often bright in [Fe II] emission, we attempted to locate the [Fe II]-emitting HH objects separately from YSO-IFOs. 454 Galactic HH objects have been compiled by Reipurth (2000), who continuously updated the SIMBAD database to include newly found HH objects. We retrieved all HH objects in SIMBAD, up to HH 1213, which includes 3140 sub-structures (e.g., HH 250A and 250B). Firstly, we search for YSO-IFOs and unknown-IFOs within a radius criterion of 10 arcmin for a given HH object. When there was a match, we looked for a possible association of the IFO with HH object structures via multi-wavelength images (mainly H$\alpha$ from IPHAS, Witham et al. 2008). For example, IFO 195 which is associated with the parsec-scale HH 803 has a very compact, small-scale structure. It was originally categorized as an unknown-IFO since we could not find any associated source just based on positional proximity. However, when we plot the IPHAS H$\alpha$ and UWISH2 molecular hydrogen emission contours together, we could associate IFO 195 with the south-western tip of the series of aligned structures of HH 803 in Hα and H$_2$ emission. Figure 1 shows the representative IFOs with respect to each counterpart.

# 3 RESULTS

## 3.1 Catalog of IFOs and their Statistical Properties

The full catalog of extended IFOs is presented in Table 1 and the description of each column of the catalog is as follows:

**Column 1. IFO identifier** : Designations of IFOs by a catalogue number in ascending order. When the IFO is identified only by a source detection algorithm, we marked them with an 'a' after its designation.

**Column 2. IFO conventional designation** : IFO full-name derived from Right Ascension and Declination (J2000) of the source center. It follows the 2MASS convention for the naming, i.e.: IFO JHH:MM:SS.SSS±DD:MM:SS.SS.

**Column 3, 4. Galactic longitude (*l*) and latitude (*b*)** : The center position of the source, in units of degree, in Galactic coordinates. For automatically identified IFOs, we adopt the geometric centre of the polygon by 2-dimensional Gaussian fitting of an ellipse.

**Column 5, 6. Semi-major axis ($r_1$) and semi-minor axis ($r_2$)** : Maximum semi-major and minor angular radius of the IFO in units of arcseconds.

**Column 7. Position angle (P.A.)** : The angle of the semi-major axis of an ellipse, in a counter-clockwise direction, from North to East in units of degree.

**Column 8. Area** : An area of an ellipse determined by the semi-major axis and semi-minor axis (Column 5, 6), in square arcseconds.

**Column 9. $F_{tot}$** : Total flux derived from summing up all flux inside an ellipse drawn from columns 5, 6. See the photometry section § 2.3 for details.

**Column 10. Counterpart** : Classification of the IFO indicating the most probable known object as follows: YSO-IFO - outflows or jets from an YSO or YSO candidate, HII-IFO - any outflows surrounding emission originated from the H II region, subdivided into HCHII, UCHII, CHII, and diffuse H II region, SNR-IFO - emission originates in SNR, PN-IFO - emission associated with PN/PN candidates, further classified into PN, PNc, and post-asymptotic giant branch (AGB), LBV-IFO - nebula structure around an LBV or LBV candidate, Unknown-IFO - multiple corresponding known object candidates or no possible known source in the vicinity.

Our IFO catalog contains 204 sources identified from 219 tiles, which is about 180 deg$^2$ in total. This number corresponds to an average surface density of ~1.1 IFOs per deg$^2$ in the first quadrant of the Galactic Plane (GP). This number should be regarded as a lower limit since our source identification methods were conservative. In general, the results of the manual and automatic search by the source detection algorithm were in good agreement. The 14 sources found only by the source detection algorithm, are marked with an 'a' after the IFO number in Table 1. They were either very faint or resembled artifacts. The majority of catalog sources are new discoveries of [Fe II] emission, and represent an order of magnitude increase in the number of extended [Fe II] sources in the first Galactic quadrant.

Table 2 presents basic statistics of IFOs for each counterpart type. We identified 100 YSO-IFO (87 YSOs, 13 HHs), 33 HII-IFO (22 CHII, 11 HII), 25 SNR-IFO, 17 PN-IFO, 4 LBV-IFO, and 25 IFOs without counterparts. Note that if a counterpart source has two distinct [Fe II] structures, they are counted as two separated IFOs which share a common counterpart (e.g., IFO 85 and 86 are from SNR G21.5-0.9 and are counted as 2 SNR-IFOs). Also, one SNR-IFO (IFO 7) is included in the number statistics in Table 2 but not used for flux statistics.

In total, 65 per cent of identified IFOs are related to star formation (49 per cent YSO- and 16 per cent HII-IFO), and 22 per cent are associated with evolved objects with 12 per cent of IFOs remaining as of unknown origin. Among them, YSO-IFO is the most frequent population showing [Fe II] emission. However, they account for only 1.6 per cent of the total [Fe II] flux. On the contrary, SNR-IFOs contribute 76 per cent of the total [Fe II] flux, though represent only 12 per cent of the IFOs by number. On average, the SNR-IFOs are 191 times brighter than the YSO-IFOs. The total flux of PN and LBV-type IFOs is similar, contributing 1 per cent of the total [Fe II] flux, albeit the number of PN-IFOs is 4 times larger. In order to understand the surface brightness of each type, the size and structure of the [Fe II] sources should be taken into account. In the next section, we will compare each counterpart's characteristics in more detail.

## 3.2 Flux and Size Distribution

In Figure 2a, we present the flux distribution of the IFOs. The flux distributions of the individual IFO types are shown in different colors. As mentioned above, some of the IFOs share the same exciting/driving source (e.g., 8 of 25 SNR-IFOs and 6 of 17 PN-IFOs). Bearing this in mind, we see that SNR-IFOs and HII-IFOs are bright with $F_{tot}$ as large as ~$10^{-14}$ W m$^{-2}$, while YSO, PN, and LBV-IFOs are much fainter, with a two-orders of magnitude smaller maximum $F_{tot}$. YSO and PN-IFOs appear in a similar flux range but the majority of PN-IFOs are brighter than YSO-IFOs. The unknown-IFOs are generally much fainter than the other types of IFOs.

Figure 2b shows the semi-major axis distribution of IFOs. IFOs appear in a wide range of sizes, from very compact, arcsecond-scale knots to large-scale objects up to ~47 arcmin in size. The distribution with respect to types is similar to that of the flux distribution, e.g., SNR-IFOs and HII-IFOs are large and bright while YSO-IFOs and PN-IFOs are small and faint. The radius range (<10 arcsec) of unknown-IFOs is similar to that of YSO-IFOs except for a few outliers. Although there are some exceptions and scatter, the overall fluxes and sizes seem to be proportional to each other. Especially for HII-IFO, the correlation coefficient of flux and size is 0.87. When divided into CHII and H II region sub-types, it is 0.52 and 0.83, respectively. The correlation coefficient of unknown IFOs is 0.99. In contrast the coefficient for SNR-IFOs is only 0.39.

Figure 2c presents the surface brightness distribution of IFOs. Unlike the flux and size distributions, the surface brightness distribution of each type shows slightly stratified distributions. Small IFOs appear to have a higher surface brightness in general, i.e., YSO-IFOs, PN-IFOs, and unknown-IFOs have higher surface brightness than HII-IFOs and SNR-IFOs. The reason for this might be due to the low surface filling factor of [Fe II]-emitting regions in the latter sources. For example, the IFO with the lowest surface brightness is SNR-IFO 117 (Kes 78). This SNR has a large size and the [Fe II] emission is patchy, apparent only around the northern and southern caps with a marginally detectable limb. For such sources, the true surface brightness of the [FeII]-emitting regions could be much greater. In Table 1, we made a note for IFOs with small surface filling factors.

## 3.3 Spatial Distribution

Figure 3 shows the distribution of IFOs in Galactic longitude and latitude. One can notice the Galactic longitude distribution is clus-tered albeit the sky coverage is more or less homogeneous. The most outstanding overdensities are seen at $l \sim 16°$ and $l \sim 51°$. At $l \sim 16°$, the dominant populations are YSO- and HII-IFOs, while at $l \sim 51°$, they are unknown- and YSO-IFOs. Including other clustered IFOs in longitude, the dominant populations responsible for these peak distributions are YSO-IFOs, followed by HII- and unknown-IFOs. A detailed description of the individual peak regions will be presented later in this section. Note that there are also voids free of IFOs at $40° \lesssim l \lesssim 50°$.

The distribution of the whole population of IFOs in Galactic latitude shows a Gaussian-like distribution. The distribution yields an average latitude at $b = -0°.12$ and standard deviation $\sigma = 0°.65$. Some concentrations of YSO-IFOs are found at $b \sim -0°.7$, $0°$ and $0°.8$. The average latitude of YSO-IFOs is $-0°.08 \pm 0°.67$. The centroids of the HII- and SNR-IFO distributions are also below the Galactic plane with an average latitude of $b = -0°.09 \pm 0°.63$ and $b = -0°.27 \pm 0°.58$, respectively. The average latitude of unknown-IFOs is also less than zero, i.e., $b = -0°.25 \pm 0°.73$. For comparison, the average latitude of PN-IFOs is $b = 0°.05 \pm 0°.57$. A similar trend has been observed in the UWISH2 survey; the average latitude of the jets and photodissociation regions (PDRs) was $-0°.18 \pm 0°.01$ and $-0°.17 \pm 0°.01$ while that of the PN group was $-0°.01 \pm 0°.01$ toward the Galactic mid-plane (Froebrich et al. 2015). The distribution of IFOs (excluding PN-IFOs) being slightly shifted to the negative latitude might be related to the 'bone' structure in the first Galactic quadrant. The bone structure refers to highly elongated, dense giant molecular filaments (GMF) that are the most probable tracer of spiral arm structure (Zucker et al. 2018). It is also worth noting the scarcity of IFOs at $0°.9 < b < 1°.5$. The number of LBV-IFOs is too small for their distribution to have any statistical meaning.

Figure 4 shows the two-dimensional distribution of IFOs in Galactic longitude and latitude along with their flux distributions. Several IFOs in the same system (e.g., jet and counter-jet of an HH object) are shown as concentric circles, as in many cases they are only a few arcsec away from each other. On average, all populations show clustered distributions with some differences from each other, though the survey coverage is homogeneous. As well as the inhomogeneous distribution of IFOs, all populations except LBV- and unknown-IFOs have more sources toward the Galactic center ($l \lesssim 30°$). About half of unknown-IFOs are located close to those of YSOs. In addition to the similar physical properties of YSO- and unknown-IFO shown in Fig 2, we suggest that at least half of the unknown-IFOs might originate from activities involved in YSOs.

The region relatively devoid of IFOs in the one-dimensional longitude and latitude distribution (Fig 3) turned out to form a large-scale 2-dimensional region; IFOs hardly exist toward $l \gtrsim 30°$, $b \gtrsim 0°.9$ and $35° \lesssim l \lesssim 50°$ near the Galactic mid-plane. This might reflect spiral arm structures and the sightline toward them, where we are seeing a shorter sightline toward the Galactic bar at $l \lesssim 30°$. Above this Galactic longitude we are seeing the local arm branching from Perseus Arm and Sagittarius-Scutum Arm (line of sight tangential to $l \sim 45°$) at a greater distance.

We have identified some areas rich in IFOs (upper panel of Fig 4), where in particular YSO and HII-IFO are major causes of overdensity. The respective regions are as follows.

(i) $l \sim 10°.2, b \sim -0°.3$: This region is coincident with the HII region G10.2−0.3, one of the three HII regions in the giant star-forming region W 31. The HII region is known to be very young (~0.6 Myr). At least four O stars are residing in it, where the brightest star W 31-1 showed permitted Fe II at 1.6878 $\mu$m and brackett lines in the NIR spectrum. In the $H$- and $K$-band spectra ($\lambda/\Delta\lambda \approx 3000$) the [Fe II] 1.644 $\mu$m emission line was not detected (Blum et al. 2001).

(ii) $l \sim 12°.8, b \sim -0°.2$: This region matches with [MDF2011b] cl1, which encircles the O4-6 (super-)giant #23 (Messineo et al. 2015). This region is immediately east of the embedded protocluster W 33 Main which is located inside the massive star-forming complex W 33. The $K_s$-band spectroscopy of #23 showed that the extinction of the region is $A_K = 1.20 \pm 0.03$ mag and the luminosity class is III-I. The Oe star #22 is located between W 33 Main and #23, with line identifications of Fe II 2.0895 $\mu$m and $H_2$, an extinction of $A_K = 2.87 \pm 0.07$ mag.

(iii) $l \sim 15°.1, b \sim -0°.7$: This over-density is coincident with one of the most massive star-forming regions, M 17. About a hundred O- and B-type stars are responsible for the emission and the system is quite young (<1 Myrs, Hanson et al. 1997). Bautista & Pradhan (1998) reported the detection of multiple iron species, including at 1.644 $\mu$m.

(iv) $l \sim 16°.9, b \sim +0°.8$: Multiple compact IFOs are connected to the apex of pillars of creation located in M 16, an active star-forming region. At the tip of the apex, there are protostars in the pillar's EGGs ('Evaporating Gaseous Globules'), which are not yet hot enough to emit X-rays. Therefore, the IFOs in M 16 might be tracing some of the youngest protostars.

(v) $l \sim 25°.4, b \sim -0°.2$: The region corresponds to W 42, an obscured giant HII region. The closest nearby source is [BCD2000] W 42 1, an O5.5 star (Blum et al. 2000). There are several point-like sources that might be true [Fe II] sources or mere variables.

(vi) $l \sim 30°.7, b \sim -0°.0$: The IFO is close to one of the closest starburst regions, W 43. This giant HII region has a central open cluster with massive stars.

(vii) $l \sim 49°.1, b \sim -0°.6$: Multiple HII- and YSO-IFOs are located in the vicinity of W 51, which is one of the most massive giant molecular clouds that is optically obscured. All large-scale representative structures, namely W 51 Main, IRS 1, and IRS 2, are bright in the [Fe II] 1.644 $\mu$m line. Each structure shows a distinct star-forming phase as follows: W 51 Main - several UCHIIs are located. IRS 1 - evolved HII region with a size of ~1 pc. IRS 2 - went through recent star formation, and an ~O3 star and a massive YSO were found (Barbosa et al. 2008). An LBV-IFO is also coincident with the region, which is a high-mass evolved star (P Cygni supergiant) with evidence for chemical enrichment (Clark et al. 2009).

(viii) $l \sim 53°.2, b \sim +0°.0$: Multiple YSO-IFOs coincide with an IRDC G53.2, which was formerly catalogued as three IRDCs in the *Midcourse Space Experiment* (*MSX*) dark cloud (MSXDC) catalog (Simon et al. 2006). The three IRDCs, viz. MSXDC G053.11+00.05, MSXDC G053.25+00.04, and MSXDC G053.31+00.00 harbor hundreds of YSO and YSO candidates, some of them in the vicinity of IFOs.

(ix) $l \sim 59°.4, b \sim -0°.2$: The IFOs are located in the central part of SH 2-87, a complex massive star-forming region. The three sub-millimeter clumps, SMM 1, SMM 2, and SMM 3 constitute this HII nebula. These three clumps are at separate evolutionary stages (Xue & Wu 2008), and two HII-IFOs were found in the vicinity of the hottest and most massive star-forming clump, SMM 1.

## 4 DISCUSSION

### 4.1 Outflows/Jets from Young Stellar Objects

Outflows/jets of YSO are composed of ejected and circumstellar swept-up material, and are recognized as an important signpost of recent star-forming activity. This phenomenon plays a key role in conventional disc accretion-outflow theories, the outflow being responsible for the removal of angular momentum and kinetic energy of accreting material that enables accreting material to overcome the centrifugal force and collapses to form a star (see theories of disk-wind; Pudritz & Norman 1983, X-wind; Shu et al. 1994, and observational studies; Ellerbroek et al. 2013 for reference).

Thanks to the development of IR instruments, previously undiscovered, highly obscured outflows have begun to be found in the near-infrared. The optical HH objects and their IR counterparts basically refer to the same phenomena, and only the conditions of jet and circumstellar matter differ. So far, molecular emission (e.g., MHO) has drawn attention in the NIR, alongside atomic/ionic lines in the optical, yet less attention has been brought to the [Fe II] lines in the NIR. The [Fe II] 1.644 $\mu$m line, the brightest iron line in the *H*–*K* band, is reported to unveil a shocked region that is denser and/or more ionized than regions where optical lines are generated (Nisini et al. 2002). In this aspect, previous studies using frequently used molecular tracers, namely SiO, CO, and HCO$^+$ in the sub-mm to mm, only revealed secondary outflows, tracing masses of low-density, distant (up to a few pc) outflows. Whereas the [Fe II] 1.644 $\mu$m line from the jet is found to extend a few AUs to parsec-scales in the form of a dense irradiated jet (Reiter et al. 2015).

Most previous [Fe II] outflow studies are confined to certain types of objects or regions: specific star-forming regions (Orion; Takami et al. 2002, Carina; Reiter et al. 2016, Shinn et al. 2013) or a certain mass range of YSOs (Caratti o Garatti et al. 2006; Caratti o Garatti et al. 2015). Recently, outflow studies toward external galaxies, namely the Large Magellanic Cloud (LMC) and the Small Magellanic Cloud (SMC), became feasible (Reiter et al. 2019). These studies showed that the [Fe II] emission tends to be observed at the tip of the bipolar outflow and is rather collimated, compared to H$_2$ and H$\alpha$ which predominantly show the morphology of a 'wake' enclosing the [Fe II] emission (Reiter et al. 2015).

We have detected 100 YSO-IFOs (Table 3). Our result provides a large and comprehensive sample for the study of [Fe II] emission associated with YSOs. Figure 5 shows the example of identified YSO-IFOs, displaying UKIDSS *KHJ*-band RGB images to show how the YSO-IFOs reveal unique structures in comparison to hot dust continuum structures. YSO-IFOs show diverse morphologies, diverse compared to traditionally observed/expected [Fe II] features that are located at the tip of bipolar outflows and/or are highly collimated toward the driving sources (Caratti o Garatti et al. 2006; Reiter et al. 2016).

We classified YSO-IFOs into four morphological categories; bipolar, cometary, knot-like, and amorphous. A representative case of each category would be IFO 13–14, 125, 122, and 4 in Figure B.1, respectively. Bipolar YSO-IFOs are a textbook case of star formation, consistent with the accretion-jet theory with the aid of a magnetic field (Konigl 1982; Shang et al. 2020; Frank 1999). They typically show two lobes located on opposite sides of a central source, but some show two wakes, tips, and collimated bow-shock shapes, distributed laterally from the apparent YSO jet axis. The prototypical bipolar YSO-IFOs are IFO 13 and 14. The [Fe II] 1.644 $\mu$m emission with bipolar morphology usually represents either the 'cap' of bow shock where an outflow collides with the ambient medium or dense, collimated jets. Cometary YSO-IFOs resemble a comet with a bright head around the driving source and a narrow faint tail-like structure. The prototypical cometary YSO-IFOs are IFO 125 and 131, both having well-defined conical structures. They are located at quite different distances, i.e., 4.7 and 1.1 kpc, and the extent of the associated conical structures has very different linear scales, i.e.,

~45000 AU (10 arcsec) and 5000 AU (5 arcsec). For the wide-angle tails of cometary morphology, it is possible that either (1) the jet is bending and/or precessing (Paron et al. 2016) (2) a cavity structure is revealed (Hsieh et al. 2017) (3) a multiple systems presence is implied (Fuente et al. 1998). Knot-like YSO-IFOs appear as knots, sometimes located symmetrically from a driving source along a certain axis. The representative knot-like YSO-IFOs, 122 and 123, are showing well-isolated compact features. These knot-like features might imply that the ejection of accreted material in the system is accompanied by sporadic bursts of accretion (Caratti o Garatti et al. 2015). Amorphous YSO-IFOs represent the remaining YSO-IFOs that are diffuse and do not have a definitive structure. The nature of the amorphous YSO is uncertain. The number of YSO-IFOs classified as bipolar, cometary, knot-like, and amorphous is 16, 18, 19, and 47, respectively.

The morphologies of YSO-IFOs are closely related to the nature of YSOs and their mass-loss histories (Caratti o Garatti et al. 2015; Paron et al. 2016). For example, the collimated and continuous jet morphology indicates a continuous ejection of accreting material from the accretion disc system (Reiter et al. 2016; Reiter et al. 2017). The overabundance of amorphous morphologies might suggest highly varying environments or multiple systems are affecting the outflow structure. But the morphology of YSO-IFOs might depend on environments as well as foreground extinction, so detailed studies are needed of the individual objects to confirm their nature. Thirteen YSO-IFOs are associated with HH objects (Table 4). Figure 6 shows a comparison of their [Fe II] and H$\alpha$ images.

YSO-IFOs constitute half the number of our cataloged sources, making YSO the most common IFO in the inner Galaxy. The number density of YSO-IFOs is 0.55 deg$^{-2}$. For comparison, the H$_2$ number density probed by UWISH2, which covered an almost identical area with a comparable surface brightness limit, is 2.15 deg$^{-2}$ (Froebrich et al. 2015). The flux density of YSO-IFOs ranges (2–820) × 10$^{-18}$ W m$^{-2}$ with a mean of 4.3 × 10$^{-17}$ W m$^{-2}$. This range can be compared with the results of other surveys. Caratti o Garatti et al. (2006) targeted H$_2$-emitting low-intermediate luminosity Class 0/I YSOs and reported that among 23, 74 per cent were also detected in [Fe II]. For the newly observed 9 [Fe II] line jets in the reference, the flux range is (2.8–27.0) × 10$^{-18}$ W m$^{-2}$. Caratti o Garatti et al. (2015) observed 18 intermediate- to massive-YSOs having H$_2$ and EGO counterparts, and the flux range is (2.5–61.9) × 10$^{-18}$ W m$^{-2}$. Note that these fluxes are obtained from spectroscopic studies using a slit of width 1 arcsec. The majority of YSO-IFOs have flux densities comparable to those of previous studies. But a few sources are exceptionally bright. The number of YSO-IFOs brighter than outflows observed in Caratti o Garatti et al. (2015) is 10 per cent of the YSO-IFOs. Since these bright YSO-IFOs do not share certain morphologies and 40 per cent of them have RMS counterparts, they might be preferentially massive YSO outflows, which have simply not yet been identified due to the limited sky coverage of past [Fe II] observations. One possible speculation is that [Fe II] brightness does not strictly scale with driving source brightness or other outflow tracers, based on the target of previous studies, which tend to be bright IRAS sources accompanying outflows discovered in other tracers. This illustrates the importance of an unbiased study to correct our understanding of the characteristics of [Fe II] emitters.

The YSO-IFOs and jet-group MHOs of the UWISH2 survey can be compared one-to-one since the UWIFE survey area was fully covered by UWISH2. The spatial distribution of YSO-IFOs in Figure 4 shows a highly clustered distribution, accompanied by the high-latitude sources in $l \sim 15$–$30°$ and the absence of YSO-IFOs in the Galactic mid-plane at $l \sim 40$–$50°$. This characteristic distribution is also shared in jet-group MHOs (see Figure 8 in Froebrich et al.

2015). As seen in Figure 5, about 85 per cent of YSO-IFOs accompany jet/PDR-group MHOs in the vicinity. For example, in the M 16 (Eagle nebula), 6 YSO-IFOs were identified, and a few hundreds of jet/PDR-group MHOs are also present. A detailed comparison of YSO-IFOs with jet/PDR-group MHOs discovered in the subsequent UWISH2 studies will be helpful for the comparison of different shock tracers (Ioannidis & Froebrich 2012a; Ioannidis & Froebrich 2012b; Froebrich & Makin 2016; Makin & Froebrich 2018; Samal et al. 2018).

We can compare our results with the results of the RMS survey where NIR spectra of YSO candidates have been obtained. In the common survey area ($10° \lesssim l \lesssim 62°, |b| \lesssim 1.5°$), there are 182 RMS sources, and among the 72 sources from which spectra have been obtained, 58 have [Fe II] line emission, though some of the detections could be confused with the Br 12 line. For comparison, only 17 of 182 RMS objects have been identified as YSO-IFO in our study (for some RMS sources, 2–3 IFOs correspond to one RMS source.) Among these 17 sources, the NIR spectra have been obtained for 8 sources, and [Fe II] lines were detected in 6 sources, i.e., [Fe II] lines were reported as non-detection for two sources in the RMS survey. We note that the non-detection for the two is based on a comparison of Br 11 and Br 12/[Fe II] line strengths (Br 11 × 0.788 > Br 12/[Fe II]) and it might be possible that a weak [Fe II] line is in fact present but missed by low spectral resolution, as the authors noted (Cooper et al. 2013). To assess this possibility, we checked the slit configuration (central position, position angle) in Cooper et al. (2013) and compared it with YSO-IFO morphology. For both IFO 72 and 141, the RMS slit intersects the driving source but does not include the bright part of extended YSO-IFO structures. Indeed, the authors tried to include extended structures inside the slit in imaging mode prior to spectroscopy mode, yet even narrow-band [Fe II] images of UWIFE without continuum subtraction turned out to severely hinder extended emission. Therefore, most YSO-IFOs apparently do not have RMS source counterparts, which is claimed to be a 90 per cent complete list of massive protostellar populations (Lumsden et al. 2013). This seems to suggest that most of YSO-IFOs are associated with low-mass star formation. It is also worthwhile to note that the majority of YSO-IFOs (87 per cent) are not associated with HHs, which suggests that the [Fe II] emission is tracing optically hidden star-forming regions.

### 4.2 Compact H II Regions

Compact and ultra-compact H II regions are the earlier stages of 'classical' H II regions. An UCHII region is a photoionized region with a diameter $\lesssim 0.1$ pc and an electron density $n_e \gtrsim 10^4$ cm$^{-3}$, embedded in a molecular cloud (Wood & Churchwell 1989). In this evolutionary stage, mass accretion of the central star is thought to be insignificant (Churchwell 2002; Zinnecker & Yorke 2007). A CHII region is a H II region in the intermediate phase between UCHII and classical H II regions, having a radius $\lesssim 0.1$ pc and $n_e \gtrsim 10^3$ cm$^{-3}$. The lifetime of UCHII and CHII is $\sim 2 - 4 \times 10^5$ yr (Davies et al. 2011, Mottram et al. 2011).

In UCHII and CHII regions, [Fe II] emission can be enhanced by the interaction of stellar wind with the ambient medium. Bloomer et al. (1998) detected enhanced shell-like [Fe II] emission along the periphery of the CHII region NGC 7538 IRS 2. The observed [Fe II] 1.644 $\mu$m / Br $\gamma$ ratio was 0.15, which is an order of magnitude greater than that of H II regions, and it implies that the [Fe II] line emission emanates from shocked stellar wind material. Shinn et al. (2014) searched for [Fe II] 1.644 $\mu$m emission associated with UCHII regions employing the CORNISH UCHII catalog and the UWIFE survey data. Among the 237 UCHII regions in the survey area, five and one candidate were found to have associated [Fe II] emission features, which were suggested to be shock-excited by outflows from central YSOs. Kim et al. (2017) also reported the detection of [Fe II] emission from UCHII Monoceros R2. Hereafter, we refer to IFOs associated with CHII/UCHII regions or with H II regions in even earlier evolutionary stages as CHII-IFOs.

We have detected 22 IFOs associated with 16 UCHII/CHIIs (Table 5). Six IFOs (IFO 24, 25, 26, 97, 107, and 156) had been previously reported by Shinn et al. (2014). We have discovered IFOs associated with an UCHII precursor (IFO 137) and an ultra-compact embedded cluster (IFO 138), which are thought to be earlier progenitor or less massive populations (Molinari et al. 1998; Alexander & Kobulnicky 2012). Among the 16 UCHII/CHII regions with [Fe II] emission features, 10 are catalogued in CORNISH, which corresponds to 4 per cent of the 237 UCHII regions in the CORNISH catalog in the survey area. The detectability might be partly due to the large extinction in UCHII/CHII regions, which is typically $A_V \sim 30-50$ or $A_K \sim 3-5$ (Hanson et al. 2002). Indeed, the $A_V$ of three UCHIIs with associated IFOs had been found to have relatively low extinction ($A_V \sim 9-20$, Shinn et al. 2014).

Figure 7 shows the 22 CHII-IFOs. CHII-IFOs have diverse morphologies, e.g., jet-like, shell-like, and amorphous morphologies. A representative IFO with jet morphology is IFO 97, which appears as a collimated beam from the center to the boundary of the H II region. The jet appears to extend beyond the radio continuum boundary (see Figure 7), which might reflect a possible correlation with the boundary of the ionization front (Goddi et al. 2020). The representative IFO with a shell-like morphology is IFO 132. An exemplary CHII-IFO of amorphous morphology would be the IFO 138, having a diffuse structure either outside or inside of the H II region in the radio. The properties of a compact H II region have been rarely investigated in [Fe II] emission. Shinn et al. (2014) proposed that some IFOs identified in the vicinity of ultra-compact H II (UCHII) regions (IFOs 24-26, 97, 107, 156) are the "footprint" outflow features of UCHIIs, i.e., the features produced by outflowing material ejected during an earlier, active accretion phase of massive young stellar objects, based on the morphological relation between the [Fe II] and 5 GHz radio features, the outflow mass-loss rate, the travel time of the [Fe II] features, and the existence of several YSO candidates near the UCHIIs. The newly discovered CHII-IFOs in this study might serve as a chance to investigate the origin of the [Fe II] emission in the vicinity of CHIIs.

### 4.3 H II Regions

H II regions are not expected to be bright in the [Fe II] lines, since in their photoionized regions, Fe atoms are predominantly in higher ionization states, and Fe atoms are thought to be mostly locked in dust grains (Koo et al. 2016). According to theoretical models of photoionized regions, the [Fe II] emission from an H II region is mainly emitted in the high-density partially ionized zones near ionization fronts, predominantly excited by electron collisions (Oliva et al. 1989; Bautista & Pradhan 1998). In the Orion nebula, for example, [Fe II] images exhibit filamentary structures and diffuse emission that might be associated with ionization fronts, together with some knotty features (Takami et al. 2002). Expanding H II regions can drive shocks, but the shock velocity is low ($\sim 10$ km s$^{-1}$) so [Fe II] line emission is not expected to be enhanced (e.g., Mouri et al. 2000). The [Fe II] 1.644 $\mu$m/Pa $\alpha$ ratio of the Orion is 0.013, which is more than two orders of magnitude smaller than those of SNRs (Oliva et al. 1989; Mouri et al. 2000). So Galactic H II regions have not been a popular target of deep and high-resolution [Fe II] imaging (Kraus et al. 2006; Bally et al. 2022). The depletion of Fe atoms in the H II region, however, is uncertain. In the Orion nebula, it has been estimated that 90 per cent of Fe is locked onto dust grains (Baldwin et al. 1991; Baldwin et al. 1996; Osterbrock et al. 1992; Rodríguez 2002). But there are studies which showed that, in many H II and star-forming regions, Fe is not depleted as heavily as in the Orion nebula (Osterbrock et al. 1992; Peimbert 1993; Rodríguez 2002; Okada et al. 2008; Peimbert & Peimbert 2010). It has been suggested that some populations of dust grains might be easily destroyed by UV radiation from OB stars and Fe atoms are released into the gas phase (Okada et al. 2008; Peißker et al. 2020). For external galaxies, Alonso-Herrero et al. (2003) did an imaging study of the starburst galaxies M 82 and NGC 253 in [Fe II] 1.644 $\mu$m and Pa $\alpha$ (1.87 $\mu$m) lines, and, by comparing their intensity ratios, concluded that 6–8 per cent of [Fe II] line fluxes are due to H II regions. Mouri et al. (2000), Riffel et al. (2016), Hennig et al. (2018), and Fazeli et al. (2019) suggested that some of the [Fe II] emission from external galaxies could be due to H II regions based on their low [Fe II] 1.257$\mu$m / Pa $\beta$ ratios.

We have identified 11 IFOs associated with 4 H II regions (Table 6). All HII-IFOs are located in the well-known star-forming complexes W 31, M 17, and W 51. Figure 8 shows the 11 HII-IFOs. We can see that some IFOs appear as thin filaments elongated along the radio structure (e.g., IFO 55, IFO 62) or as diffuse amorphous emission structures within the radio structure (e.g., IFO 11 and 12), so the association of IFOs with H II regions is very likely. The filamentary structures might correspond to ionization fronts and/or boundaries of PDRs as in the Orion nebula. On the other hand, some IFOs are faint and diffuse, and they extend beyond the radio boundary of the H II regions, e.g., IFO 159 and 161, so their association is uncertain and needs to be confirmed.

### 4.4 Planetary Nebulae

Planetary nebulae represent a short-lived phase near the endpoint of low- to intermediate-mass star (1-8 $M_\odot$) evolution which is preceded by the AGB, post-AGB, and pre-PN phases. The circumstellar envelope of the AGB carbon star is considered highly Fe-depleted (Mauron & Huggins 2010), though Fe becomes abundant with time (Fe abundance is negatively correlated with the C/O ratio, Delgado-Inglada & Rodríguez 2014). In turn, PNe are not expected to be strong [Fe II] emitters, also having a Fe-deficit nature with <10 per cent existing in gas and the remaining probably enshrouded in dust grains (Delgado-Inglada & Rodríguez 2014). Meanwhile, in the context of environmental factors, PNe could be a [Fe II] emitter since it has a partially ionized zone where Fe$^+$ is apt to exist, and at a certain point of its evolution, a low-velocity shock is expected to occur. In short, suitable ionization conditions and energy to excite Fe (Greenhouse et al. 1991) can be established in PNe, and its iron-depleted nature is a key factor to determine the existence of [Fe II] emission.

Besides the theoretical expectation, previous studies reported the detection of [Fe II] emission towards stellar objects in a variety of evolutionary stages: post-AGB (IRAS 16594-4656; Van de Steene & van Hoof 2003), pre-PN (M 1-92; Davis et al. 2005), and PN (Hubble 12; Welch et al. 1999, M 2-9; Smith et al. 2005, NGC 2440; Hora et al. 1999). Some authors suggested a circumstellar origin (e.g., Clark et al. 2014, Smith et al. 2005), especially Baan et al. (2021) reported the detection of [Fe II] emission revealing the interaction of an accretion inflow, which is composed of material ejected in earlier post-AGB and pre-PN circumstellar material, and stellar outflow.

Table 7 shows PN-IFOs. They are IFOs spatially coincident with PNe, PN candidates, and sources in earlier evolutionary stages such as post-AGBs. Seventeen PN-IFOs are associated with 14 PNe; 5 PNe, 8 PN candidates, and one post-AGB candidate. For comparison, in a previous study, Lee et al. (2014) reported the detection of [Fe II] emission in six PNe among 29 known PNe. In the survey area, there are 296 HASH 'true' (131), likely (40), and possible (125) PNe, so that the detection rate is 4.7 per cent. If we limit the sample to the 'true' PN, the detection rate slightly drops to 3.8 per cent (i.e., 5 out of 131). This very low detection rate of PNe in [Fe II] emission (4.7 and 3.8 per cent) contrasts with the results in H$_2$, where detection rates are 30 and 21 per cent, respectively (for $10° < l < 66°$, $|b| < 1°.5$, Gledhill et al. 2018). It is interesting that even with an order of magnitude larger sample of PNe in this study, our result is somewhat consistent with the former [Fe II] and H$_2$ detection rates of 7 and 39 per cent derived from 41 PNe (Hora et al. 1999). The slightly higher detection rate of Hora et al. (1999) could be because their samples are either moderately sized or optically bright.

The number density of PN-IFOs is 0.07 deg$^{-2}$ within 180 deg$^2$ whereas it is 1.25 deg$^{-2}$ within 209 deg$^2$ in UWISH2 (Froebrich et al. 2015). However, unlike the previous argument (Kastner et al. 1996), not all [Fe II]-emitting PNe are seen in H$_2$ emission; we found 3 of our 14 PNe in Table 7 were absent from the list of PNe with H$_2$ emission. Also, the median flux of PN-IFOs is greater than that of the H$_2$-emitting PNe, i.e., 6.46 $\times 10^{-17}$ W m$^{-2}$ versus 4.53 $\times 10^{-17}$ W m$^{-2}$. Therefore, our result shows that the H$_2$-emitting PNe are not necessarily brighter than the non-H$_2$-emitting PNe in [Fe II] emission.

Figure 9 presents [Fe II]-H images of PN-IFOs. Note that there are three bipolar PNe, each of which possesses two associated IFOs (IFO 5 and IFO 6, IFO 8 and IFO 9, and IFO 129 and IFO 130). We classified the morphologies of PN-IFOs using the basic 'ERBIAS' classifier following Parker et al. (2006), where 'E'=elliptical, 'R'=round, 'B'=bipolar, 'I'=irregular, 'A'=asymmetric, and 'S'=quasi-stellar. Their sub-classifiers 'amprs' are also adopted to describe detailed morphology; the main object has a one-sided enhancement/asymmetry 'a', has multiple shells or external structure 'm', exhibits point symmetry 'p', has a well-defined ring structure or annulus 'r', or resolved internal structure 's'. An IFO can have several 'amprs' sub-classifications. The results are summarized in Table 7, where their morphologies in H$\alpha$ and H$_2$ are also listed (Parker et al. 2016; Gledhill et al. 2018). The H$\alpha$ morphologies are from the HASH survey, while the H$_2$ morphologies are from the UWISH2 survey. For the PN-IFOs without a counterpart in the UWISH2 PN catalog (IFO 50, 95, 188), we inspected the UWISH2 data and classified their morphology in the same format (see Table 7). Some PNe have different morphologies in the [Fe II], H$_2$, and H$\alpha$ emission, which implies a complex surrounding environment and/or complex mass-loss history.

The physical sizes of PN-IFOs have been derived for 10 PNe that have previously estimated distances (Table 7). The sizes of 4 IFOs associated with 'true' PNe range from 0.13 pc to 0.92 pc, and three of the PN-IFOs are larger than 0.9 pc. This contrasts with the majority of PNe in H$\alpha$ being $\leq$0.2 pc (González-Santamaría et al. 2020). This seems to suggest that the [Fe II] emission preferentially traces large, bright PNe. For example, in PNG 050.4+00.7, the size of the associated IFO (IFO 157) substantially exceeds the previously known size of the counterparts (2 arcmin and 19 arcsec, respectively). The IFO has a partial 'S' shape elongated along the east-west direction, with IRAS 19194+1548 superposed at the western part. The structure becomes gradually fainter toward the west, therefore the angular size of the partial 'S' shape should be considered as lower limit. The implied physical scale of 3.1 pc largely surpasses the generally accepted size of PNe (one of the oldest and largest PNe, the Helix nebula has an outermost size of 1.76 pc). The driving source is suspected to be in a symbiotic star system (Akras et al. 2019) and the updated size is compatible with the sizes of large shells/nebulae around symbiotic stars (McCollum et al. 2008).

### 4.5 Nebulae of Luminous Blue Variables

Infrared [Fe II] 1.644 $\mu$m emission around prominent nebulosity of LBVs is thought to be ubiquitous. Smith (2002) (henceforth, S02) searched for [Fe II] 1.644 $\mu$m emission in nine well-known LBVs and found the emission in 7 of them, resulting in a detection rate of 77 per cent. This high detection rate surpasses that of SNRs (i.e., 24 per cent, Lee et al. 2019), the population that is thought to provide the most adequate environment for the existence of [Fe II] 1.644 $\mu$m emission. S02 could not pinpoint the essential condition needed for strong [Fe II] emission to exist. Shock heating and radiative heating as possible excitation mechanism of [Fe II] emission were suggested by the author.

Shock-excited [Fe II] emission can arise when the LBV's environment meets requirements such as (i) a large difference in the outflow speed between the stellar wind and pre-existing LBV nebula and (ii) a difference of velocities between the stellar wind and ejected shell created during S Doradus outbursts or giant eruption phases. This velocity difference of $50-150$ km s$^{-1}$ (S02) is ascribed to weaker gravity in an active phase. When LBV evolves toward a cooler temperature (to a local temperature lower than 30,000 K), Hydrogen atoms and opacity-enhancing ions start to emerge on the surface, which is known as the 'modified' Eddington limit (Humphreys & Davidson 1994). The elevated opacity makes the outward radiation pressure stronger and overpowers the inward gravity force. The resultant lower effective gravity helps LBVs easily induce the aforementioned mass loss. In these S Doradus outbursts and giant eruption phases, the weaker gravity results in an ejected shell having a lower expansion velocity than normal stellar winds. The following post-eruption wind has a velocity higher than that of the aforementioned high mass-loss phase and eventually overtakes the ejected shell. Meanwhile, photoionized [Fe II] emission was reported from two hot (30,000 K) LBVs (AG Car and R 127, S02) which was attributed to their stronger UV flux.

We detected [Fe II] emission features associated with 3 LBVs (Table 8). So the [Fe II] detection rate of LBV nebula in our study is 14 per cent. If we include the 9 LBV samples of S02, the detection rate would be 29 per cent, i.e., 9 out of 30 LBV nebulae (HD 168625 duplicated in both studies). This new detection rate with a three-fold sample is lower than the previous study, making the general physical conditions of LBVs not particularly suitable for the [Fe II] 1.644 $\mu$m line to arise but comparable to those of SNRs. The discrepancy in detection rates between this study and S02 might be due to the biased sample S02 used, which includes confirmed LBVs and candidate LBVs showing nebulosity in the Galaxy and the two most famous LBVs in LMC.

The [Fe II]-H images of identified LBV-IFOs are shown in Figure 10. Brief information about them is listed in Table 8. In the [Fe II] emission, all identified LBV-IFOs share an elliptical/circular morphology. This is similar to their morphologies at 8 $\mu$m but the extent appears smaller. We note that for G26.47+0.02 (IFO 102, 103) the south-eastern diffuse structure was noticed in the [Fe II]-H image. But the possibility of it being an artifact prevented us from assigning it as an IFO. The morphological coincidence of this South-East structure, IFO 102 and 103 with respect to the prominent part of the 8 $\mu$m nebula (Paron et al. 2012) implies the possibility of more extended, diffuse [Fe II] emission than seemingly identified. There are some new features revealed by [Fe II] emission: (1) IFO 65 - HD 168625 is located at the center of optical/IR elliptical structures that are broken toward the north-east. In the [Fe II] emission, we see a complete circular structure, the center of which is offset toward the northeast. (2) IFO 162 - [KW97] 37-17 shows multiple shells in [Fe II] emission, forming together a much brighter elliptical structure than those in 8 $\mu$m or optical. This possibly indicates that the LBV had several active erupting phases that manufactured bright [Fe II]-emitting shells one by one.

We found that all [Fe II]-detected LBVs in the UWIFE survey also accompany nebulosities at 8 $\mu$m, but not vice versa. For example, we could not detect [Fe II] emission in three LBVs with 8 $\mu$m nebulosity (HD 168607, AFGL 2298, and GAL 024.73+00.69). Thus, the question of whether the LBV nebula, on account of the preceding giant eruption, is a prerequisite for the [Fe II] emission remains unanswered (S02). More comprehensive LBV samples and constrained physical properties of LBVs are needed to understand the possible relationship between the existence of the [Fe II] 1.644 $\mu$m line in LBV nebula and their past eruption histories.

### 4.6 Supernova Remnants

SNRs are the brightest objects in [Fe II] emission. In SNRs this line is mostly emitted from cooling gas behind radiative shocks. [Fe II] lines are strong in shocked gas because Fe abundance could be enhanced by shocks owing to grain destruction (Dinerstein 1995; Nisini 2008; Koo et al. 2016 and references therein). Before the UWIFE survey, a dozen Galactic and LMC SNRs had been observed in the NIR [Fe II] lines. The SNRs that are bright in [Fe II] emission lines may be divided into two groups: (1) middle-aged SNRs interacting with dense molecular (or atomic) clouds such as W 44 (Reach et al. 2005), 3C 391 (Reach et al. 2002), and (2) young SNRs interacting with the dense circumstellar medium (CSM) such as Cas A (Koo et al. 2018), G11.2-0.3 (Moon et al. 2009), RCW 103 (Burton & Spyromilio 1993), and W49B (Lee et al. 2019). Then Lee et al. (2019) (hereafter, L19) searched for [Fe II] emission at the positions of the SNRs in the catalog of Green (2014) using the UWIFE survey data and detected [Fe II] emission features toward 19 SNRs, more than half of which were new detections. In external galaxies, [Fe II] emission is used as a tracer of SNRs (Blair et al. 2014; Bruursema et al. 2014; Long et al. 2020), although strong [Fe II] lines may originate from sources ionized by X-rays, e.g., in active galactic nuclei (Mouri et al. 2000; Morel et al. 2002).

We detected 25 IFOs associated with SNRs. All these SNR-IFOs belong to the 19 SNRs in Lee et al. (2019) except one (Table 9). It is worthwhile to point out that Lee et al. (2019) searched [Fe II] emission at 79 SNRs of the Green's catalog that are fully covered by the UWIFE survey. Four SNRs partially observed in the survey (i.e., G7.0−0.1, G13.3−1.3, G28.8+1.5, G38.7−1.3) were not investigated, and our unbiased search resulted in the identification of a small [Fe II]-emitting patch inside the region of G28.8+1.5. Meanwhile, the Green's catalog of Galactic SNRs has been updated (Green 2019), adding a new SNR G53.4+0.0 (partially covered in the UWIFE) and rejecting four (G20.4+0.1, G21.5-0.1, G23.6+0.3, G59.8+1.2 that were reclassified as H II regions) in the survey area. None of the new or rejected SNRs showed [Fe II] emission features. So the new detection rate for fully covered SNRs is 25 per cent (19/75). We note that Lee et al. (2019) compensated the [Fe II] line flux for the flux subtracted from the H-band by multiplying with 1.15, whereas the fluxes in Table 9 are observed fluxes. As presented in Lee et al. (2019), IFO 147 that matches W49B is the brightest SNR-IFO. The detailed results for the 19 SNRs can be found in Lee et al. (2019).

# 5 SUMMARY

We have presented the first comprehensive catalog of Galactic IFOs discovered in the UKIRT Widefield Infrared Survey for [Fe II] (UWIFE). It is the first Galactic catalog of extended [Fe II] line emission sources using an unbiased, large-scale survey. We have discovered many previously unreported [Fe II] 1.644 $\mu$m line sources. Therefore, this catalog provides an opportunity to broaden the horizons of the study of the shocked regions of our Galaxy, especially with the synergy of the UWISH2 survey.

We have searched for extended IFOs in the inner Galactic plane ($7° < l < 62°$; $|b| \lesssim 1°.5$). In order for the search to be efficient, we removed point-like continuum sources from the [Fe II] 1.644 $\mu$m images using $H$-band images taken as part of the UKIDSS GPS survey. We identified most of the IFOs by visual inspection and added several faint IFOs with an automatic source identification which uses the same source detection algorithm as in UWISH2 (Froebrich et al. 2015). In total, 204 IFOs were identified. We measured the sizes and fluxes of these 204 IFOs and presented their properties. We have searched for the counterparts of the IFOs via positional cross-matching with previously known sources and found that the majority of IFOs are associated with supernova remnants, young stellar objects, H II regions, planetary nebulae, and luminous blue variables. We group IFOs by their counterpart types and discuss their statistical and morphological properties. The main results are summarized as follows.

(1) In the 180 deg$^2$ Galactic plane area of the 1st Galactic quadrant covered by the UWIFE survey ($7° < l < 62°$; $|b| \lesssim 1°.5$), we identified 204 IFOs. The identified IFOs are classified according to their counterparts: IFOs associated with young stellar objects (YSO-IFOs), H II regions (HII-IFOs), compact H II regions (CHII-IFOs), planetary nebulae (PN-IFOs), luminous blue variables (LBV-IFOs), and supernova remnants (SNR-IFOs). There are 100 YSO-IFOs, 11 HII-IFOs, 22 CHII-IFOs, 17 PN-IFOs, 4 LBV-IFOs, and 25 SNR-IFOs. We could not identify counterparts for 25 IFOs, and they are classified as 'unknown-IFOs'. The majority of IFOs are new discoveries that have never been revealed in previous [Fe II] line studies.

(2) The SNR-IFOs and HII-IFOs are the brightest IFOs, and they dominate the [Fe II] 1.644 $\mu$m line flux in the Galactic plane. They contribute 96 per cent of the total [Fe II] 1.644 $\mu$m line flux of the IFOs ($2.6 \times 10^{-13}$ W m$^{-2}$); 76 per cent by SNR-IFOs and 20 per cent by HII/CHII-IFOs. The YSO-IFOs, PN-IFOs, and LBV-IFOs are generally orders of magnitude fainter, while the unknown-IFOs are the faintest.

(3) The average number density of IFOs is $\sim 1.1$ deg$^{-2}$. The number density is highly variable spatially, especially for the IFOs associated with objects in the early-evolutionary phase, e.g., IFOs associated with H II regions and YSOs. In Galactic longitude, there are prominent peaks at $l \sim 16°$ and $51°$, while there is a 'void' at $l \sim 40°-50°$ where the number of IFOs is very small. The spatial distribution in Galactic latitude is centered at $b = -0°.12$ with a standard deviation of $0°.65$.

(4) The results on the individual types of IFOs are summarized below.

(i) YSO-IFOs

We detected 100 YSO-IFOs, which constitutes half of the IFOs in our catalog. Only seventeen of those are associated with the RMS sources, which represent massive YSOs. The YSO-IFOs might be preferentially tracing low-mass YSOs. On the other hand, the majority (87 per cent) of YSO-IFOs are not associated with HH objects, suggesting that the YSO-IFOs are revealing previously hidden, optically obscured outflows in star-forming regions. YSO-IFOs have diverse morphologies, and we have classified them into four categories; bipolar, cometary, knot-like, and amorphous.

(ii) HII-IFOs and CHII-IFOs

We have identified 11 IFOs associated with 4 H II regions (Table 6). Almost all HII-IFOs are located in the well-known star-forming complexes, W 31, M 17, and W 51. Some HII-IFOs appear as either thin filaments or diffuse amorphous emission structures within the radio structure, so their association with the H II regions is very likely. But some are faint and diffuse and extend beyond the radio boundary of the H II regions, so their association is uncertain and needs to be confirmed. We also detected 22 IFOs associated with 16 CHIIs, including 6 previously reported (Table 5). Among the 16 CHII regions, ten are catalogued in CORNISH, which corresponds to 4 per cent of the 237 CHII regions in the CORNISH catalog in the survey area. CHII-IFOs have diverse morphologies: jet-like, shell-like, and amorphous.

(iii) PN-IFOs

We detected 17 PN-IFOs. They are associated with 14 PNe (i.e., 5 PNe, 8 PN candidates, and one post-AGB candidate; Table 7), which correspond to about 4.7 per cent of the PNe in the survey area. We have classified the morphologies of PN-IFOs following Parker et al. (2006) and compared them with those in H$\alpha$ and H$_2$. Some PNe have [Fe II] morphologies different from the H$\alpha$ and H$_2$ morphologies, which implies that the [Fe II] line reveals new substructures, possibly probing additional mass-loss histories. The physical sizes of some PN-IFOs are larger than 0.9 pc.

(iv) LBV-IFOs

We detected 4 LBV-IFOs. They are associated with 3 LBVs out of 22 LBVs and their candidates in the survey area (Table 8), so the detection rate of [Fe II] emission associated with LBVs in this study is 14 per cent. All LBV-IFOs share an elliptical or circular morphology. Some show multiple shells. We found that all [Fe II]-detected LBVs in the UWIFE survey also accompany nebulosity at 8 $\mu$m, but not vice versa.

(v) SNR-IFOs

We detected 25 SNR-IFOs. They are associated with 20 SNRs, which corresponds to 25 per cent of the 75 known SNRs in the survey area. The SNR-IFOs occupy 76 per cent of the total [Fe II] flux of IFOs, and the four brightest IFOs are SNR-IFOs. On the other hand, the lowest surface brightness IFOs are also SNR-IFOs, showing the patchy [Fe II] emission in SNRs. All SNRs with [Fe II] emission features except one (G28.8+1.5) have been previously reported by Lee et al. (2019). The detailed results on the [Fe II] emission on the 19 SNRs can be found in Lee et al. (2019).

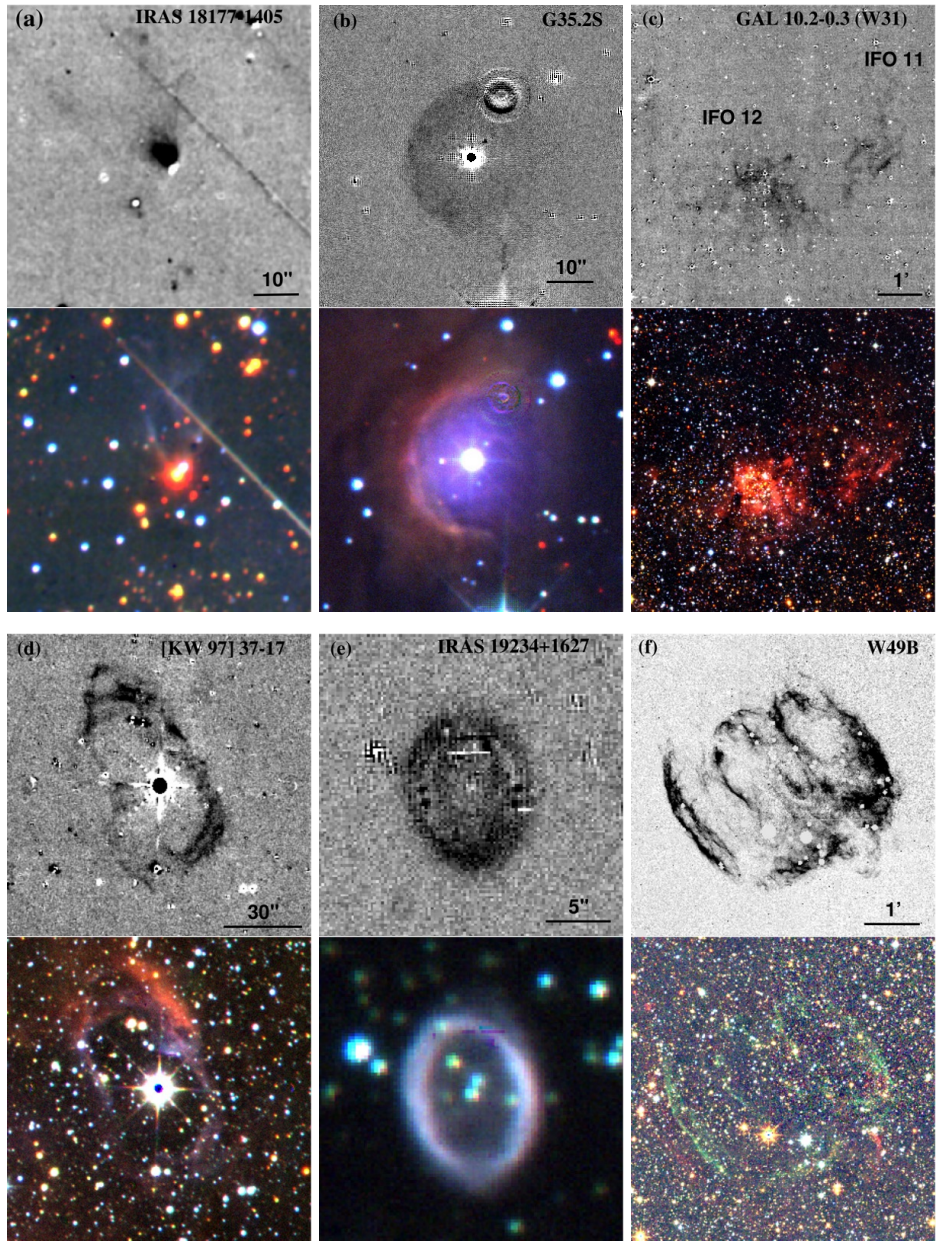

**Figure 1.** Continuum-subtracted [Fe II] and NIR three-color images of IFOs with various origins: (a) YSO outflow: IRAS 18177-1405; (b) compact H II region: G35.2S; (c) diffuse H II region: GAL 10.2-0.3; (d) LBV nebula: [KW 97] 37-17; (e) PN: IRAS 19234+1627; and (f) SNR: W49B. Grey-scale images in the upper rows are UWIFE [Fe II]-H images; Colour-composite images in the lower rows are R/G/B = *KHJ*-band images from the UKIDSS GPS survey. The units of the UWIFE [Fe II]-H images are DNs, with the darker colour denoting a higher DN. The UWIFE images of the panels (a) IRAS 18177−1405 and (d) [KW97] 37-17 are smoothed with a two-pixel Gaussian. In all images, North is at the top, and east to the left side. Note the following artifacts: panel (a) IRAS 18177−1405: diffraction spike from southwest to northeast; (b) G35.2S: crosstalk on the northwest edge of the source; diffraction spikes and an airy disk at the south; (e) IRAS 19234+1627: dead pixels on the north and southwestern part at the boundary of the source; (f) W49B: masked bright stars.

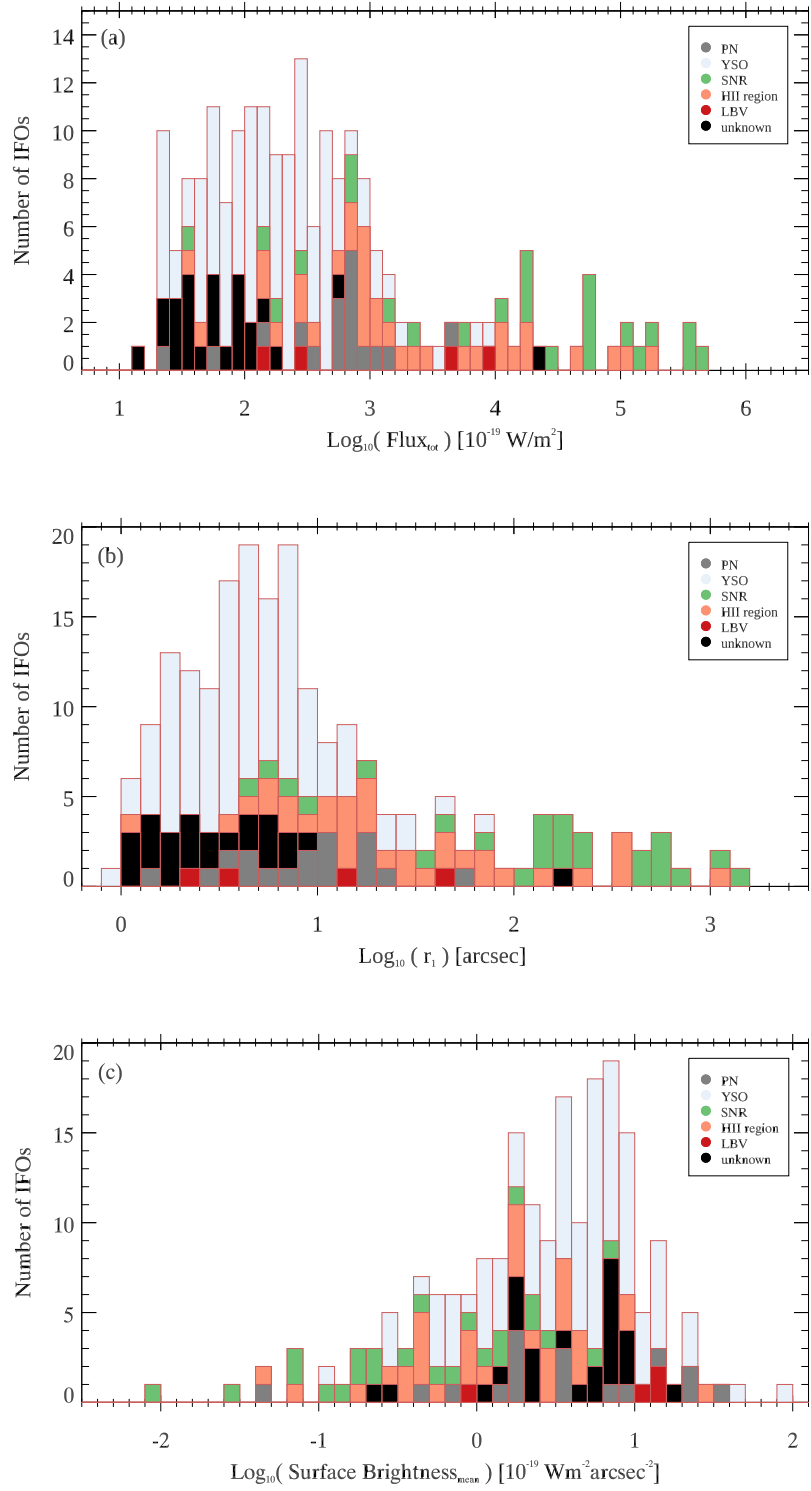

**Figure 2.** (a) $F_{tot}$ distribution of IFOs. Note that the flux of a large-scale IFO 7 is excluded in this figure. (b) Semi-major axis $r_1$ distribution of IFOs. The semi-major axis of automatically-identified IFOs is the best estimate of the coordinate, semi-major, and minor axes from the best-fitting ellipse from IDL procedure 2dgaussfit. (c) Surface brightness distribution of IFOs. The counterparts of IFOs are presented in different colors.

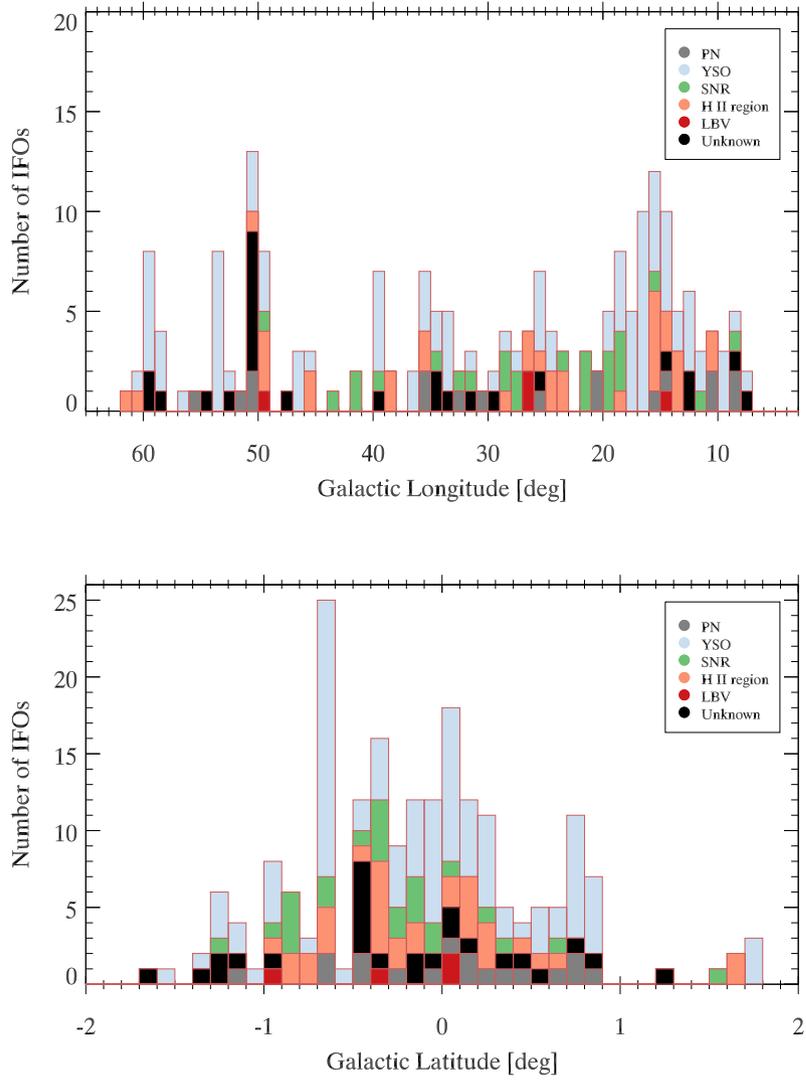

**Figure 3.** The spatial distribution of IFOs in Galactic longitude and latitude. The top panel shows the distribution of IFOs in Galactic longitude. The bottom panel shows the distribution of IFOs in Galactic latitude. IFOs are shown in accordance with their counterparts: YSO, H II region, PN, SNR, LBV, and unknown-IFOs.

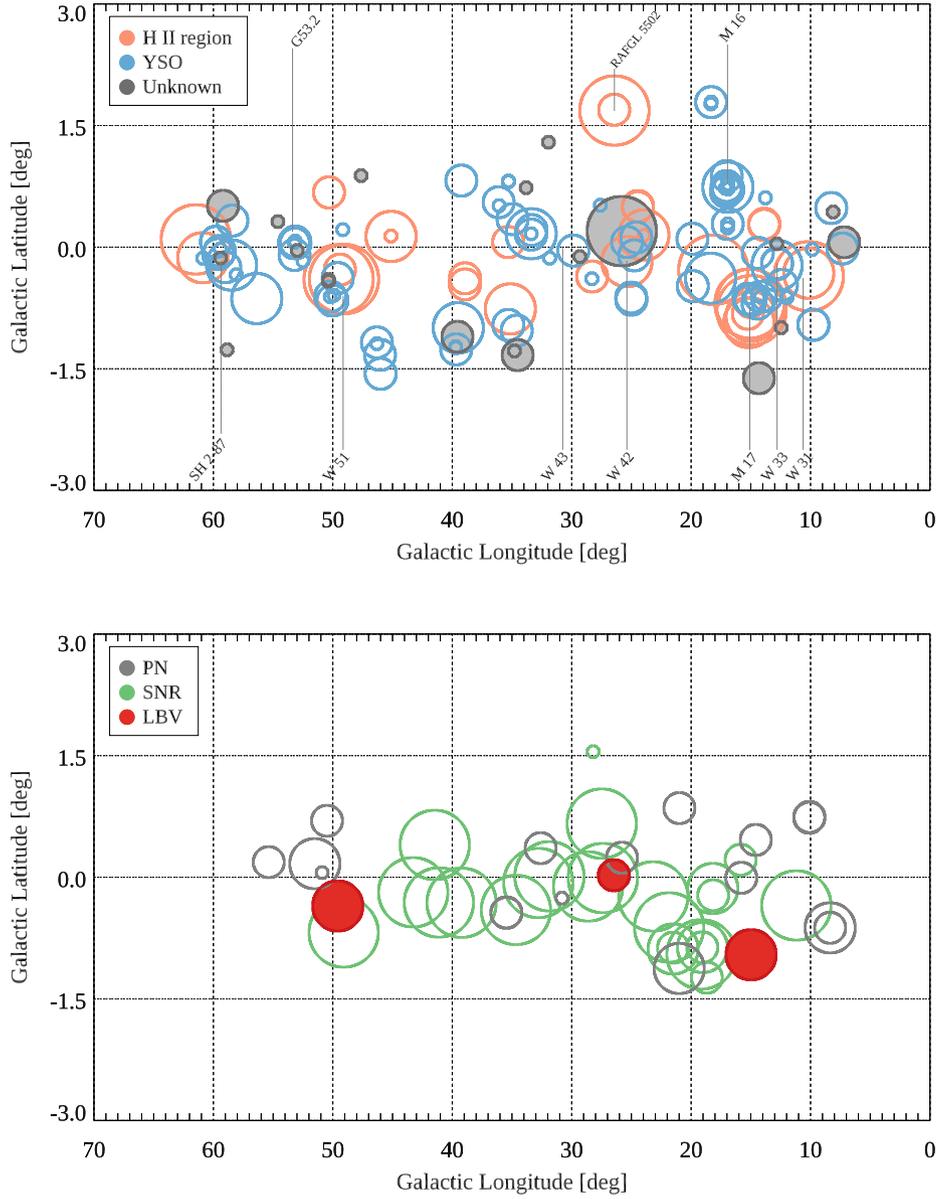

**Figure 4.** Two-dimensional distribution of IFOs. The top panel shows the spatial distribution of IFOs having counterparts in the H II region (orange), YSO (blue), and unknown (gray) categories. Each circle represents an IFO, and the size of each circle is proportional to its logarithmic $F_{tot}$ (in order of $10^{-17}$, $10^{-16}$, $10^{-15}$ W m$^{-2}$). Star-forming regions whose positions match those of IFOs in the distribution are labelled. Due to clustered IFOs, many circles overlap. The bottom panel shows the spatial distribution of IFOs with counterparts of PNe (gray), SNRs (green), and LBVs (red). Note that the flux of IFO 7 (i.e., SNR G8.7-0.1) is not provided, therefore excluded here.

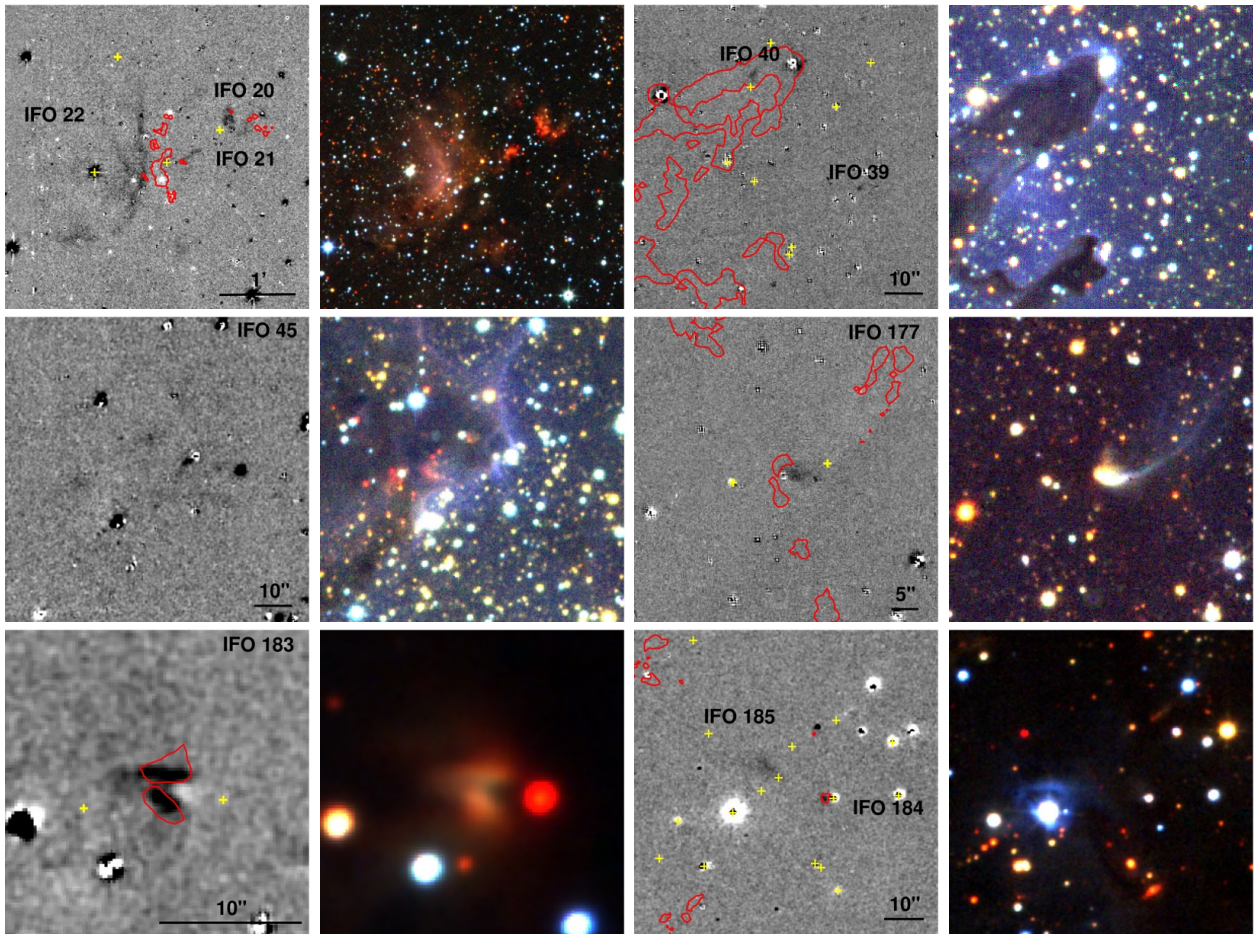

**Figure 5.** IFOs with YSO counterpart candidates in continuum-subtracted [Fe II] images as in Fig. 1. Only six representative IFOs are shown. The yellow crosses denote adjacent YSOs in the field of view, while the red contours are H$_2$ 2.12 $\mu$m emission contours adopted from UWISH2. The right frames are three-color *KHJ* UKIDSS images of the same field of view. This figure is available in its entirety in the Appendix B.1.

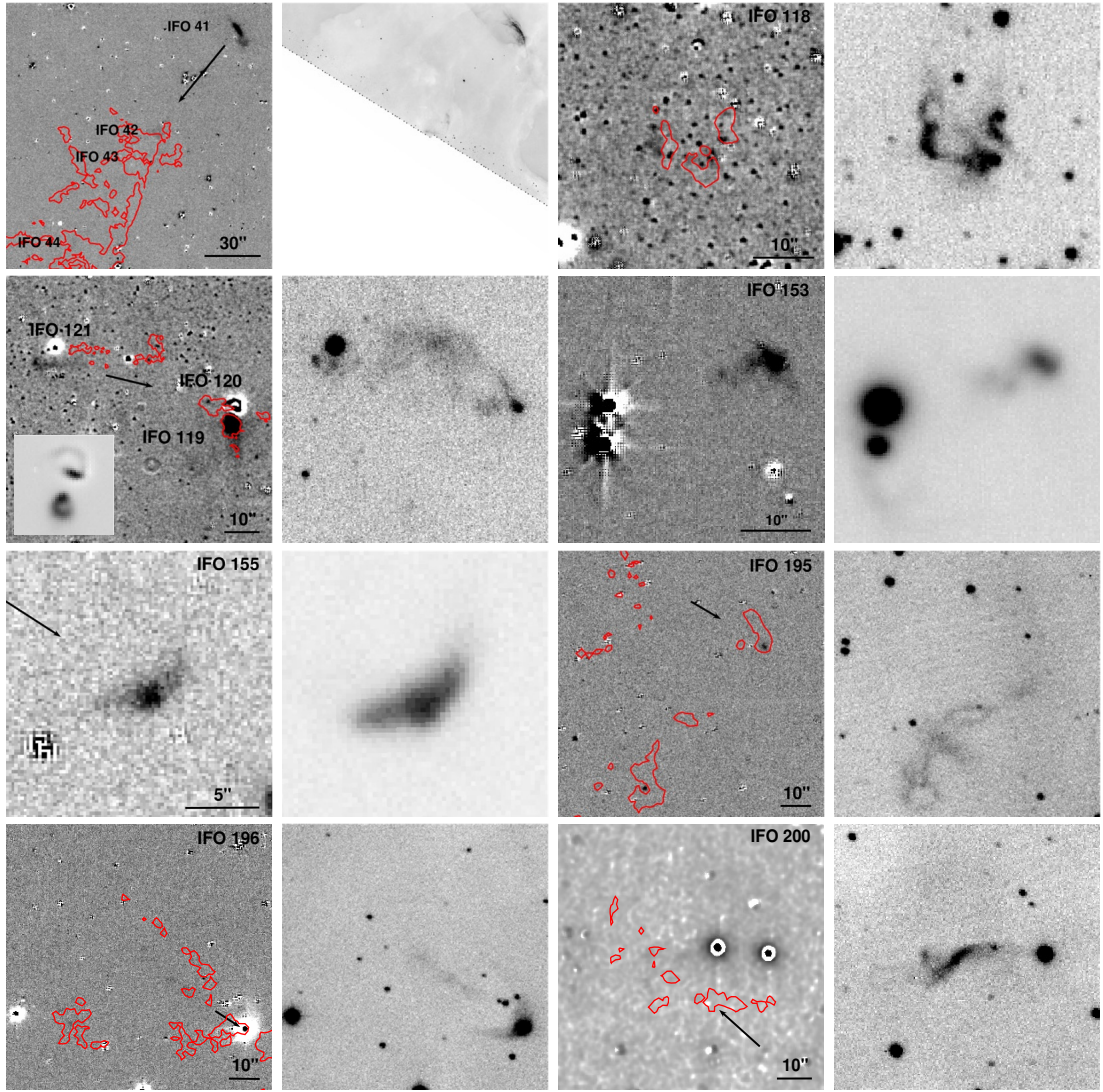

**Figure 6.** IFOs with HH counterpart candidates. The left panels show continuum-subtracted [Fe II] images as in Fig. 1. The right panels show H$\alpha$ in the same field of view. IPHAS images were used except for IFO 41–44 where the *Hubble Space Telescope (HST)* F657N image was used. Inset on the IFO 119–121 is a magnified [Fe II]-H image of the saturated star (West of IFO 120, North in the inset) and bright part of IFO 119 (South in the inset). Red contours are H$_2$ 2.12 $\mu$m emission adopted from UWISH2. The black arrow points to the driving source of the HH object. When the driving source is out of the image field of view, the arrow points from the driving source to the IFO.

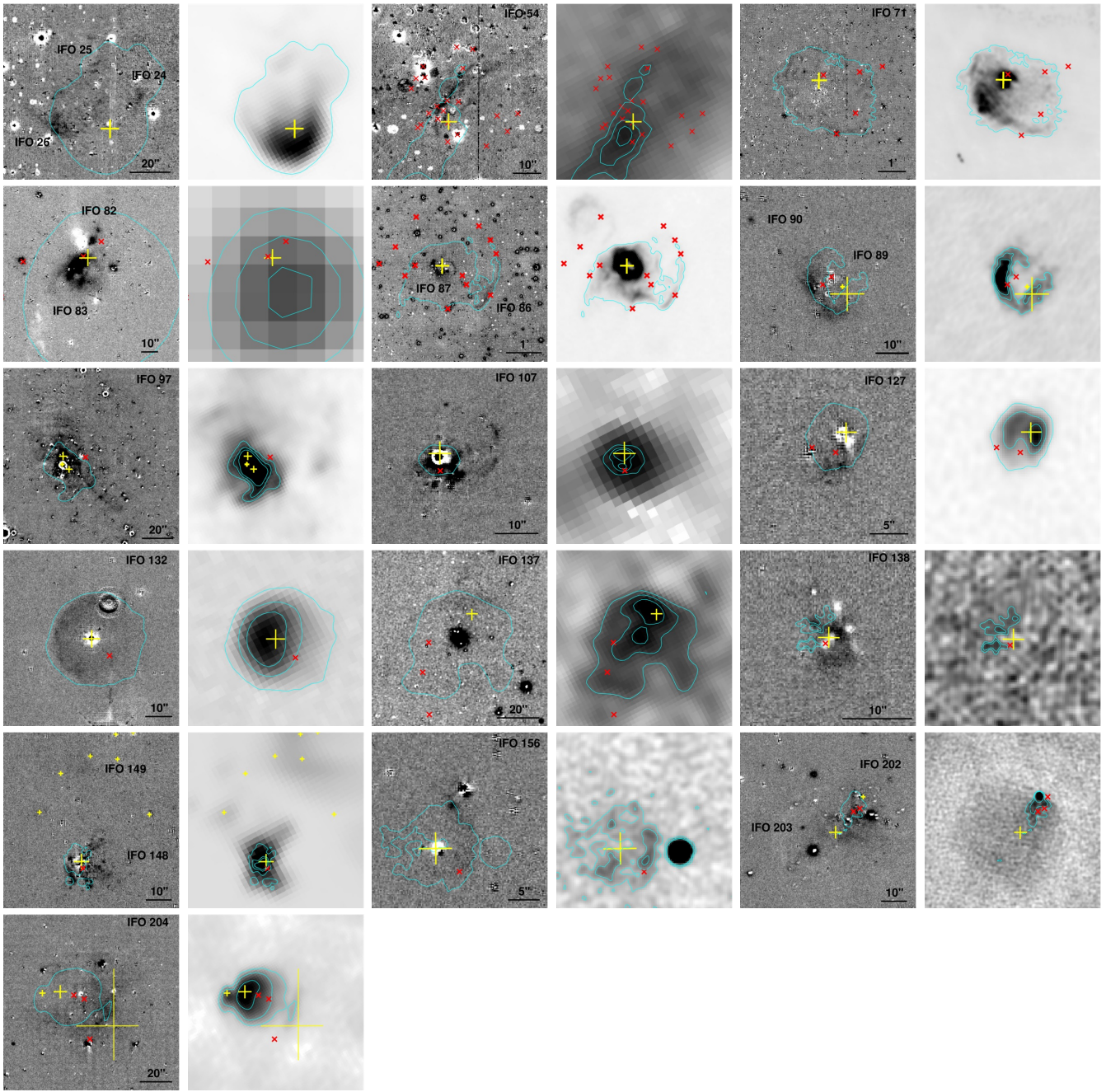

**Figure 7.** IFOs with CHII counterpart candidates. The left panels show continuum-subtracted [Fe II] images as in Fig. 1. The right panels show a radio continuum in the same field of view. Cyan contours on both images are the boundaries of CHIIs in the radio. The contours of IFO 24–26, 54, 71, 86–87, 97, 132, and 137 are from New-GPS 20 cm, IFO 89–90, 107, 127, 138, 148–149, 156, and 202–203 are from CORNISH 5 GHz, and IFO 82–83 are from the National Radio Astronomy Observatory (NRAO) Very Large Array (VLA) Sky Survey (NVSS). Only the radio image and contour of IFO 204 are from the old-GPS 20 cm. IFO 203 is a YSO-IFO inside the field of view. The yellow crosses in both panels are the same as in Fig. 5. The red cross shows the central position of the UWISH2 $H_2$ emission.

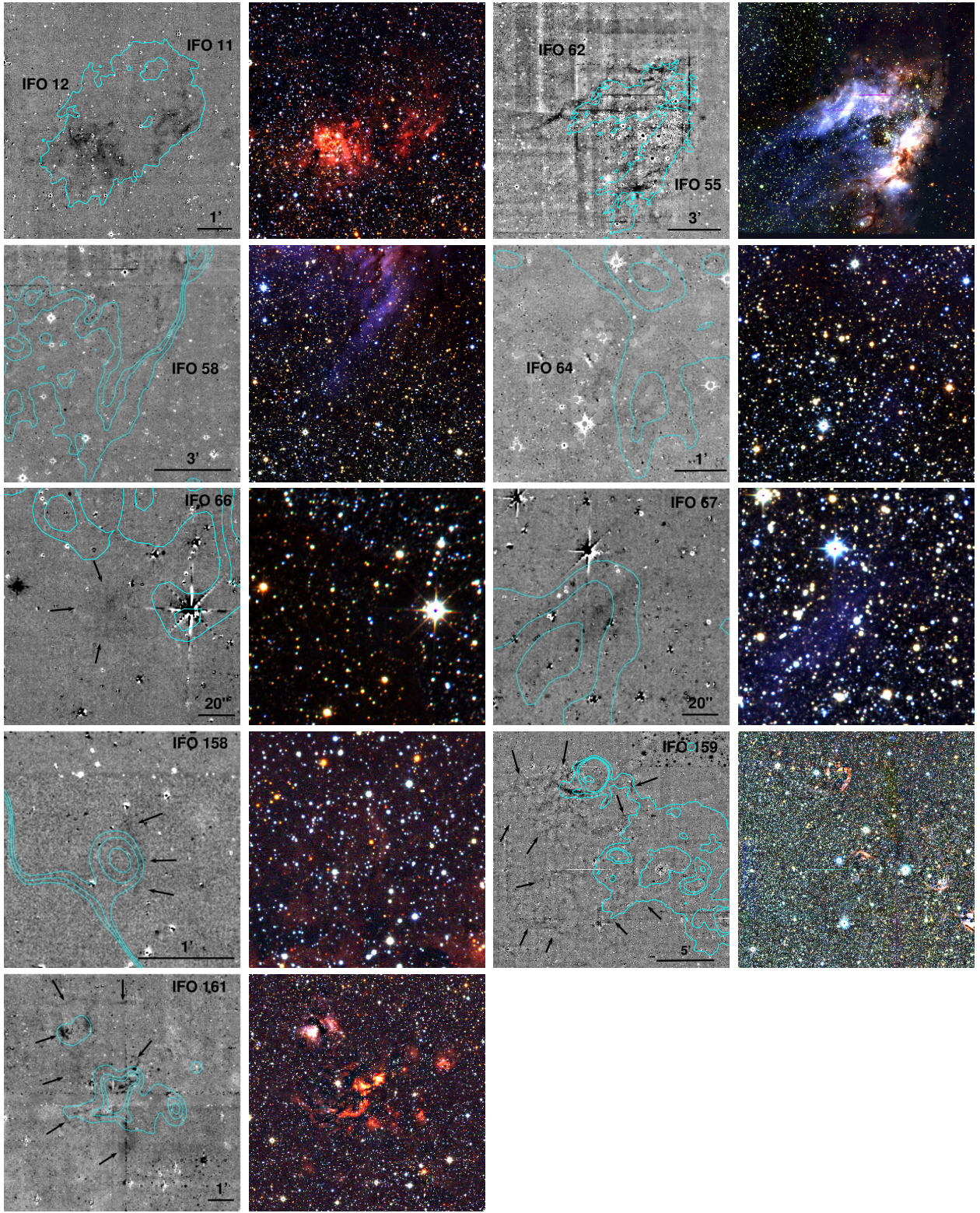

**Figure 8.** IFOs with H II region counterpart candidates in continuum-subtracted [Fe II] images as in Fig. 1. Cyan contours are boundaries of H II regions in the radio continuum: IFO 11, 12, 55, 62 with New-GPS 20 cm data; IFO 58, 64, 66, 67 with GPS 90 cm data; 158, 159, 161 with THOR 1420 MHz continuum + VLA Galactic Plane Survey (VGPS) H I data. Black arrows point to the boundaries of IFO structures. The format for these images is the same as that of Fig. 1.

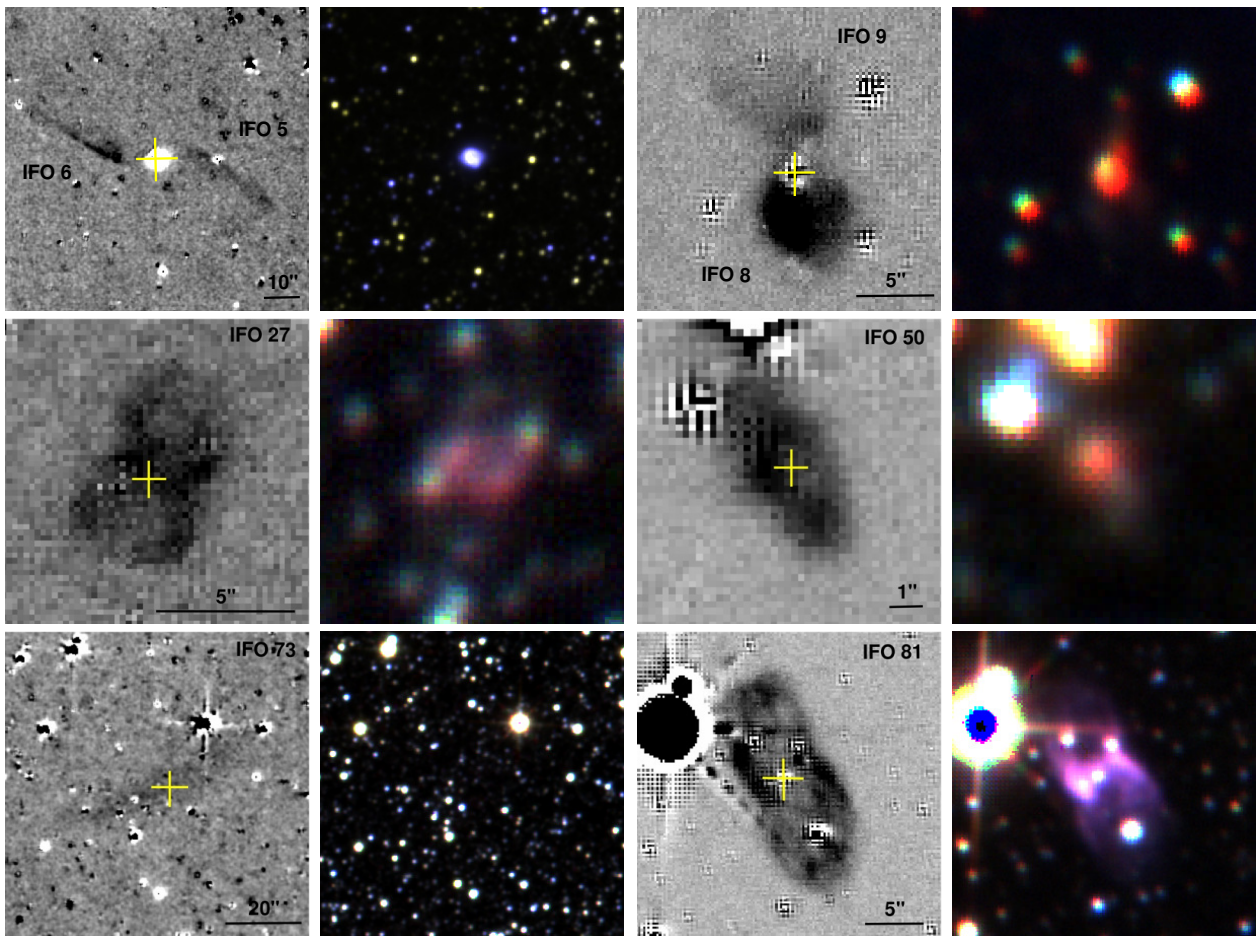

**Figure 9.** Continuum-subtracted [Fe II] images of IFOs with PN counterpart candidates. Only six representative IFOs are shown. The units on the UWIFE [Fe II]-H are DNs, with the darker colour denoting a higher DN. The right frames are three-color *KHJ* UKIDSS images of the same field of view. The corresponding source names for each IFO are shown. The yellow cross marks the central position of the counterpart. The images of IFO 73, 95, 112, 129, 157, and 164 are smoothed with a two-pixel Gaussian. In all images, North is at the top and east is on the left side. This figure is available in its entirety in Appendix B.2.

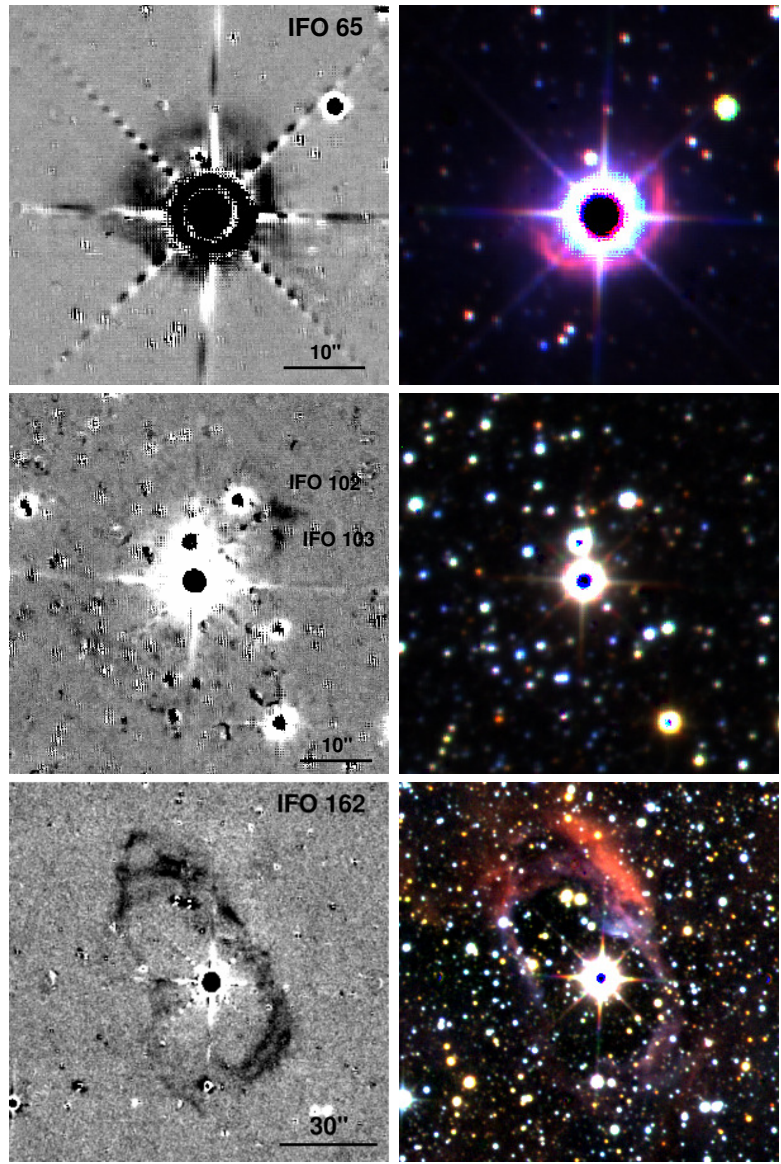

**Figure 10.** Continuum-subtracted [Fe II] images of IFOs having LBV counterparts. The right frames are three-color *KHJ* UKIDSS images of the same field of view. Note that there is a spike pattern around a bright star, coincident with IFO 65. The format for these images is the same as that of Fig. 1.

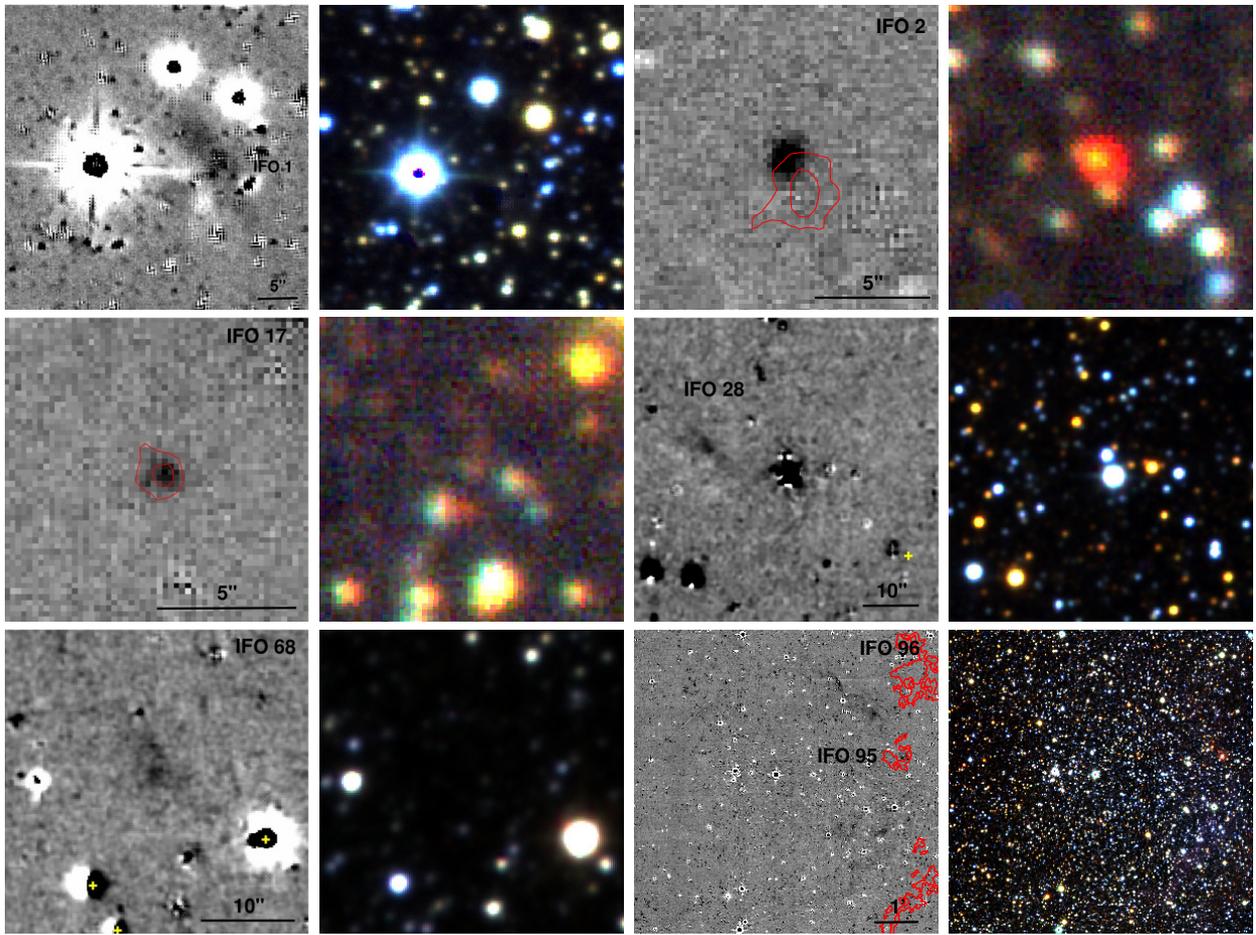

**Figure 11.** IFOs with counterpart candidates unknown in continuum-subtracted [Fe II] images as in Fig. 1. Only six representative IFOs are presented. The right frames are three-color *KHJ* UKIDSS images of the same field of view. The format for these images is the same as that of Fig. 1. This figure is available in its entirety in Appendix B.3.

Table 1. Catalog of identified IFOs

| IFO # | UWIFE designation | $l$ [deg] | $b$ [deg] | $r_1$ [arcsec] | $r_2$ [arcsec] | PA [deg] | Area [arcsec$^2$] | $F_{tot}$ [10$^{-17}$ W m$^{-2}$] | Counterpart |
|---|---|---|---|---|---|---|---|---|---|
| IFO 001[a] | J180136.927−224228.03 | 7.19190 | +0.06028 | 9.3 | 2.9 | 40 | 84.7 | 5.88 | - |
| IFO 002 | J180210.565−214326.69 | 8.11119 | +0.43376 | 2.3 | 2.0 | 0 | 14.4 | 0.89 | - |
| IFO 003[a] | J180212.398−223720.49 | 7.33358 | −0.01604 | 2.6 | 1.8 | 157 | 14.7 | 2.25 | YSO |
| IFO 004 | J180219.380−213350.56 | 8.26721 | +0.48309 | 5.4 | 4.2 | 120 | 71.2 | 4.27 | YSO |
| IFO 005 | J180511.698−195040.56 | 10.09481 | +0.74344 | 18.0 | 7.2 | 60 | 407.1 | 3.15 | PN |
| IFO 006 | J180514.644−195027.77 | 10.10356 | +0.73511 | 16.0 | 7.3 | 60 | 366.9 | 6.04 | PN |
| IFO 007[b] | J180627.378−213227.20 | 8.75903 | −0.34293 | 1400.0 | 1400.0 | 0 | 6157521.8 | - | SNR |
| IFO 008 | J180640.397−220136.63 | 8.35934 | −0.62390 | 4.5 | 4.2 | 20 | 59.3 | 13.80 | PN |
| IFO 009 | J180640.585−220126.95 | 8.36205 | −0.62322 | 6.1 | 4.9 | 25 | 93.9 | 6.10 | PN |
| IFO 010 | J180732.633−202606.06 | 9.84849 | −0.02588 | 3.4 | 3.4 | 0 | 36.3 | 0.91 | YSO |
| IFO 011 | J180916.373−201852.96 | 10.15033 | −0.32179 | 68.0 | 39.0 | 150 | 8331.5 | 72.10 | HII |
| IFO 012 | J180925.793−201934.10 | 10.15813 | −0.35954 | 67.3 | 62.5 | 0 | 13214.3 | 166.00 | HII |
| IFO 013 | J181050.844−205738.76 | 9.76257 | −0.95642 | 4.2 | 3.4 | 160 | 44.8 | 1.59 | YSO |
| IFO 014 | J181051.015−205748.39 | 9.76056 | −0.95829 | 5.9 | 3.8 | 170 | 70.4 | 2.81 | YSO |
| IFO 015 | J181129.632−192515.52 | 11.18511 | −0.34766 | 131.4 | 131.1 | 0 | 54118.7 | 1090.00 | SNR |
| IFO 016 | J181312.424−164111.54 | 13.77907 | +0.60864 | 3.5 | 2.8 | 90 | 30.7 | 0.94 | YSO |
| IFO 017 | J181322.096−174758.22 | 12.82064 | +0.04152 | 1.3 | 1.1 | 0 | 5.3 | 0.35 | - |
| IFO 018 | J181407.762−185101.73 | 11.98433 | −0.62012 | 3.3 | 2.1 | 110 | 21.7 | 0.64 | YSO |
| IFO 019 | J181408.073−185058.01 | 11.98582 | −0.62071 | 1.7 | 1.5 | 60 | 8.0 | 0.44 | YSO |
| IFO 020 | J181413.148−175528.03 | 12.80784 | −0.19604 | 8.3 | 7.4 | 90 | 192.9 | 4.79 | YSO |
| IFO 021 | J181415.121−175557.61 | 12.80436 | −0.20684 | 10.7 | 8.3 | 120 | 279.0 | 1.57 | YSO |
| IFO 022 | J181419.929−175616.05 | 12.80898 | −0.22602 | 67.6 | 42.7 | 160 | 9068.2 | 69.50 | YSO |
| IFO 023 | J181422.928−182508.51 | 12.39193 | −0.46647 | 3.0 | 2.7 | 60 | 25.4 | 1.95 | YSO |
| IFO 024 | J181434.822−164514.38 | 13.87718 | +0.28764 | 8.0 | 4.0 | 160 | 100.3 | 3.18 | UCHII |
| IFO 025[b] | J181436.683−164507.70 | 13.88234 | +0.28200 | 24.0 | 20.0 | 0 | 1507.9 | 6.60 | UCHII |
| IFO 026[b] | J181437.731−164526.78 | 13.87970 | +0.27580 | 13.0 | 8.0 | 60 | 326.7 | 9.40 | UCHII |
| IFO 027 | J181521.196−160255.94 | 14.58529 | +0.46128 | 4.5 | 3.9 | 160 | 55.1 | 7.16 | PN |
| IFO 028 | J181627.693−183653.67 | 12.45471 | −0.99317 | 7.6 | 4.3 | 50 | 102.6 | 0.30 | - |
| IFO 029 | J181658.050−162710.24 | 14.41440 | −0.07172 | 3.8 | 3.5 | 90 | 41.7 | 4.66 | YSO |
| IFO 030 | J181724.551−172216.03 | 13.65638 | −0.60071 | 4.0 | 3.1 | 40 | 38.9 | 2.01 | YSO |
| IFO 031 | J181750.609−120805.80 | 18.31683 | +1.78972 | 1.7 | 1.4 | 90 | 7.4 | 0.43 | YSO |
| IFO 032 | J181750.953−120802.39 | 18.31835 | +1.78895 | 2.7 | 2.5 | 0 | 21.2 | 2.11 | YSO |
| IFO 033 | J181758.445−120724.48 | 18.34208 | +1.76705 | 2.3 | 2.0 | 90 | 14.45 | 0.98 | YSO |
| IFO 034 | J181828.251−165525.05 | 14.17066 | −0.61193 | 2.0 | 1.3 | 0 | 8.1 | 0.54 | YSO |
| IFO 035 | J181828.590−165523.72 | 14.17162 | −0.61295 | 3.1 | 2.4 | 120 | 23.3 | 4.89 | YSO |
| IFO 036 | J181837.058−134248.28 | 17.01527 | +0.87651 | 6.5 | 4.1 | 100 | 83.7 | 1.51 | YSO |
| IFO 037 | J181839.539−134237.23 | 17.02271 | +0.86910 | 2.8 | 2.5 | 120 | 21.9 | 0.83 | YSO |
| IFO 038 | J181845.167−150257.21 | 15.85388 | +0.21557 | 6.4 | 5.1 | 130 | 102.5 | 7.34 | SNR* |
| IFO 039 | J181847.449−135022.70 | 16.92393 | +0.77973 | 1.2 | 1.1 | 0 | 4.5 | 0.23 | YSO |
| IFO 040 | J181849.365−134952.55 | 16.93498 | +0.77687 | 2.0 | 1.5 | 120 | 9.4 | 0.49 | YSO |
| IFO 041 | J181855.428−135145.51 | 16.91893 | +0.74041 | 10.2 | 5.6 | 30 | 179.4 | 16.80 | HH |
| IFO 042 | J181858.301−135236.39 | 16.91199 | +0.72350 | 10.0 | 6.0 | 150 | 188.4 | 2.65 | HH |
| IFO 043 | J181858.835−135252.81 | 16.90897 | +0.71943 | 4.0 | 3.0 | 140 | 37.6 | 0.23 | HH |
| IFO 044 | J181901.895−135346.30 | 16.90173 | +0.70150 | 15.0 | 10.0 | 140 | 471.2 | 2.52 | HH |
| IFO 045 | J181905.871−134522.91 | 17.03256 | +0.75343 | 25.1 | 19.0 | 90 | 1498.2 | 4.14 | YSO |
| IFO 046[a] | J181914.708−164949.13 | 14.34049 | −0.73101 | 1.6 | 1.3 | 30 | 6.5 | 0.33 | YSO |
| IFO 047 | J181917.916−164355.78 | 14.43321 | −0.69611 | 4.0 | 3.0 | 90 | 37.6 | 2.24 | YSO |
| IFO 048 | J181922.591−134114.45 | 17.12557 | +0.72624 | 7.1 | 3.0 | 30 | 66.9 | 0.70 | YSO |
| IFO 049 | J181925.259−134542.71 | 17.06480 | +0.68168 | 5.4 | 4.0 | 0 | 67.8 | 1.01 | YSO |
| IFO 050 | J181927.118−151211.16 | 15.79925 | −0.00624 | 3.5 | 2.3 | 30 | 25.2 | 8.84 | PN |
| IFO 051[b] | J182019.871−161031.33 | 15.04081 | −0.65134 | 2.2 | 2.0 | 0 | 13.8 | 1.33 | YSO |
| IFO 052[b] | J182020.767−161018.45 | 15.04566 | −0.65282 | 2.8 | 2.0 | 0 | 17.5 | 4.05 | YSO |
| IFO 053[b] | J182021.725−161015.05 | 15.04831 | −0.65575 | 2.3 | 1.7 | 30 | 12.2 | 1.05 | YSO |
| IFO 054[b] | J182024.436−161126.80 | 15.03583 | −0.67472 | 18.7 | 9.4 | 140 | 552.2 | 5.11 | HCHII |
| IFO 055[b] | J182028.170−161245.10 | 15.02369 | −0.69815 | 250.0 | 100.0 | 130 | 78539.8 | 1257.00 | HII |
| IFO 056 | J182032.784−160124.98 | 15.19905 | −0.62538 | 7.6 | 3.9 | 0 | 93.1 | 0.63 | YSO |
| IFO 057 | J182034.306−160158.97 | 15.19359 | −0.63521 | 8.0 | 2.0 | 120 | 50.2 | 1.12 | YSO |
| IFO 058 | J182035.196−161942.63 | 14.93464 | −0.77759 | 330.0 | 120.0 | 160 | 124407.0 | 473.00 | HII |
| IFO 059 | J182035.224−140436.84 | 16.92057 | +0.28355 | 1.9 | 1.8 | 90 | 10.7 | 0.85 | YSO |
| IFO 060 | J182035.656−140409.72 | 16.92803 | +0.28556 | 5.8 | 5.1 | 0 | 92.9 | 7.82 | YSO |
| IFO 061 | J182036.014−140344.82 | 16.93481 | +0.28754 | 3.3 | 3.3 | 0 | 34.2 | 0.54 | YSO |
| IFO 062[b] | J182037.224−160828.36 | 15.10369 | −0.69649 | 341.8 | 197.3 | 115 | 211860.0 | 875.00 | HII |
| IFO 063 | J182049.312−140353.86 | 16.95793 | +0.23896 | 2.6 | 2.4 | 30 | 19.6 | 0.60 | YSO |
| IFO 064 | J182056.997−161934.88 | 14.97762 | −0.85347 | 100.0 | 50.0 | 40 | 15707.9 | 58.27 | HII |
| IFO 065[b] | J182119.587−162224.78 | 14.97848 | −0.95541 | 15.0 | 13.5 | 140 | 636.2 | 81.16 | LBV |
| IFO 066 | J182121.701−160424.16 | 15.24737 | −0.82163 | 18.2 | 16.2 | 0 | 926.2 | 8.96 | HII |
| IFO 067 | J182134.867−161209.80 | 15.15799 | −0.92897 | 25.5 | 8.1 | 135 | 648.8 | 11.40 | HII |
| IFO 068 | J182228.548−171548.32 | 14.32188 | −1.61616 | 3.5 | 2.3 | 30 | 25.2 | 1.07 | - |
| IFO 069 | J182432.827−130950.81 | 18.17823 | −0.13740 | 101.0 | 56.0 | 70 | 17768.8 | 21.50 | SNR |
| IFO 070 | J182448.037−131345.15 | 18.14955 | −0.22238 | 6.0 | 5.0 | 90 | 94.2 | 1.38 | SNR |
| IFO 071 | J182459.449−131552.08 | 18.14002 | −0.27977 | 130.0 | 90.0 | 60 | 36756.6 | 110.10 | UCHII |
| IFO 072[a] | J182548.520−130629.85 | 18.37107 | −0.38288 | 7.0 | 4.5 | 120 | 98.9 | 82.00 | YSO |

Table 1 (cont'd)

| IFO # | UWIFE designation | $l$ | $b$ | $r_1$ | $r_2$ | PA | Area | $F_{tot}$ | Counterpart |
|---|---|---|---|---|---|---|---|---|---|
| | | [deg] | | [arcsec] | | [deg] | [arcsec$^2$] | [10$^{-17}$ W m$^{-2}$] | |
| IFO 073 | J182619.105−101318.67 | 20.98261 | +0.85302 | 19.9 | 7.4 | 120 | 462.6 | 6.64 | PN |
| IFO 074 | J182656.992−113210.92 | 19.89167 | +0.10337 | 4.2 | 3.4 | 60 | 44.8 | 4.51 | YSO |
| IFO 075 | J182851.018−124415.55 | 19.04409 | −0.86640 | 67.4 | 27.4 | 15 | 5801.7 | 14.10 | SNR |
| IFO 076 | J182852.671−124351.10 | 19.06306 | −0.86404 | 16.4 | 5.5 | 0 | 283.3 | 2.55 | SNR |
| IFO 077[a] | J182859.486−115026.08 | 19.85474 | −0.48056 | 6.1 | 2.3 | 80 | 44.0 | 3.45 | YSO |
| IFO 078 | J182919.659−124153.90 | 19.13295 | −0.95127 | 132.0 | 89.4 | 140 | 37073.3 | 192.00 | SNR |
| IFO 079 | J182930.563−131350.84 | 18.68160 | −1.23735 | 8.7 | 3.6 | 70 | 98.3 | 1.89 | SNR |
| IFO 080[b] | J183314.281−100831.20 | 21.84283 | −0.61852 | 500.0 | 200.0 | 40 | 314159.2 | 627.00 | SNR |
| IFO 081[b] | J183328.975−110726.68 | 20.99907 | −1.12478 | 11.8 | 6.4 | 25 | 237.2 | 49.30 | PN |
| IFO 082[a] | J183330.115−050050.30 | 26.43603 | +1.69464 | 6.8 | 3.5 | 165 | 74.7 | 6.73 | UCHII |
| IFO 083[a] | J183330.673−050110.94 | 26.42214 | +1.68483 | 15.4 | 12.0 | 114 | 580.5 | 150.00 | UCHII |
| IFO 084 | J183331.327−103257.93 | 21.51352 | −0.86841 | 35.0 | 17.0 | 0 | 1869.2 | 6.59 | SNR |
| IFO 085 | J183333.430−103402.86 | 21.50149 | −0.88437 | 44.8 | 42.1 | 90 | 5925.2 | 59.70 | SNR |
| IFO 086 | J183404.384−071820.28 | 24.45479 | +0.50637 | 60.0 | 30.0 | 160 | 5654.8 | 9.42 | UCHII |
| IFO 087 | J183408.045−071801.82 | 24.46632 | +0.49530 | 12.7 | 7.6 | 30 | 303.2 | 1.72 | UCHII |
| IFO 088 | J183420.390−084722.27 | 23.16829 | −0.23606 | 600.0 | 400.0 | 10 | 753982.2 | 584.00 | SNR |
| IFO 089 | J183425.284−075448.33 | 23.95514 | +0.14969 | 11.0 | 9.3 | 20 | 321.3 | 11.20 | UCHII |
| IFO 090 | J183426.772−075428.56 | 23.96285 | +0.14677 | 1.1 | 1.1 | 0 | 3.8 | 0.32 | UCHII |
| IFO 091 | J183541.856−072203.61 | 24.58525 | +0.12015 | 15.0 | 8.0 | 140 | 376.9 | 3.06 | YSO |
| IFO 092 | J183648.912−071850.94 | 24.76014 | −0.10123 | 2.3 | 1.9 | 20 | 13.7 | 2.13 | YSO |
| IFO 093 | J183716.440−032958.39 | 28.20160 | +1.54899 | 4.0 | 2.3 | 10 | 28.9 | 0.39 | SNR |
| IFO 094 | J183720.713−064200.69 | 25.36590 | +0.06398 | 2.0 | 1.7 | 30 | 10.6 | 1.18 | YSO |
| IFO 095 | J183730.398−061412.49 | 25.79595 | +0.24115 | 8.0 | 8.0 | 0 | 201.0 | 3.93 | PN/UCHII |
| IFO 096[b] | J183740.829−061452.41 | 25.80594 | +0.19768 | 180.0 | 180.0 | 0 | 101787.6 | 221.00 | - |
| IFO 097 | J183813.600−064815.32 | 25.37390 | −0.17819 | 25.2 | 8.0 | 45 | 633.3 | 25.60 | UCHII |
| IFO 098[b] | J183907.168−043230.84 | 27.48618 | +0.66204 | 500.0 | 400.0 | 20 | 628318.5 | 167.00 | SNR |
| IFO 099 | J183909.562−071927.89 | 25.01779 | −0.62238 | 4.0 | 3.0 | 0 | 37.6 | 2.05 | YSO |
| IFO 100 | J183911.798−072019.31 | 25.00933 | −0.63714 | 3.4 | 2.8 | 90 | 29.9 | 2.56 | YSO |
| IFO 101 | J183913.302−072057.12 | 25.00284 | −0.64748 | 1.8 | 1.5 | 130 | 8.4 | 1.12 | YSO |
| IFO 102 | J183931.338−054409.74 | 26.47082 | +0.02555 | 3.3 | 2.3 | 80 | 23.8 | 3.07 | LBV |
| IFO 103 | J183931.437−054414.64 | 26.46980 | +0.02456 | 2.4 | 1.4 | 0 | 10.5 | 1.27 | LBV |
| IFO 104 | J183950.426−043037.60 | 27.59648 | +0.51675 | 4.5 | 2.0 | 0 | 28.2 | 0.20 | YSO |
| IFO 105 | J184120.333−045606.47 | 27.38989 | −0.00960 | 150.0 | 120.0 | 70 | 56548.6 | 274.00 | SNR |
| IFO 106 | J184358.299−035306.11 | 28.62385 | −0.11304 | 216.2 | 170.1 | 0 | 115534.0 | 197.00 | SNR |
| IFO 107 | J184414.391−041754.32 | 28.28667 | −0.36139 | 7.0 | 3.6 | 165 | 79.1 | 2.70 | UCHII |
| IFO 108 | J184422.810−041734.78 | 28.30748 | −0.39003 | 3.9 | 2.8 | 90 | 34.3 | 0.57 | YSO |
| IFO 109 | J184501.647−001716.48 | 31.94472 | +1.29493 | 2.6 | 2.1 | 90 | 17.1 | 0.37 | - |
| IFO 110 | J184515.462−031604.01 | 29.31952 | −0.11656 | 4.3 | 3.5 | 120 | 47.2 | 0.85 | - |
| IFO 111 | J184559.282−024502.58 | 29.86281 | −0.04271 | 7.3 | 3.7 | 135 | 84.8 | 5.47 | YSO |
| IFO 112[a] | J184829.526−021003.18 | 30.81541 | −0.25716 | 2.9 | 1.5 | 150 | 13.6 | 0.52 | PN |
| IFO 113 | J184927.068−005638.37 | 31.86530 | +0.01156 | 240.0 | 210.0 | 120 | 158336.2 | 3423.55 | SNR |
| IFO 114[ab] | J184933.121−003810.21 | 32.59821 | +0.35877 | 8.9 | 6.3 | 156 | 176.1 | 6.46 | PN/HII |
| IFO 115 | J184955.670−010153.39 | 31.84176 | −0.13438 | 1.9 | 0.9 | 130 | 5.3 | 0.25 | YSO |
| IFO 116[ab] | J185026.138+012739.08 | 33.81379 | +0.73328 | 5.5 | 2.1 | 68 | 36.2 | 0.83 | - |
| IFO 117[b] | J185125.777−000930.42 | 32.78995 | −0.07040 | 740.0 | 530.0 | 170 | 1232132.7 | 114.00 | SNR |
| IFO 118 | J185128.102+002840.15 | 33.36065 | +0.21110 | 8.6 | 8.0 | 170 | 216.1 | 1.02 | HH |
| IFO 119 | J185140.619+002850.89 | 33.38708 | +0.16604 | 7.0 | 3.5 | 5 | 70.6 | 34.60 | HH |
| IFO 120 | J185141.136+002900.71 | 33.39049 | +0.16537 | 1.4 | 1.2 | 90 | 5.2 | 0.24 | HH |
| IFO 121 | J185144.114+002911.43 | 33.39880 | +0.15568 | 7.0 | 4.0 | 90 | 87.9 | 1.64 | HH |
| IFO 122 | J185249.992+022802.16 | 35.28688 | +0.81455 | 1.4 | 0.9 | 90 | 3.9 | 0.56 | YSO |
| IFO 123 | J185251.980+022804.91 | 35.29134 | +0.80753 | 1.2 | 1.0 | 90 | 3.7 | 0.74 | YSO |
| IFO 124 | J185353.538+015714.24 | 34.95065 | +0.34503 | 3.9 | 2.8 | 60 | 34.3 | 1.87 | YSO |
| IFO 125 | J185516.571+030512.10 | 36.11639 | +0.55408 | 8.0 | 4.8 | 160 | 120.6 | 9.03 | YSO |
| IFO 126 | J185521.357+030154.61 | 36.07665 | +0.51134 | 1.5 | 1.4 | 0 | 6.5 | 0.42 | YSO |
| IFO 127[a] | J185534.205+021209.31 | 35.31833 | +0.06246 | 6.3 | 4.2 | 175 | 83.1 | 1.46 | UCHII |
| IFO 128 | J185602.316+012139.70 | 34.66768 | −0.40273 | 1120.0 | 860.0 | 150 | 3025982.1 | 3942.00 | SNR |
| IFO 129[b] | J185737.231+020350.02 | 35.47351 | −0.43365 | 11.0 | 11.0 | 0 | 380.1 | 6.54 | PN |
| IFO 130[b] | J185738.014+020332.58 | 35.47069 | −0.43876 | 11.0 | 11.0 | 0 | 380.1 | 7.11 | PN |
| IFO 131 | J185808.531+010048.26 | 34.59832 | −1.02921 | 4.5 | 4.5 | 0 | 63.6 | 3.22 | YSO |
| IFO 132 | J185810.531+013656.88 | 35.13815 | −0.76162 | 17.5 | 14.7 | 0 | 808.1 | 24.60 | CHII |
| IFO 133 | J185905.114+004833.51 | 34.52432 | −1.33211 | 2.1 | 1.4 | 0 | 9.2 | 1.60 | - |
| IFO 134 | J185910.539+014013.31 | 35.30084 | −0.95905 | 5.0 | 3.1 | 130 | 48.7 | 1.56 | YSO |
| IFO 135 | J185923.204+010413.33 | 34.79106 | −1.27998 | 2.8 | 2.8 | 0 | 24.6 | 0.39 | - |
| IFO 136 | J190003.046+055926.63 | 39.24352 | +0.82130 | 26.6 | 13.0 | 170 | 1086.3 | 3.29 | YSO |
| IFO 137 | J190347.070+050946.79 | 38.93288 | −0.38355 | 35.0 | 18.0 | 100 | 1979.2 | 8.31 | UCHII |
| IFO 138 | J190403.633+050753.38 | 38.93638 | −0.45909 | 5.2 | 4.4 | 30 | 71.8 | 2.74 | UCHII |
| IFO 139[b] | J190404.180+052703.51 | 39.22135 | −0.31459 | 197.3 | 150.1 | 0 | 93037.4 | 618.00 | SNR |
| IFO 140 | J190540.231+074634.49 | 41.46973 | +0.39939 | 550.0 | 450.0 | 100 | 777544.2 | 1421.45 | SNR |
| IFO 141 | J190659.919+052253.12 | 39.49429 | −0.99412 | 5.4 | 5.2 | 90 | 88.2 | 12.20 | YSO |
| IFO 142 | J190731.328+052333.18 | 39.56407 | −1.10471 | 5.5 | 4.4 | 130 | 76.0 | 1.17 | - |
| IFO 143 | J190734.278+070829.06 | 41.12259 | −0.31090 | 170.0 | 120.0 | 115 | 64088.4 | 1691.00 | SNR |
| IFO 144 | J190813.552+052757.00 | 39.70977 | −1.22650 | 1.6 | 1.4 | 90 | 7.0 | 0.23 | YSO |

Table 1 (cont'd)

| IFO # | UWIFE designation | $l$ [deg] | $b$ [deg] | $r_1$ [arcsec] | $r_2$ [arcsec] | PA [deg] | Area [arcsec$^2$] | $F_{tot}$ [$10^{-17}$ W m$^{-2}$] | Counterpart |
|---|---|---|---|---|---|---|---|---|---|
| IFO 145 | J190816.446+052726.79 | 39.70784 | −1.24102 | 3.7 | 1.6 | 125 | 18.5 | 0.40 | YSO |
| IFO 146 | J190816.782+052506.10 | 39.67378 | −1.26024 | 3.9 | 3.6 | 0 | 44.1 | 1.69 | YSO |
| IFO 147 | J191106.846+090604.55 | 43.26595 | −0.18486 | 160.0 | 150.0 | 120 | 75398.2 | 4739.00 | SNR |
| IFO 148[a] | J191327.650+105334.62 | 45.12018 | +0.13279 | 11.2 | 8.5 | 137 | 299.0 | 14.80 | UCHII |
| IFO 149[a] | J191327.754+105413.66 | 45.13129 | +0.13745 | 3.2 | 2.4 | 105 | 24.1 | 0.48 | UCHII |
| IFO 150 | J191530.963+132747.36 | 47.63089 | +0.88215 | 1.7 | 1.4 | 120 | 7.4 | 0.29 | - |
| IFO 151 | J192026.005+111955.24 | 46.30789 | −1.17527 | 5.6 | 3.9 | 120 | 68.6 | 2.25 | YSO |
| IFO 152 | J192029.411+111942.04 | 46.31118 | −1.18928 | 1.3 | 1.1 | 90 | 4.5 | 0.27 | YSO |
| IFO 153 | J192029.485+110159.44 | 46.05061 | −1.32806 | 7.5 | 5.3 | 90 | 124.8 | 8.71 | HH |
| IFO 154 | J192054.201+143031.29 | 49.16624 | +0.21581 | 1.6 | 1.6 | 0 | 8.0 | 0.75 | YSO |
| IFO 155 | J192113.714+105232.92 | 45.99659 | −1.56168 | 4.2 | 2.8 | 110 | 36.9 | 2.34 | HH |
| IFO 156 | J192127.938+154426.63 | 50.31742 | +0.67543 | 4.5 | 2.6 | 30 | 36.7 | 1.58 | UCHII |
| IFO 157 | J192142.900+155351.18 | 50.48401 | +0.69629 | 60.0 | 20.0 | 90 | 3769.9 | 1.55 | PN |
| IFO 158 | J192309.835+142912.63 | 49.40475 | −0.27709 | 40.0 | 15.0 | 10 | 1884.9 | 8.92 | HII |
| IFO 159[b] | J192255.023+140745.93 | 49.06144 | −0.39309 | 1200.0 | 1000.0 | 160 | 3769911.3 | 1647.09 | HII |
| IFO 160[b] | J192401.145+140105.48 | 49.08966 | −0.68118 | 580.0 | 430.0 | 30 | 783513.2 | 582.37 | SNR |
| IFO 161 | J192348.822+143137.35 | 49.51449 | −0.39670 | 350.0 | 210.0 | 115 | 230907.0 | 182.47 | HII |
| IFO 162 | J192348.169+143641.50 | 49.58771 | −0.35441 | 50.0 | 30.0 | 25 | 4712.3 | 45.50 | LBV |
| IFO 163 | J192354.032+143548.00 | 49.58825 | −0.38096 | 45.0 | 35.0 | 0 | 4948.0 | 5.76 | YSO |
| IFO 164 | J192451.838+155729.06 | 50.89493 | +0.05763 | 1.5 | 1.5 | 0 | 7.0 | 0.24 | PN |
| IFO 165 | J192516.759+144625.72 | 49.89914 | −0.59204 | 2.3 | 1.9 | 0 | 13.7 | 0.72 | YSO |
| IFO 166 | J192529.675+151646.36 | 50.36959 | −0.39785 | 1.1 | 0.9 | 0 | 3.1 | 0.20 | - |
| IFO 167 | J192531.202+151603.90 | 50.36214 | −0.40884 | 2.1 | 1.3 | 0 | 8.5 | 0.79 | - |
| IFO 168 | J192531.399+151556.79 | 50.36075 | −0.41049 | 0.8 | 0.8 | 0 | 2.0 | 0.13 | - |
| IFO 169 | J192532.882+151538.18 | 50.35903 | −0.41819 | 1.5 | 1.3 | 40 | 6.1 | 0.50 | - |
| IFO 170 | J192533.417+151616.62 | 50.36947 | −0.41501 | 1.1 | 1.1 | 0 | 3.8 | 0.26 | - |
| IFO 171 | J192534.199+151612.08 | 50.36983 | −0.41838 | 1.2 | 0.8 | 90 | 3.0 | 0.23 | - |
| IFO 172 | J192534.538+151632.38 | 50.37544 | −0.41690 | 1.6 | 1.4 | 90 | 7.0 | 0.59 | - |
| IFO 173[a] | J192540.546+163305.18 | 51.50973 | +0.16761 | 7.0 | 5.8 | 16 | 127.5 | 11.50 | PN |
| IFO 174 | J192547.157+145145.84 | 50.03612 | −0.65762 | 4.0 | 2.5 | 140 | 31.4 | 4.22 | YSO |
| IFO 175 | J192557.625+150231.65 | 50.21401 | −0.60951 | 1.8 | 1.6 | 90 | 9.0 | 0.36 | YSO |
| IFO 176 | J192557.848+150243.23 | 50.21727 | −0.60877 | 10.0 | 5.5 | 0 | 172.2 | 1.98 | YSO |
| IFO 177 | J192852.403+171458.61 | 52.48804 | −0.17205 | 3.3 | 2.8 | 90 | 29.0 | 0.92 | YSO |
| IFO 178 | J192918.342+175615.42 | 53.14142 | +0.06679 | 15.0 | 10.0 | 140 | 471.2 | 4.95 | YSO |
| IFO 179 | J192918.796+175723.68 | 53.15891 | +0.07429 | 3.7 | 1.9 | 130 | 22.0 | 0.31 | YSO |
| IFO 180 | J192920.127+175716.54 | 53.15971 | +0.06871 | 25.0 | 8.0 | 80 | 628.3 | 4.41 | YSO |
| IFO 181 | J192920.506+175458.14 | 53.12668 | +0.04898 | 7.0 | 5.0 | 110 | 109.9 | 1.11 | YSO |
| IFO 182 | J192922.491+174442.54 | 52.98034 | −0.03983 | 4.6 | 3.5 | 0 | 50.5 | 0.57 | - |
| IFO 183 | J192931.617+175951.30 | 53.21927 | +0.04934 | 4.4 | 3.0 | 90 | 41.4 | 1.08 | YSO |
| IFO 184 | J192931.871+180058.11 | 53.23604 | +0.05734 | 1.7 | 1.4 | 90 | 7.4 | 0.25 | YSO |
| IFO 185 | J192932.874+180106.35 | 53.23994 | +0.05495 | 7.0 | 4.0 | 45 | 87.9 | 1.06 | YSO |
| IFO 186[b] | J193001.921+175455.44 | 53.20473 | −0.09547 | 5.4 | 3.3 | 20 | 55.9 | 3.15 | YSO |
| IFO 187 | J193120.744+192014.92 | 54.60141 | +0.31496 | 1.5 | 1.3 | 90 | 6.1 | 0.35 | - |
| IFO 188 | J193323.546+195647.07 | 55.36730 | +0.18676 | 20.0 | 18.0 | 40 | 1130.9 | 5.22 | post-AGBc/YSO |
| IFO 189 | J193831.665+202519.19 | 56.36978 | −0.63373 | 7.0 | 4.3 | 150 | 94.5 | 10.50 | YSO |
| IFO 190 | J193914.355+224021.52 | 58.41141 | +0.32789 | 7.0 | 6.0 | 40 | 131.9 | 3.75 | YSO |
| IFO 191 | J194014.058+232652.51 | 59.19889 | +0.51050 | 5.5 | 4.2 | 80 | 72.5 | 1.57 | - |
| IFO 192 | J194103.922+220340.80 | 58.08778 | −0.34083 | 3.2 | 1.7 | 0 | 17.1 | 0.62 | YSO |
| IFO 193 | J194127.149+222739.58 | 58.47940 | −0.22095 | 14.2 | 6.8 | 130 | 303.3 | 14.70 | YSO |
| IFO 194 | J194241.016+225417.72 | 59.00574 | −0.24738 | 6.5 | 3.0 | 60 | 61.2 | 2.76 | YSO |
| IFO 195 | J194244.693+232250.36 | 59.42558 | −0.02322 | 3.0 | 3.0 | 0 | 28.2 | 0.65 | HH |
| IFO 196 | J194256.665+232435.17 | 59.47362 | −0.04848 | 29.0 | 18.5 | 40 | 1685.4 | 5.20 | HH |
| IFO 197 | J194306.295+231810.63 | 59.39926 | −0.13356 | 1.9 | 1.3 | 90 | 7.7 | 0.52 | - |
| IFO 198 | J194310.286+234358.03 | 59.77970 | +0.06707 | 4.9 | 2.9 | 90 | 44.6 | 2.83 | YSO |
| IFO 199 | J194310.930+234402.64 | 59.78203 | +0.06557 | 5.3 | 2.5 | 90 | 41.6 | 2.23 | YSO |
| IFO 200 | J194320.930+232952.89 | 59.59633 | −0.08502 | 10.5 | 8.0 | 150 | 263.9 | 1.34 | HH |
| IFO 201 | J194610.902+221559.08 | 58.85575 | −1.26581 | 7.1 | 2.5 | 120 | 55.7 | 0.96 | - |
| IFO 202 | J194620.335+243520.73 | 60.88253 | −0.13043 | 14.0 | 12.9 | 0 | 567.3 | 17.20 | UCHII |
| IFO 203 | J194621.675+243516.78 | 60.88413 | −0.13538 | 1.3 | 1.3 | 0 | 5.3 | 0.45 | YSO |
| IFO 204 | J194646.921+251241.33 | 61.47104 | +0.09568 | 40.5 | 36.5 | 140 | 4644.0 | 111.00 | UCHII / HII |

[a] IFOs marked with '*a*' are identified only by an automatic detection method.

[b] Note on the individual sources. IFO 7: Due to the complexity of the region, flux is not provided. IFO 25–26: The flux of the superposed part is allocated only to IFO 26. IFO 51–55: IFO 51–54 are located inside IFO 55. IFO 62, 98: The flux is derived for a partial region free from severe artifacts. IFO 65, 81, 114, 116, 186: missing flux due to 2MASS-bright star mask. IFO 80, 96, 117: Contaminated by an instrumental artifact. The pixels with DN > ±3$\sigma$ are masked for the flux measurement. IFO 129–130: Contaminated by an instrumental artifact. The superposed part is excluded from the flux measurement. IFO 139: Contaminated by an instrumental artifact. The pixels with DN < −2$\sigma$ are masked for the flux measurement. IFO 159–160: The flux of the superposed part is allocated only to IFO 160. Note that there is an astrometry problem with certain continuum-subtracted [Fe II] images, where IFO 4, 8, 9, 73, 114, 155, 165, and 186 are located. Therefore, we determined the central positions of the IFOs based on the UKIDSS NIR image.

* IFO 38 is located within the SNR G15.9+0.2 domain but highly confined to a southwestern region (see fig. 1 of Sasaki et al. 2018 for an X-ray image of the SNR). Since there is no other possible counterpart in the SIMBAD query and the X-ray emission is coincident, we concluded that the SNR origin cannot be ruled out.

Table 2. Statistics of IFOs

|  | N | Flux$_{total}$ | Flux$_{min}$ | Flux$_{max}$ | Flux$_{mean}$ | F$^{sb}_{mean}$ |
|---|---|---|---|---|---|---|
| YSO | 100 | 4.3 (-15) | 2.0 (-18) | 8.2 (-16) | 4.3 (-17) | 6.6 (-19) |
| CHII | 22 | 5.2 (-15) | 3.2 (-18) | 1.5 (-15) | 2.4 (-16) | 4.0 (-19) |
| H II | 11 | 4.8 (-14) | 8.9 (-17) | 1.6 (-14) | 4.3 (-15) | 0.7 (-19) |
| PN | 17 | 1.4 (-15) | 2.4 (-18) | 4.9 (-16) | 8.5 (-17) | 7.6 (-19) |
| SNR | 25 | 2.0 (-13) | 3.9 (-18) | 4.7 (-14) | 8.2 (-15) | 1.3 (-19) |
| LBV | 4 | 1.3 (-15) | 1.3 (-17) | 8.1 (-16) | 3.3 (-16) | 9.7 (-19) |
| Unknown | 25 | 2.4 (-15) | 1.3 (-18) | 2.2 (-15) | 9.7 (-17) | 5.0 (-19) |
| total/mean | 204 | 2.6 (-13) | 1.6 (-17) | 4.7 (-14) | 1.3 (-15) | 5.3 (-19) |

Note. — N: Number of IFOs in each type. Flux units are in W $m^{-2}$. F$^{sb}_{mean}$: Mean surface brightness of each type (flux divided by area) in W $m^{-2}$ $arcsec^{-2}$. Note that one SNR-type (IFO 7) was not used for statistics of fluxes.

Table 3. IFOs Associated with YSO or YSO candidates

| IFO # | YSO / YSOc Name | Morphology | d [kpc] | Reference YSO Counterpart | Dist |
|---|---|---|---|---|---|
| IFO 003 | YSO AGAL G007.333-00.016 | k | 2.96 | ro8/u18 | u18 |
| IFO 004 | YSO candidate ALLWISE J180219.38-213351.9 | a | - | ro8 | - |
| IFO 010 | Class I YSO [RBG2009] G009.86-0.04 4 | k | $2.36^{+0.78}_{-0.88}$ | ro8/r09 | r09 |
| IFO 013 | Northern lobe of YSO candidate SSTGLMC G009.7612-00.9575 | b | - | ro8 | - |
| IFO 014 | Southern lobe of YSO candidate SSTGLMC G009.7612-00.9575 | b | - | ro8 | - |
| IFO 016 | YSO AGAL G013.779+00.609 | a | 2.90 | ro8/u18 | u18 |
| IFO 018 | YSO candidate 2MASS J18140816-1850560 | a | - | ro8 | - |
| IFO 019 | YSO candidate 2MASS J18140816-1850560 | a | - | ro8 | - |
| IFO 020 | W 33, IRS 3 having an O6.5 star | a | $2.40^{+0.17}_{-0.15}$ | b98 | i13 |
| IFO 021 | W 33, IRS 1 having an O6.5 + an O7.5 or O8 star | a | $2.40^{+0.17}_{-0.15}$ | b98 | i13 |
| IFO 022 | W 33, [MDF2011b] cl1 which encircles O6-7 star #23 | a | $2.40^{+0.17}_{-0.15}$ | m15 | i13 |
| IFO 023 | 1. IRAS 18114-1825: Class I, 2. J181421.71-182459.0: Class I/IIc | c | 2.41 | ro8/yu12/m16 | yu12 |
| IFO 029 | YSO AGAL G014.414-00.069 | c | 3.1 | u18 | u18 |
| IFO 030 | IRAS 18144-1723, Class I/II binary | k | 4.33 | c13/v18 | v18 |
| IFO 031 | 1. YSO IRAS 18151-1208 2. 2MASS J18175094-1208028, Class I/II YSO | a | 3.00 | m16 | m13 |
| IFO 032 | 1. 2MASS J18175094-1208028: Class I/II, 2. ALLWISE J181749.45-120751.1 | a | 3.00 | m16 | m13 |
| IFO 033 | YSO IRAS 18151-1208 | a | 3.00 | v10 | m13 |
| IFO 034 | [PW2010] 236, Class 0/I | a | 2.10 | ro8/c13/p10 | p10 |
| IFO 035 | [PW2010] 236, Class 0/I | a | 2.10 | ro8/c13/p10 | p10 |
| IFO 036 | In the middle of multiple YSOs in M 16 | c | 2.14 | c13 | b99 |
| IFO 037 | In the middle of multiple YSOs in M 16 | k | 2.14 | c13 | b99 |
| IFO 039 | Tip of column 3 of M 16, either T-Tauri star [TSH2002] S-1 or S-2 | k | 2.14 | t02 | b99 |
| IFO 040 | Edge of column 2 of M 16 | k | 2.14 | t02 | b99 |
| IFO 045 | Near the edge of M 16 Pillar V, RMS massive YSO G017.0332+00.7476A | a | 2.14 | ro8/c13 | b99 |
| IFO 046 | Proximity of Class 0/I YSO [PW2010] 378 | a | 2.10 | ro8/p10 | p10 |
| IFO 047 | Feature connected to Class 0/I YSO [PW2010] 411 | c | 2.10 | c13/p10 | p10 |
| IFO 048 | In the vicinity of YSOs in M 16 | a | 2.14 | g07 | b99 |
| IFO 049 | Spatially connected to massive YSO G017.0666+0.6826 | c | 2.14 | c13 | b99 |
| IFO 051 | Compact feature in the crowded region of YSOs in M 17 | k | 1.60±0.30 | si | n01 |
| IFO 052 | Ditto | k | 1.60±0.30 | si | n01 |
| IFO 053 | Ditto | k | 1.60±0.30 | si | n01 |
| IFO 056 | Multiple YSO candidates in the northern region of M17, EB (extended bubble) | c | 1.98 | ro8/p09 | c16 |
| IFO 057 | Ditto | a | 1.98 | ro8/p09 | c16 |
| IFO 059 | Southern jet of IFO 060 | b | 1.85±0.2 | ro8/c13 | x19 |
| IFO 060 | IRAS 18177-1405 aligned with IFO 059, 061, in M 16 | b | 1.85±0.2 | ro8/c13 | x19 |
| IFO 061 | Northern jet of IFO 060 | b | 1.85±0.2 | ro8/c13 | x19 |
| IFO 063 | Located at the edge of IRDC HEC G016.93+00.24 | a | 2.40 | si | r10 |

Table 3  (cont'd)

| IFO # | YSO / YSOc Name | Morphology | d [kpc] | Reference YSO Counterpart | Dist |
|---|---|---|---|---|---|
| IFO 072 | Class I/II YSO IRAS 18229-1308 | c | 3.40 | ro8/c13/m16 | u22 |
| IFO 074 | Massive YSO IRAS 18241-1134. [Fe II] 1.64 μm detected | c | 12.60 | c13 | c13 |
| IFO 077 | Class I YSO candidate J182859.53-115009.6 | a | - | k21 | - |
| IFO 091 | Biconical structure coincident with FIR clumps, new PN in Froebrich et al. (2015) | b | 3.42 | e17 | t15 |
| IFO 092 | Located in IRDC 24.764-0.12. Proto-stellar clumps in the vicinity | k | 3.57 | si | t15 |
| IFO 094 | Coincident with UKIDSS source UGPS J183720.81-064158.4. Multiple nearby YSOs | c | - | si | - |
| IFO 099 | North-western jet, aligned with IFO 100 & 101 | b | 3.50 | ro8/k21 | t15 |
| IFO 100 | Class I YSO candidate, previously reported as AGB candidate | b | 3.50 | ro8/k21 | t15 |
| IFO 101 | South-eastern jet, aligned with IFO 99 & 100 | b | 3.50 | ro8/k21 | t15 |
| IFO 104 | In the middle of ALLWISE J183951.16-043113.8 Class III or more evolved YSO and semi-regular variable ASASSN-V J183948.07-043015.9 | a | 2.20 | ro8/m16/j18 | j18 |
| IFO 108 | Proximity of pre-main sequence star candidate GaiaDR2 4258232818679065216 | a | 2.01 | v20 | b18 |
| IFO 111 | Massive protostellar object [VEN2013] G029.8623-0.0437, [Fe II] detection reported | c | 6.21 | c13/a20 | l16 |
| IFO 115 | Spatially coincident with H$_2$, which is connected to MSX6C G031.8380-00.1284, YSO candidate SSTGLMC G031.8361-00.1408 in the vicinity | a | - | ro8/e03 | - |
| IFO 122 | Shares a similar compact structure with IFO 123, aligned East to West | k | 4.80 | ro8/k21 | u18 |
| IFO 123 | Coincident with flat-spectrum YSO candidate SSTGLMC G035.2913+00.8076 | k | 4.80 | ro8/k21 | u18 |
| IFO 124 | Matches to proto-stellar clump 34.93+0.338 1 | a | 2.90 | t15 | t15 |
| IFO 125 | Cometary structure coincident with massive YSO IRAS 18527+0301 | c | 4.70 | m96 | u18 |
| IFO 126 | Ultrawide binary Gaia2 4280756726686953984 is the closest, Class III or more evolved YSO ALLWISE J185522.49+030130.3 in 30" distance | a | - | t20/m16 | - |
| IFO 131 | Class I/II massive YSO IRAS 18555+0056, [Fe II] reported by p16 | c | 1.10 | c13/p16 | l13 |
| IFO 134 | Close to YSO candidate SSTGLMC G035.2868-00.9528, proto-stellar clump is coincident | a | 2.48 | ro8 | t15 |
| IFO 136 | Flat spectrum YSO SSTGLMC G039.2199+00.8638. G039.2060+00.8818 in West | a | - | ro8/k21 | - |
| IFO 141 | Compact component matches to massive YSO IRAS 19045+0518 | c | 3.60 | c13 | c13 |
| IFO 144 | Aligned with YSO AGAL G039.708-01.237 | b | 0.60 | u18 | u18 |
| IFO 145 | Elongated and pointing toward YSO AGAL G039.708-01.237 and IFO 145 | b | 0.60 | u18 | u18 |
| IFO 146 | Hα PN candidate, yet aligned with IFO 144, 145, H$_2$ knots | a | 0.60 | s14 | u18 |
| IFO 151 | Coincident with class I and flat-SED YSOs, [TBP2010] L673 10 and 13 | a | 0.60 | t10 | u18 |
| IFO 152 | Close to HH 1186, 42 arcsec away from [TBP2010] L673 YSO 15 & IFO 151 | k | 0.60 | t10 | u18 |
| IFO 154 | Class II/photosphere YSO SSTOERC G049.1662+0.2159 at W 51 (or foreground, see k09) | k | 5.40 | s17 | s17 |
| IFO 163 | Surrounding YSOs, e.g., class III/photosphere SSTOERC G049.5851-0.3814 | a | 5.40 | s17 | l13 |
| IFO 165 | Connected to Class I YSO SSTOERC G049.9010-00.5922, spectral index of flat (FS) according to k21 | c | 5.40 | s17 | l13 |
| IFO 174 | Emerges from Class I YSO [RML2017] MC2 M105 | c | 3.09 | r17 | t15 |
| IFO 175 | A southern compact jet of massive YSO IRAS 19236+1456 | k | 3.39 | c13 | t15 |
| IFO 176 | Southern diffuse emission from massive YSO IRAS 19236+1456 | a | 3.39 | c13 | t15 |

Table 3 (cont'd)

| IFO # | YSO / YSOc Name | Morphology | d [kpc] | Reference YSO Counterpart | Reference Dist |
|---|---|---|---|---|---|
| IFO 177 Diffuse structure in contact with YSO AGAL G052.488-00.172 | | c | 1.60 | u18 | u18 |
| IFO 178 Class I YSO 1 (~10M$_\odot$) or 2 (~5M$_\odot$) in k18 | | a | 1.60 | ro8/c13/k18 | u18 |
| IFO 179 In the proximity of Class I YSO SSTGLMC G053.1570+00.0735 | | a | 1.60 | ro8/k15 | u18 |
| IFO 180 Close to Class I YSO SSTGLMC G053.1612+00.0668 and multiple YSOs | | a | 1.60 | ro8/k15 | u18 |
| IFO 181 Surrounding Class I YSO SSTGLMC G053.1266+00.0499 | | k | 1.60 | ro8/k15 | u18 |
| IFO 183 Coincident with MSXDC G053.25+00.04 MM6 & ISOGAL-P J192931.1+175954 (Class I) | | a | 1.60 | k15 | u18 |
| IFO 184 Flat-spectrum YSO 2MASS J19293167+1800581 | | k | 1.60 | ro8/k15 | u18 |
| IFO 185 SSTGLMC G053.2389+00.0552 (Class I), 2MASS J19293167+1800581 (FS) | | a | 1.60 | ro8/k15 | u18 |
| IFO 186 2MASS J19300219+1755001 (FS) | | c | - | k15 | - |
| IFO 189 Multiple compact structures surrounding ES-NW of Massive YSO MSX6C G056.3694-00.6333 | | a | 6.40 | c13 | c13 |
| IFO 190 Diffuse structure on South of Class II YSO SSTGLMC G058.4098+00.3279 | | a | 2.80 | ro8/k21 | v13 |
| IFO 192 The head of the cometary structure matches EGO G058.09-0.34, one of the low-mass EGOs | | k | 0.74 | ro8/cy13 | cy13 |
| IFO 193 Two biconical structures, tails toward SE & W, Class I YSO SSTGLMC G058.4801-00.2205 at the center | | b | 6.15 | ro8/k21 | m21 |
| IFO 194 The head of cometary structure corresponds to YSO candidate SSTGLMC G059.0069-00.2481 | | c | - | ro8 | - |
| IFO 198 Aligned with star-forming region IRAS 19410+2336 and IFO 199 | | a | 2.20 | c13 | l13 |
| IFO 199 Amorphous IFO points toward star-forming region IRAS 19410+2336 | | a | 2.20 | c13 | l13 |
| IFO 203 Compact IFO at the East of biconical outflow S87, emerging from ~20 M$_\odot$ pre-main-sequence object | | a | 2.20 | b89 | l13 |

*Column 3. Morphology categories : b - bipolar, c - cometary, k - knot-like, a - amorphous. Column 4. Distance of counterpart in kpc. Column 5. References of counterpart classification and distance.

*References : a20 - Areal et al. (2020), b18 - Bailer-Jones et al. (2018), b89 - Barsony (1989), b98 - Beck et al. (1998), b99 - Belikov et al. (1999), c13 - Cooper et al. (2013), c16 - Csengeri et al. (2016), cy13 - Cyganowski et al. (2013), e03 - Egan et al. (2003), e17 - Elia et al. (2017), g07 - Guarcello et al. (2007), i13 - Immer et al. (2013), j18 - Jayasinghe et al. (2018), k09 - Kang et al. (2009), k15 - Kim et al. (2015), k18 - Kim et al. (2018), k21 - Kuhn et al. (2021), l13 - Lumsden et al. (2013), l16 - Li et al. (2016), m13 - Sánchez-Monge et al. (2013), m15 - Messineo et al. (2015), m16 - Marton et al. (2016), m21 - Mège et al. (2021), m96 - Molinari et al. (1996), n01 - Nielbock et al. (2001), p09 - Povich & Whitney (2009), p10 - Povich et al. (2009), p16 - Paron et al. (2016), r09 - Ragan et al. (2009), r10 - Rygl et al. (2010), r17 - Retes-Romero et al. (2017), ro8 - Robitaille et al. (2008), s14 -Sabin et al. (2014), s17 - Saral et al. (2017), t02 - Thompson et al. (2002), t10 - Tsitali et al. (2010), t15 - Traficante et al. (2015), t20 - Tian et al. (2020), u18 - Urquhart et al. (2018), u22 - Urquhart et al. (2022), v13 - Veneziani et al. (2013), v18 - Varricatt et al. (2018), v20 - Vioque et al. (2020), x19 - Xu et al. (2019), yu12 - Yuan et al. (2012)

Table 4. IFOs Associated with Herbig-Haro objects

| IFO # | HH Name | d [kpc] | Exciting Source | Region | Comment | Morp. | Reference Type | Dist |
|---|---|---|---|---|---|---|---|---|
| IFO 041 | | | | | Parsec-scale HH | b | m82 | b99 |
| IFO 042 | HH 216 | 2.14 | HH-N | Eagle nebula | Parsec-scale HH | b | m82 | b99 |
| IFO 043 | | | | | Parsec-scale HH | b | m82 | b99 |
| IFO 044 | | | | | Parsec-scale HH | b | m82 | b99 |
| IFO 118 | HH 722 | | | | Wrongly identified as HH 172 (Nikogossian et al. 2007) | a | c94 | d92 |
| IFO 119 | | | | | Binary. One of them is Class I YSO, $A_V\sim25$ mag, bipolar envelope SE-NW side | a | s07 | s87 |
| IFO 120 | GGD 30 | 1.70 | GGD30IR | GM 2-30 | Binary. One of them is Class I YSO, $A_V\sim25$ mag, bipolar envelope SE-NW side | a | s07 | s87 |
| IFO 121 | | | | | Binary. One of them is Class I YSO, $A_V\sim25$ mag, bipolar envelope SE-NW side | a | s07 | s87 |
| IFO 153 | HH 32 | 0.20±0.03 | AS353 | Aquila Rift | T-Tauri binary, H$\alpha$ is coincident, [Fe II] detection reported (d03) | a | h74 | r06 |
| IFO 155 | HH 250A | 0.30 | HH 250-IRS | Aquila Rift | Class I binary, bow-shock, H$\alpha$ detected. Launched 3500 yrs ago, adjacent to IFO 153 | a | d97 | s97 |
| IFO 195 | HH 803 | | | | SW of parsec-scale HH (7.5 pc). H$\alpha$ detected. $L_{bol}=580\ L_\odot$ | k | c04 | d92 |
| IFO 196 | HH 165 | 2.40 | 1548C27 IRS1 | | NE counterpart of the HH 803, very faint | a | c81 | d92 |
| IFO 200 | HH 365 | | | | Central structure of parsec-scale HH, both perpendicular jet and curvature are identified | a | a97 | d92 |

*Column 3. Distance of counterpart in kpc. Column 7. Morphology categories : b - bipolar, c - cometary, k - knot-like, a - amorphous. Column 8. References of counterpart classification and distance.: a97 -Alten et al. (1997), b99 -Belikov et al. (1999), c04 -McGroarty et al. (2004), c81 -Craine et al. (1981), c94 -Cappellaro et al. (1994), d03 -Davis et al. (2003), d92 -Dent & Aspin (1992), d97 -Devine et al. (1997), h74 -Herbig (1974), m82 -Meaburn & White (1982), r06 -Rice et al. (2006), s07 -Smith et al. (2007), s87 -Scoville et al. (1987), s97 -Sakamoto et al. (1997)

Table 5. IFOs Associated with UCHII / CHII or UCHII / CHII Candidates

| IFO # | UCHII, CHII Name | d [kpc] | Type | Comment | Reference Type | Dist |
|---|---|---|---|---|---|---|
| IFO 024 | IRAS 18116-1646 | 4.50 | UCHII | Outer arc, East of cometary H II region | k18 | u18 |
| IFO 025 | IRAS 18116-1646 | 4.50 | UCHII | Diffuse filament filling inside, coincident with the boundary of cometary H II region | k18 | u18 |
| IFO 026 | IRAS 18116-1646 | 4.50 | UCHII | Outer arc, West of cometary H II region | k18 | u18 |
| IFO 054 | M 17 UC1 | 1.60±0.30 | HCHII | Multiple shell-like features around hypercompact H II region | s04 | n01 |
| IFO 071 | IRAS 18222-1317 | 4.04 | UCHII | Partial shell-like + amorphous structure around UCHII | k18 | l14 |
| IFO 082 | IRAS 18308-0503 | 2.90 | UCHII | Amorphous, compact feature North of UCHII | b96 | l13 |
| IFO 083 | IRAS 18308-0503 | 2.90 | UCHII | Amorphous, diffuse feature South of UCHII | b96 | l13 |
| IFO 086 | IRAS 18314-0720 | 9.30±0.40 | UCHII | Shell-like structures, West of UCHII | k18 | k03 |
| IFO 087 | IRAS 18314-0720 | 9.30±0.40 | UCHII | Multiple shell-like structures encircling UCHII | k18 | k03 |
| IFO 089 | IRAS 18317-0757 | 4.60 | UCHII | Amorphous structures enveloping the South-East of UCHII | k18 | l13 |
| IFO 090 | IRAS 18317-0757 | 4.60 | UCHII | Compact knot-like structure North-East of UCHII | k18 | l13 |
| IFO 097 | G025.3809-00.1815 | 3.80 | UCHII | Amorphous structures in the central region of UCHII + South-Western jet | k18 | a09 |
| IFO 107 | IRAS 18416-0420 | 3.30 | UCHII | Shell-like structure stretching from the center of UCHII to West | k18 | u18 |
| IFO 127 | IRAS 18530+0215 | 9.00 | UCHII | Amorphous + knot-like feature South of UCHII | k18 | l13 |
| IFO 132 | G35.2S | 2.40±0.50 | CHII | Oval structure, slightly limb-brightened, brighter at East. Compact H II region (radius : 0.14 pc) | f11 | z13 |
| IFO 137 | [CPA2006] N75 | 2.80 | UCHII? | Northern shell-like structure. Single O9.5V star. UCHII precursors, Mol 91C1–5, along the structure | m98 | z13 |
| IFO 138 | UCEC 8 | 2.80 | UCHII? | Ultra-compact embedded cluster (UCEC) with a radius of 0.09 pc. Eight possible ionizing stars, most of them are early-B. | a12 | l13 |
| IFO 148 | IRAS 19111+1048 | 4.40 | UCHII | About 19 central sources of spectral type earlier than B0.5 (v06) | k18 | l13 |
| IFO 149 | IRAS 19111+1048 | 4.40 | UCHII | Diffuse emission North of IFO 148 | k18 | l13 |
| IFO 156 | IRAS 19191+1538 | 2.10 | UCHII | Partial shell-like structure in the West, compact structure superposed in the North | k18 | l13 |
| IFO 202 | IRAS 19442+2427 | 2.20 | UCHII | Diffuse emission South of cometary UCHII, the spectral type of O9-B0 (d20) | k18 | l13 |
| IFO 204 | IRAS 19446+2505 | 2.20 | UCHII/HII | CORNISH UCHII G061.4763+00.0892 in East, H II region G061.4758+00.0913 in West | k18 | l13 |

*Column 3. Distance of counterpart in kpc. Column 6 : References of counterpart classification and distance : a09 - Anderson & Bania (2009), a12 - Alexander & Kobulnicky (2012), b96 - Bronfman et al. (1996), d20 - de la Fuente et al. (2020), f11 - Froebrich & Ioannidis (2011), k03 - Kolpak et al. (2003), k18 - Kalcheva et al. (2018), l13 - Lumsden et al. (2013), l14 - Leahy et al. (2014), m98 - Molinari et al. (1998), n01 - Nielbock et al. (2001), s04 - Sewiło et al. (2004), u18 - Urquhart et al. (2018), v06 - Vig et al. (2006), z13 - Zhu et al. (2013)

Table 6. IFOs Associated with HII Region or HII Region Candidates

| IFO # | HII / HIIc Name | d [kpc] | Type | Comment | Reference Type | Dist |
|---|---|---|---|---|---|---|
| IFO 011 | G10.2−0.3 (W 31) | 3.55$^a$ | HII | Diffuse structure in West of H II region | r07 | u12 |
| IFO 012 | G10.2−0.3 (W 31) | 3.55$^a$ | HII | Diffuse structure coincident with O stars/YSOs in the central region. UCHII G10.15−0.34 is embedded | r07 | u12 |
| IFO 055 | M 17 S | 1.60±0.30 | HII | The South-Western large-scale structure of M 17 | p09 | n01 |
| IFO 058 | M 17 S (South) | 1.60±0.30 | HII | Diffuse filaments at the South of M 17 | p09 | n01 |
| IFO 062 | M 17 N | 1.60±0.30 | HII | The North-Eastern large-scale structure of M 17 | p09 | n01 |
| IFO 064 | M 17 S (South) | 1.60±0.30 | HII | Diffuse filaments at the South of M 17 | p09 | n01 |
| IFO 066 | M 17 EB$^b$ | 1.60±0.30 | HII?/YSO? | Amorphous structure located in the extremity of M 17, star Gaia2 4097840529185958272 in West | p09 | n01 |
| IFO 067 | [KC97c] G015.1−00.9 | 1.98 | HII | The elongated structure along the radio emission of H II region [KC97c] G015.1−00.9 | k97 | r03 |
| IFO 158 | W 51A | 5.40 | HII | Shell-like diffuse emission at the Western boundary of compact radio | f21 | l13 |
| IFO 159 | W 51B | 6.00 | HII | Large-scale, multiple filamentary structures in W 51B | f21 | k95 |
| IFO 161 | W 51A | 5.50 | HII | Large-scale, miscellaneous amorphous structures in W 51A | f21 | c09 |

$^a$IFO distances marked with 'a' are controversial values. Column 3. Distance of counterpart in kpc. Column 6 : References of counterpart classification and distance : c09 - Clark et al. (2009), f21 - Fujita et al. (2021), k97 - Kuchar & Clark (1997), k95 - Koo et al. (1995), n01- Lumsden et al. (2013), l13 - Nielbock et al. (2001), p09 - Povich et al. (2009), r03 - Russeil (2003), r07 - Roshi et al. (2017), u12 - Urquhart et al. (2012)

$^b$The region M 17 EB (extended bubble) is defined in detail by Povich et al. (2009).

Table 7. IFOs Associated with PN or PNc Candidates

| IFO # | PN / PNc Name | Type | d [kpc] | Size [pc] | Morphology [Fe II] | Morphology H$_2$ | Morphology H$\alpha$ | Reference Distance | Reference Morphology | Reference Type |
|---|---|---|---|---|---|---|---|---|---|---|
| IFO 005 | NGC 6537 | PN | 2.81±0.45 | 0.49$^a$ | Bp | Bmps | Bmps | 3 | 1,2 | - |
| IFO 006 | NGC 6537 | PN | 2.81±0.45 | 0.43$^a$ | Bp | Bmps | Bmps | 3 | 1,2 | - |
| IFO 008 | HRDS G008.362−0.623 | PNc | - | −$^a$ | Bas | Bmps | - | - | 1 | 4 |
| IFO 009 | HRDS G008.362−0.623 | PNc | - | −$^a$ | Bas | Bmps | - | - | 1 | 4 |
| IFO 027 | PNG G014.5+00.4 | PNc | - | - | B | B | S | - | 1,2 | 5 |
| IFO 050 | SSTGLMC G015.7993−00.0063 | PN | - | - | Ia | I? | - | - | 6 | - |
| IFO 073 | G020.9+00.8 | PNc | - | - | A | Is | Aa | - | 1,2 | 7 |
| IFO 081 | M 1-51 | PN | 1.59 | 0.18 | Bp$^b$ | - | B | 8 | 2 | - |
| IFO 095 | IRAS 18348−0616 | PNc/UCHII? | 6.5 | 0.50 | Ema | Ema | - | 18 | 6 | 20,19 |
| IFO 112 | IRAS 18458−0213 | PN | 4.90 | 0.13 | Bsa | Bs | - | 9 | 1 | - |
| IFO 114 | PNG 032.1+00.1 | PNc/HII? | 4.90 | 0.42 | Ia | - | - | 9 | - | 10,2 |
| IFO 129 | IRAS 18551+0159 | PNc | 4.30±0.50 | 0.45$^a$ | Br | Bs | B | 11 | 1 | 4 |
| IFO 130 | IRAS 18551+0159 | PNc | 4.30±0.50 | 0.45$^a$ | Br | Bs | B | 11 | 1 | 4 |
| IFO 157 | PNG 050.4+00.7 | PNc/sym? | 5.30 | 3.08 | Bp | Bs | B | 9 | 1 | 12,2 |
| IFO 164 | PNG 050.8+00.0 | PNc | 18.46±2.59 | 0.26 | I | - | - | 17 | - | 17 |
| IFO 173 | IRAS 19234+1627 | PN | 4.70−9.50 | 0.31−0.64 | Ears | Er | Emrs | 13 | 1,2 | - |
| IFO 188 | IRAS 19312+1950 | pAGBc/YSO? | 2.42 | 0.46 | Ams | As | - | 14 | 6 | 15,16 |

*For the morphological classification, we adopted an 'ERBIAS' classifier (Parker et al. 2006 and references therein), and when listed, we showed the morphological classification of H$\alpha$ counterpart in HASH (Hong Kong/AAO/Strasbourg/H$\alpha$; Parker et al. 2016). For the morphological classification of the H$_2$ counterpart and H$\alpha$ classification of IFO 129, 130, and 173, we made use of Gledhill et al. (2018). The 'ERBIAS' represents the main structure of PN by its elliptical, round, bipolar, irregular, and asymmetric or quasi-stellar morphology. The substructure is further classified into the sub-classifier 'amprs', which explains asymmetry, multiple structures, point symmetry, ring, and internal structure. When no counterpart was detected, we marked '-'. PNc: PN candidate, sym: symbiotic star candidate, and pAGBc: post-AGB candidate. Column 4. Distance of counterpart in kpc. Column 5. Physical scale of an IFO in the adopted distance. Column 7. References of distance, H$_2$/H$\alpha$ morphology, counterpart type in Column 3, respectively. Ref. 1 - Gledhill et al. (2018), 2 - Parker et al. (2016), 3 - Navarro et al. (2012), 4 - Froebrich et al. (2015), 5 - Miszalski et al. (2008), 6 - this study (UWISH2), 7 - Boissay et al. (2012): Miscellaneous Emission Nebulae (MEN) list, 8 - Phillips (2004), 9 - Lumsden et al. (2013), 10 - Ferrero et al. (2015), 11 - Zhu et al. (2013), 12 - Sabin et al. (2014), 13 - Yang et al. (2016), 14 - Vickers et al. (2015), 15 - Nakashima et al. (2016), 16 - Cordiner et al. (2016), 17 - Irabor et al. (2018), 18 - Cichowolski et al. (2018), 19 - Bronfman et al. (1996), 20 - Kanarek et al. (2015)

$^a$Part of bipolar lobes.

$^b$PN with quadrupolar morphology is included in the bipolar category considering the formation mechanisms of the two could be analogous; see Manchado et al. (1996) for details.

Table 8. IFOs Associated with LBV nebula or LBV nebula candidates

| IFO # | LBV / LBVc Name | d [kpc] | Reference Type | Reference d |
|---|---|---|---|---|
| IFO 065 | HD 168625 | 1.55 | 5 | 4 |
| IFO 102 | 26.47+0.02 | ≤6.5 | 1 | 1 |
| IFO 103 | 26.47+0.02 | ≤6.5 | 1 | 1 |
| IFO 162 | [KW97] 37-17 (=LS1) | $6.0, 2.5^{+2.4}_{-1.3}$ | 2 | 3,4 |

*Column 3. Distance of counterpart in kpc. Column 4. References of counterpart classification and distance : 1 - Clark et al. (2003) : assuming 1.8 mag kpc$^{-1}$, 2 - Okumura et al. (2000), 3 - Clark et al. (2009) : observational and theoretical constraints + W51 membership + parallax, 4 - Bailer-Jones et al. (2018) : Gaia DR2 parallax, 5 - Hutsemékers et al. (1994)

Table 9. IFOs Associated with SNRs

| IFO # | SNR Name G-Name | Other Name(s) | d [kpc] | $F_{tot}$ [$10^{-17}$ W m$^{-2}$] |
|---|---|---|---|---|
| IFO 007 | G8.7-0.1 | W 30 | 4.5 | —[a] |
| IFO 015 | G11.2-0.3 | — | 4.4 | 1090.00 |
| IFO 038 | G15.9+0.2 | — | 10.0 | 7.34 |
| IFO 069 | G18.1-0.1 | — | 5.6 | 21.50 |
| IFO 070 | G18.1-0.1 | — | 5.6 | 1.38 |
| IFO 075 | G18.9-1.1 | — | 2.0 | 14.10 |
| IFO 076 | G18.9-1.1 | — | 2.0 | 2.55 |
| IFO 078 | G18.9-1.1 | — | 2.0 | 192.00 |
| IFO 079 | G18.9-1.1 | — | 2.0 | 1.89 |
| IFO 080 | G21.8-0.6 | Kes 69 | 5.2 | 627.00 |
| IFO 084 | G21.5-0.9 | — | 4.6 | 6.59 |
| IFO 085 | G21.5-0.9 | — | 4.6 | 59.70 |
| IFO 088 | G23.3-0.3 | W 41 | 4.2 | 584.00 |
| IFO 093[b] | G28.8+1.5 | — | [c]3.4 | 0.39 |
| IFO 098 | G27.8+0.6 | — | 2.0 | 167.00 |
| IFO 105 | G27.4+0.0 | Kes 73 | 8.5 | 274.00 |
| IFO 106 | G28.6-0.1 | — | 9.6 | 197.00 |
| IFO 113 | G31.9+0.0 | 3C 391 | 7.1 | 3423.55 |
| IFO 117 | G32.8-0.1 | Kes 78 | 4.8 | 114.00 |
| IFO 128 | G34.7-0.4 | W 44 | 2.8 | 3942.00 |
| IFO 139 | G39.2-0.3 | 3C 396 | 8.5 | 618.00 |
| IFO 140 | G41.5+0.4 | — | 4.1 | 1421.45 |
| IFO 143 | G41.1-0.3 | 3C 397 | 10.0 | 1691.00 |
| IFO 147 | G43.3-0.2 | W49B | 10.0 | 4739.00 |
| IFO 160 | G49.2-0.7 | W51C | 6.0 | 582.37 |

*Distances are from Lee et al. (2019). See Lee et al. (2019) for original references.

[a]Due to significant background errors, flux was not derived and not included in the statistics of fluxes.

[b]Distance from Shan et al. (2018).

[c]Partially covered in the UWIFE survey.

Table 10. IFOs Associated with Unknown-type

| IFO # | Comment | Reference |
|---|---|---|
| IFO 001 | Amorphous structure elongated from NE to SW. The bright star 15 arcsec East to the IFO is ALLWISE J180138.06−224227.8, a luminosity class of giant | s19 |
| IFO 002 | Compact IFO, H II region in the vicinity, $H_2$ at South is slightly more extended | - |
| IFO 017 | Located in between late-type star [MHS2002] 196 and dark cloud SDC G12.804+0.055 | - |
| IFO 028 | Biconical structure in the North-East of IRAS 18134-1838, however, relation is uncertain | - |
| IFO 068 | An amorphous source in the middle of a possible evolved star J182227.66−171556.6 and the two stars having a high proper motion at South | m19 r10 |
| IFO 096 | Spatially coincident with large-scale radio shell G25.8+0.2 | c18 |
| IFO 109 | IFO at the border of AllWISE J184501.48−001716.3, $H_2$ emission at North | - |
| IFO 110 | Amorphous IFO, possibly surrounding MIR source AllWISE J184515.66−031606.9 | - |
| IFO 116 | Amorphous IFO, no well-known sources in the vicinity | - |
| IFO 133 | Cometary IFO elongated North to South, AllWISE J185905.19+004837.9 at the North | - |
| IFO 135 | Amorphous IFO adjoining $K$-band source UGPS J185923.30+010415.0 which was not detected in $J$- and $H$-bands | s19 |
| IFO 142 | Bow-shock morphology with apex on NW side, in between possible giant star at South and evolved/Class III YSO candidate at North | m16 |
| IFO 150 | Compact structure coincident with $H_2$, ALLWISE J191530.41+132737.2 showing MIR photometric anomaly is located in the vicinity | s17 |
| IFO 166 | Knot-like small-scale IFO | - |
| IFO 167 | Coincident with extended $H_2$, but with the shifted peak position | - |
| IFO 168 | A compact, faint knot in between IFO 167 and 169 | - |
| IFO 169 | Knot-like diffuse IFO | - |
| IFO 170 | Knot-like IFO in the middle of IFO 166-169 and IFO 172 | - |
| IFO 171 | Knot-like IFO in the middle of IFO 166-169 and IFO 172 | - |
| IFO 172 | Compact IFO, small-scale $H_2$ in the vicinity | - |
| IFO 182 | Amorphous IFO with the bright northern component coincident with MIR source GLIMPSE G052.9805−00.0394 | - |
| IFO 187 | Knot-like IFO. Possible evolved star WISE J193120.72+192006.1 and dwarf star Gaia DR2 1825442262813750912 in South | m19 s19 |
| IFO 191 | Diffuse elongated IFO from East to West. MIPSGAL source MG059.1989+00.5106 at North, MIR source ALLWISE J194013.97+232657.5 superposed to the IFO, red giant star 2MASS J19401620+2326577 at East | - |
| IFO 197 | Slightly elongated compact IFO, the center of NGC 6823 is located at East | - |
| IFO 201 | Cometary IFO with its South-Eastern head part coincident with ALLWISE J194611.21+221554.3, which is a luminosity class of dwarf candidate | s19 |

*Column 3. References of counterpart candidates : c18 - Cichowolski et al. (2018), m16 - Marton et al. (2016), m19 - Marton et al. (2019), r10 - Roeser et al. (2010), s17 - Solarz et al. (2017), s19 - Stassun et al. (2019)

*Counterpart candidates around IFO 166−172: 1: YSO candidate SSTGLMC G050.3746-00.4149 2: AGB candidate SSTGLMC G050.3756-00.4214 3: MHO 2624 4: EGO G050.36-0.42 5: YSO candidate SSTGLMC G050.3666-00.3944 6: YSO candidate SSTGLMC G050.3647-00.3979 7: YSO candidate SSTGLMC G050.3587-00.4123 8: YSO candidate SSTGLMC G050.3675-00.4089 9: YSO candidate SSTGLMC G050.3691-00.4096 10: YSO candidate SSTGLMC G050.3704-00.4095 11: YSO candidate SSTGLMC G050.3741-00.4083 12: YSO AGAL G050.376-00.421 13: YSO candidate SSTGLMC G050.3762-00.4205 14: YSO candidate SSTGLMC G050.3575-00.4182

**ACKNOWLEDGEMENTS**

The authors are grateful to the anonymous referee for the comments and suggestions that improved the clarity of this paper. We also acknowledge the Cambridge Astronomical Survey Unit and the WFCAM Science Archive for the reduction and ingest of the survey data. This research has made use of the SIMBAD database, operated at CDS, Strasbourg, France. This research has made use of the VizieR catalogue access tool, CDS, Strasbourg, France (DOI : 10.26093/cds/vizier). The original description of the VizieR service was published in 2000, A&AS 143, 23. The UWIFE survey was supported by the K-GMT Science Program funded through the Korea GMT Project operated by the Korea Astronomy and Space Science Institute (KASI). B.-C. Koo was supported by a National Research Foundation of Korea (NRF) grant funded by the Korea Government (MSIP) (no. 2012R1A4A1028713). Y. Kim acknowledges support from Kim In-Ha Scholarship.

**DATA AVAILABILITY**

The data underlying this article are available to download from https://uwife.gems0.org/.

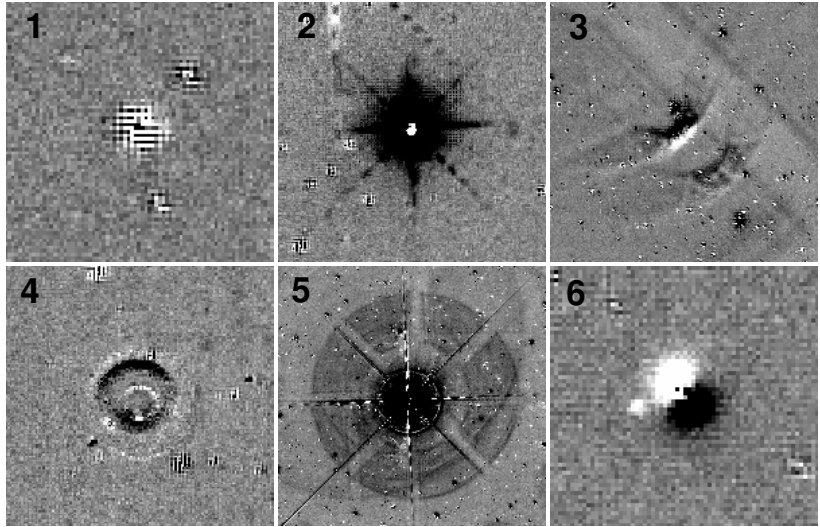

**Figure A.1.** Example of artifacts in the [Fe II]-H image. 1) The residual of star subtraction shown as cross stripes 2) A variable and saturated star with a diffraction pattern 3) Arc-shape ghosts near a very bright star. The diffraction pattern of a bright star is superposed. 4) Electronic cross-talk near a bright star 5) Diffraction pattern of a bright star 6) High proper-motion star. (Black: bright, white: dark.)

## APPENDIX A: EXAMPLES OF ARTIFACTS
## APPENDIX B: THE UWIFE [FEII]-H AND GPS 3-COLOR IMAGES OF IFOS

This paper has been typeset from a T<sub>E</sub>X/L<sup>A</sup>T<sub>E</sub>X file prepared by the author.

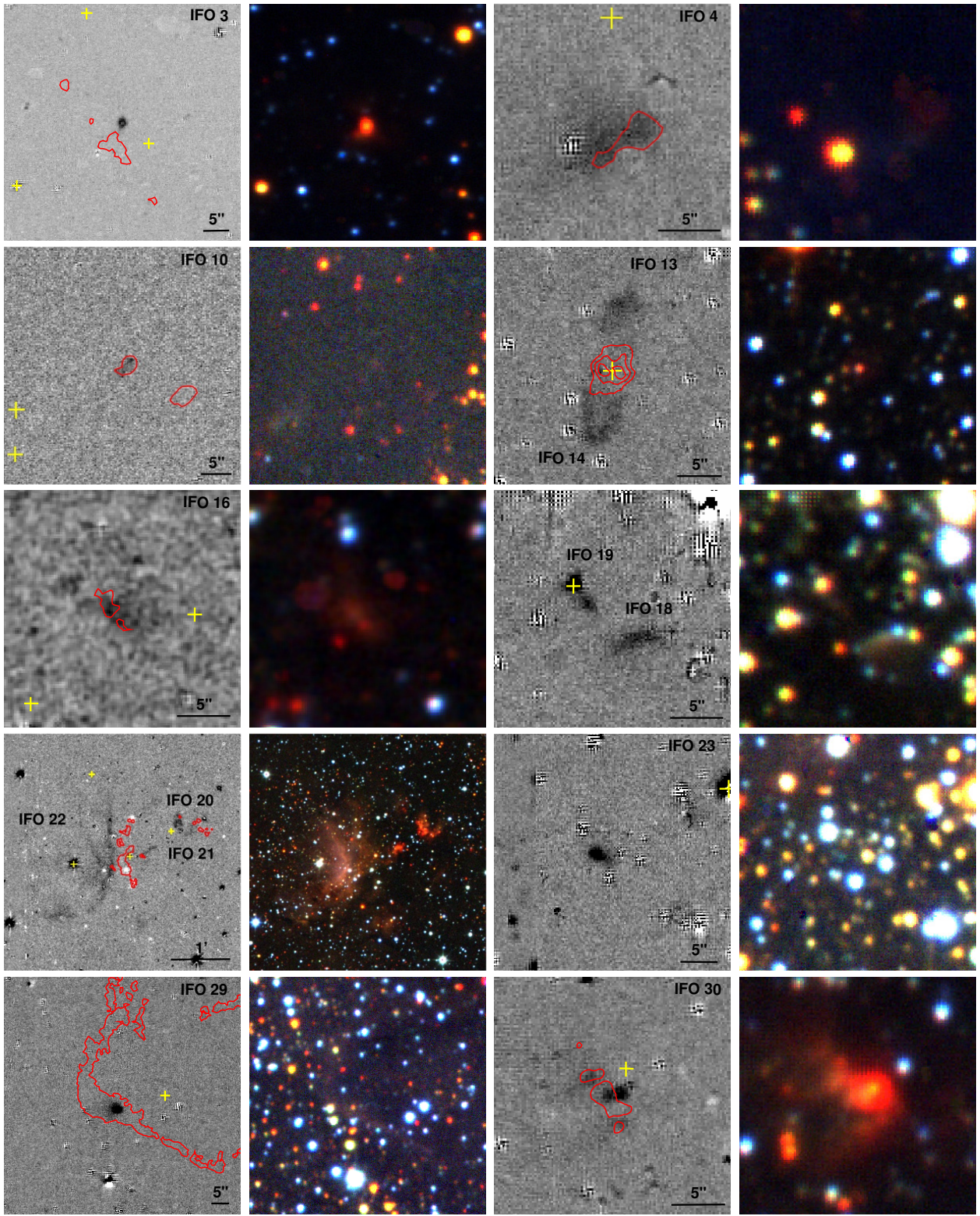

**Figure B.1.** IFOs with YSO counterpart candidates in continuum-subtracted [Fe II] and three-color *KHJ* UKIDSS images in the same field of view. The format for these images is the same as that of Fig. 1. Yellow crosses denote adjacent YSOs in the field of view. Red contours are H$_2$ 2.12 $\mu$m emission adopted from UWISH2.

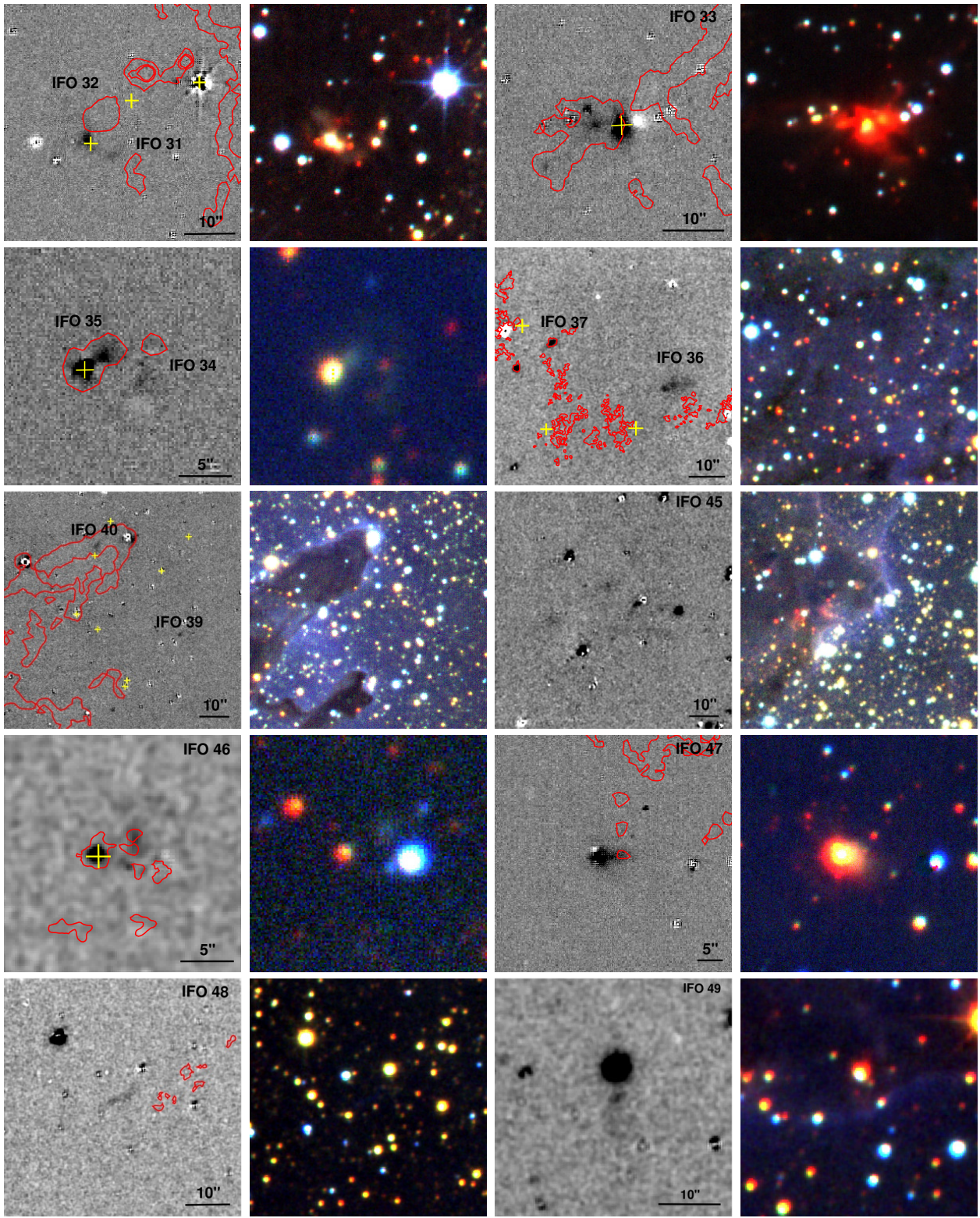

**Figure B.1.** – *Continued*

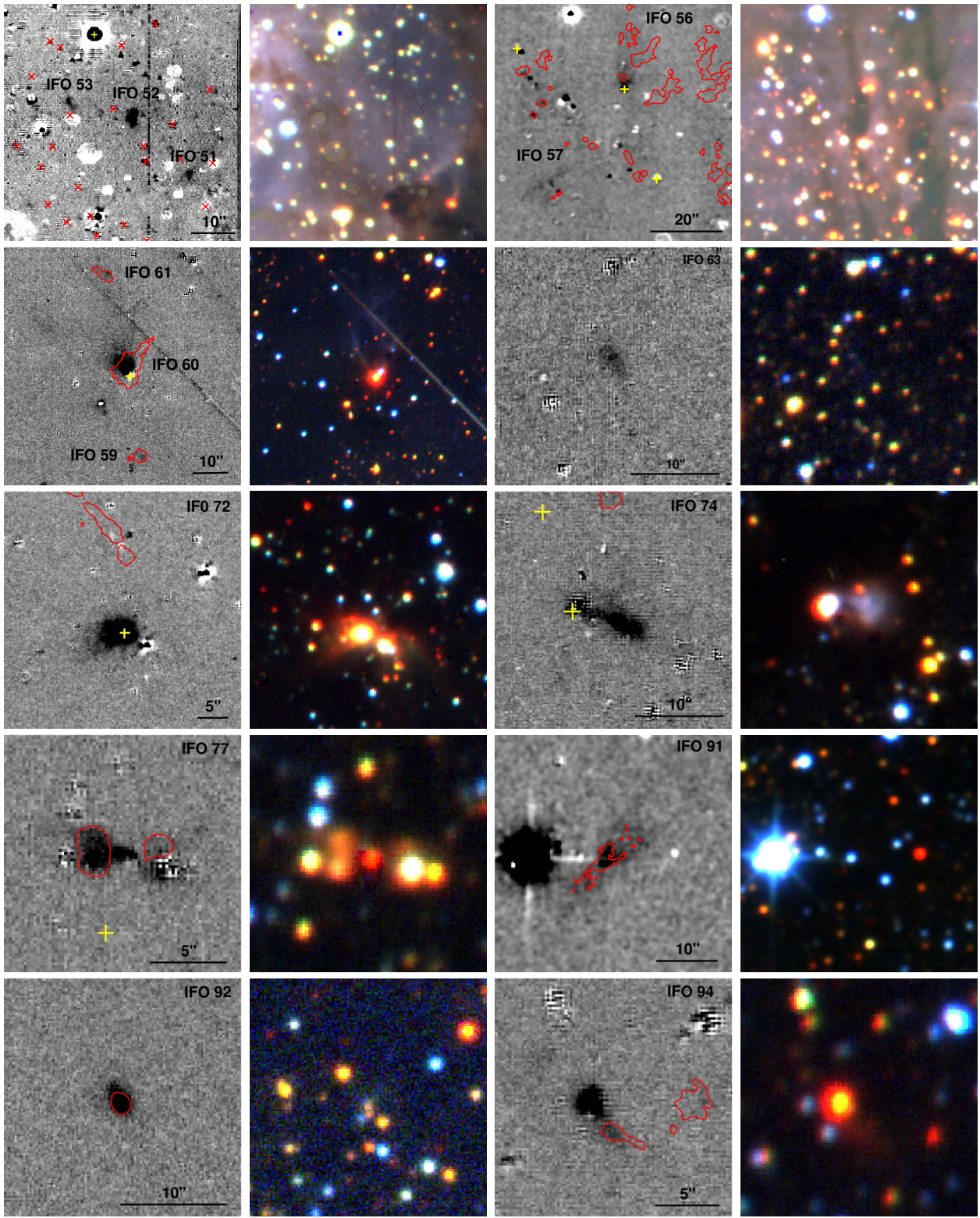

**Figure B.1.** – *Continued*

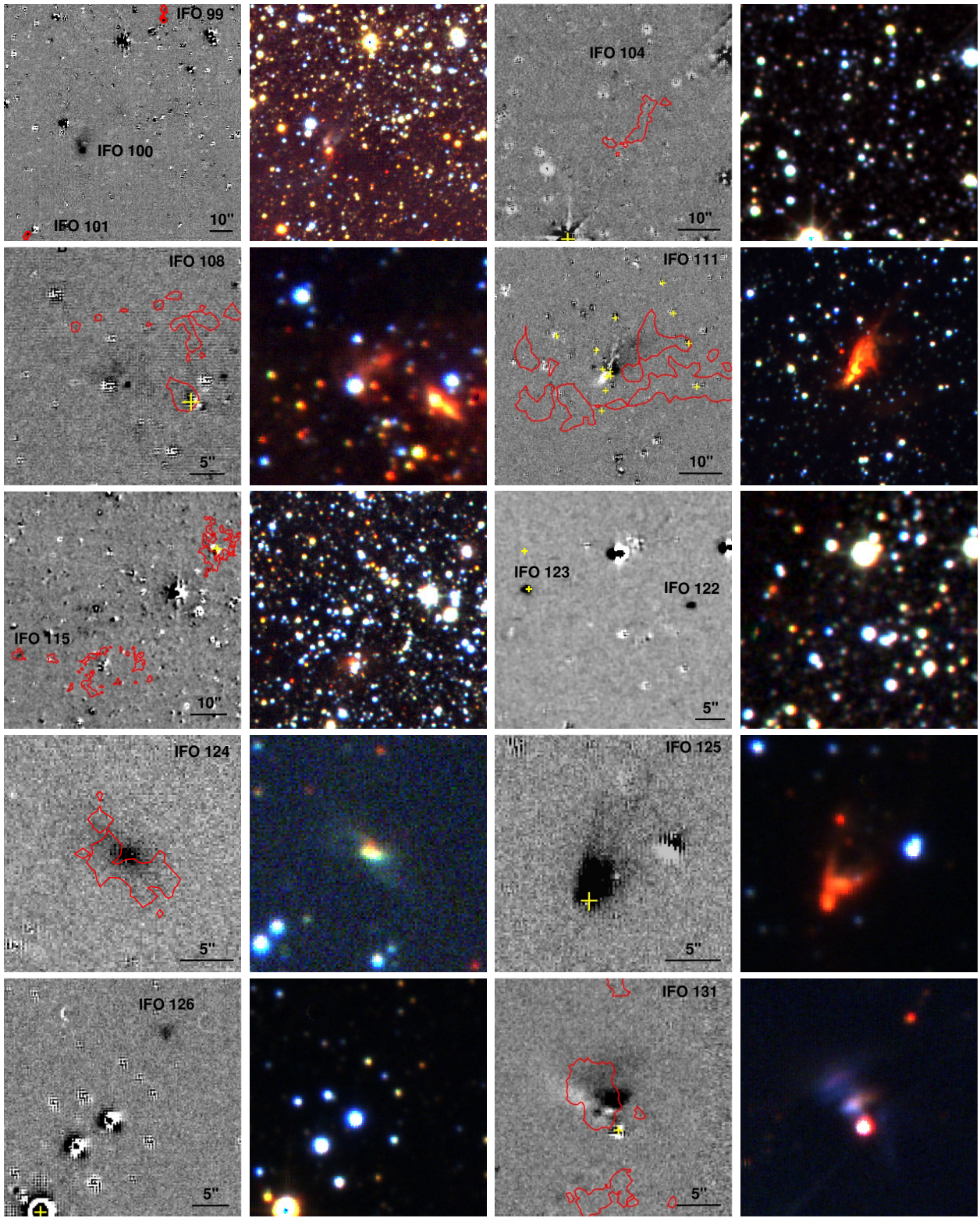

**Figure B.1.** – *Continued*

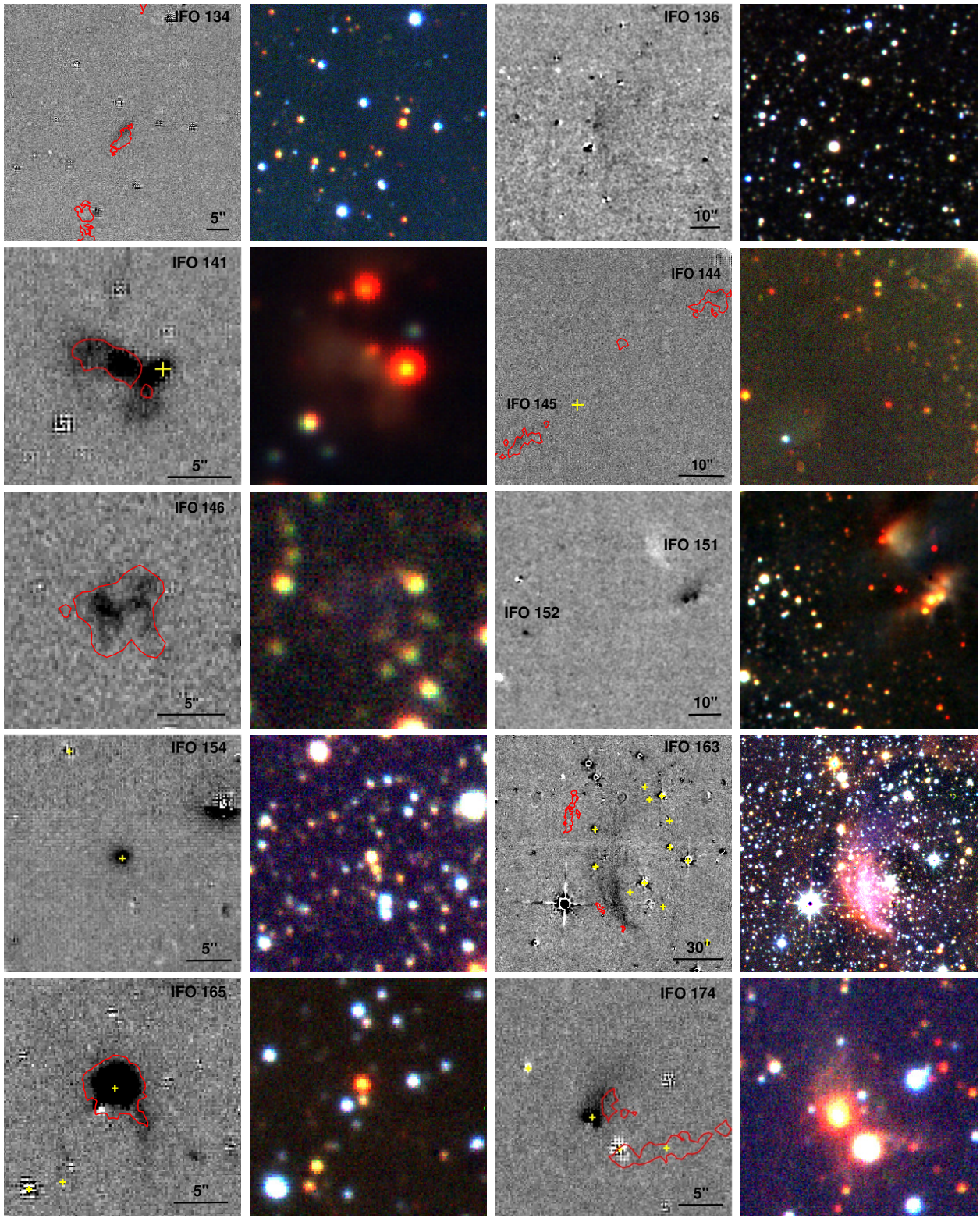

**Figure B.1.** – *Continued*

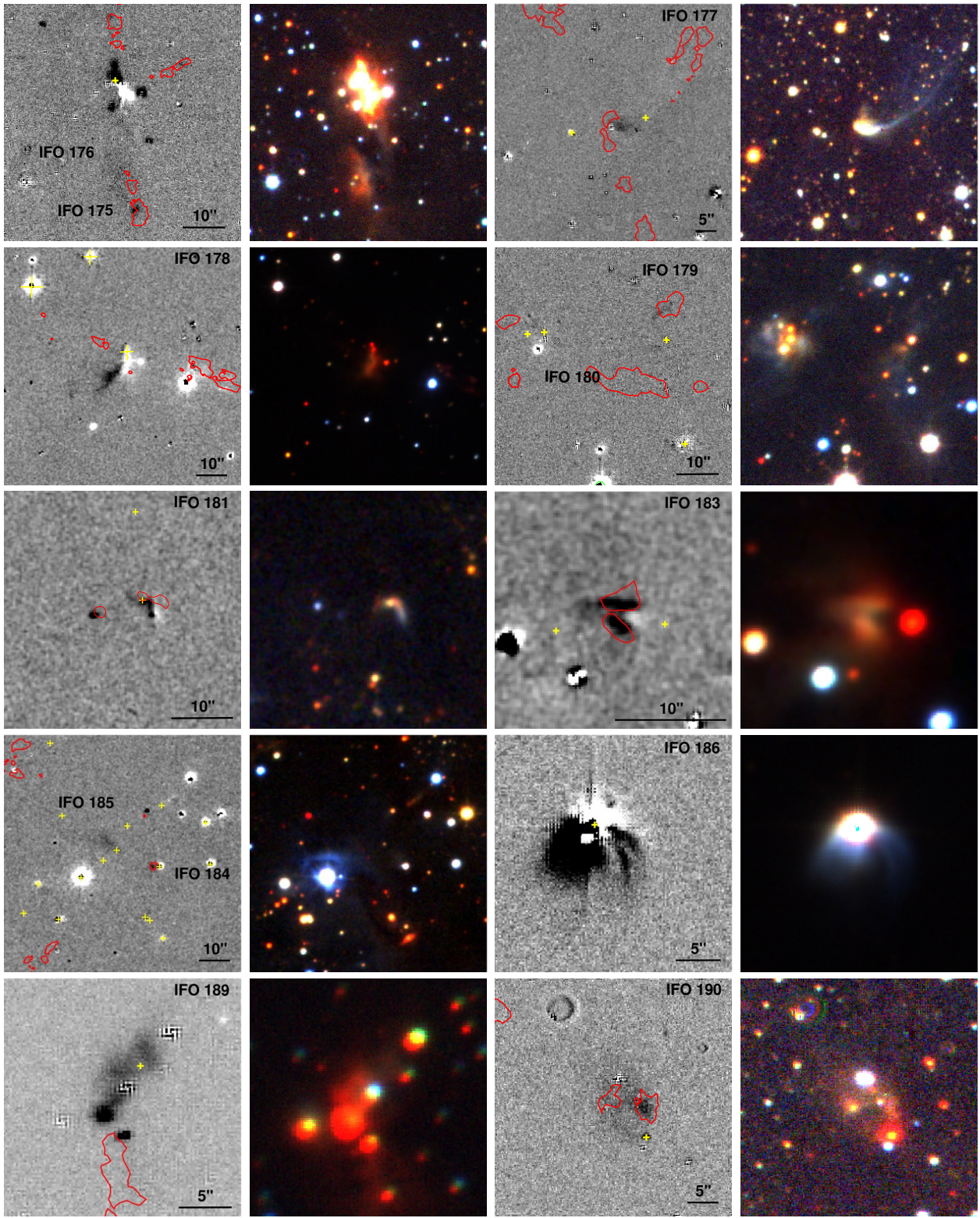

**Figure B.1.** – *Continued*

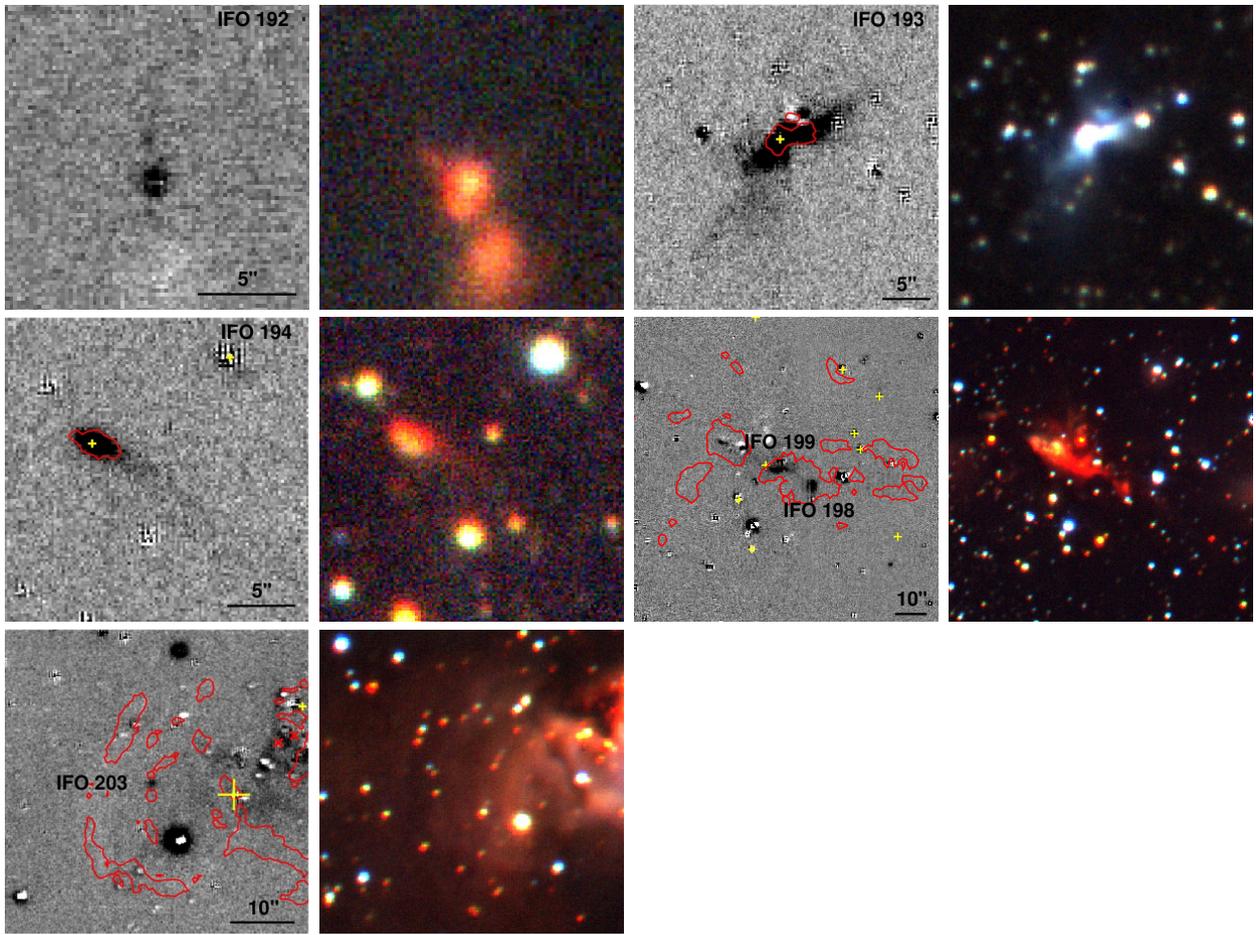

**Figure B.1.** – *Continued*

**Figure B.2.** Continuum-subtracted [Fe II] images of IFOs with PN counterpart candidates. The units on the UWIFE [Fe II]-H are DNs, with the darker colour denoting a higher DN. The corresponding source names for each IFO are shown. The yellow cross marks the central position of the counterpart. The images of IFO 73, 095, 112, 129, 157, and 164 are smoothed with a two-pixel Gaussian. In all images, North is at the top and east is on the left side.

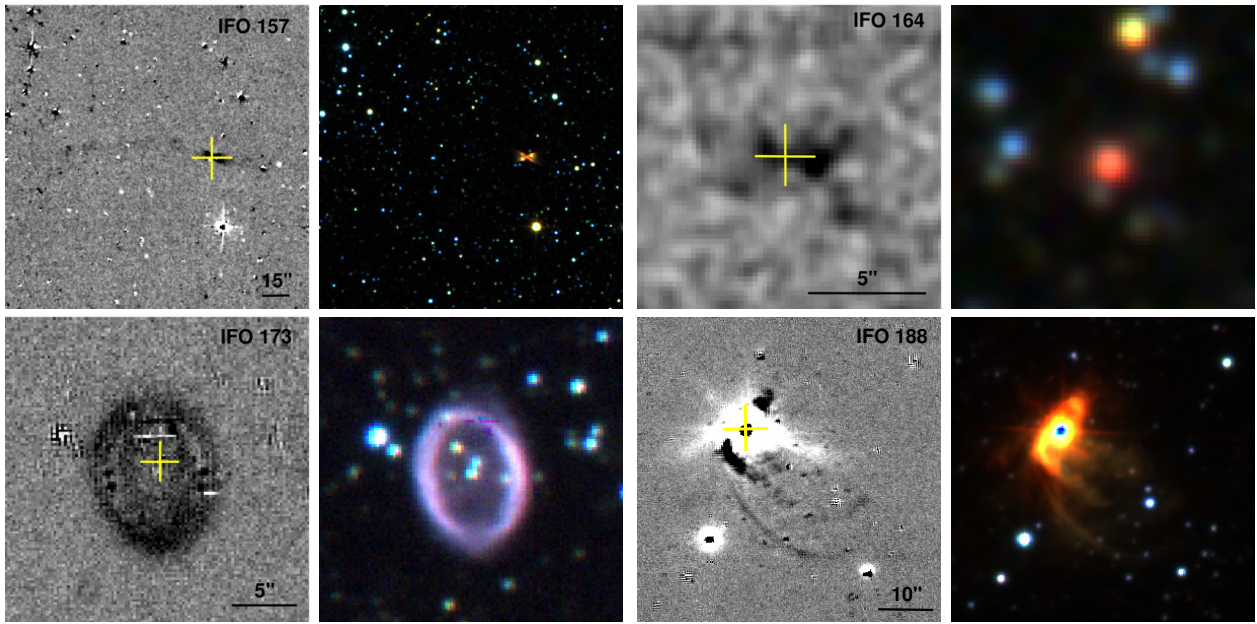

**Figure B.2.** – *Continued*

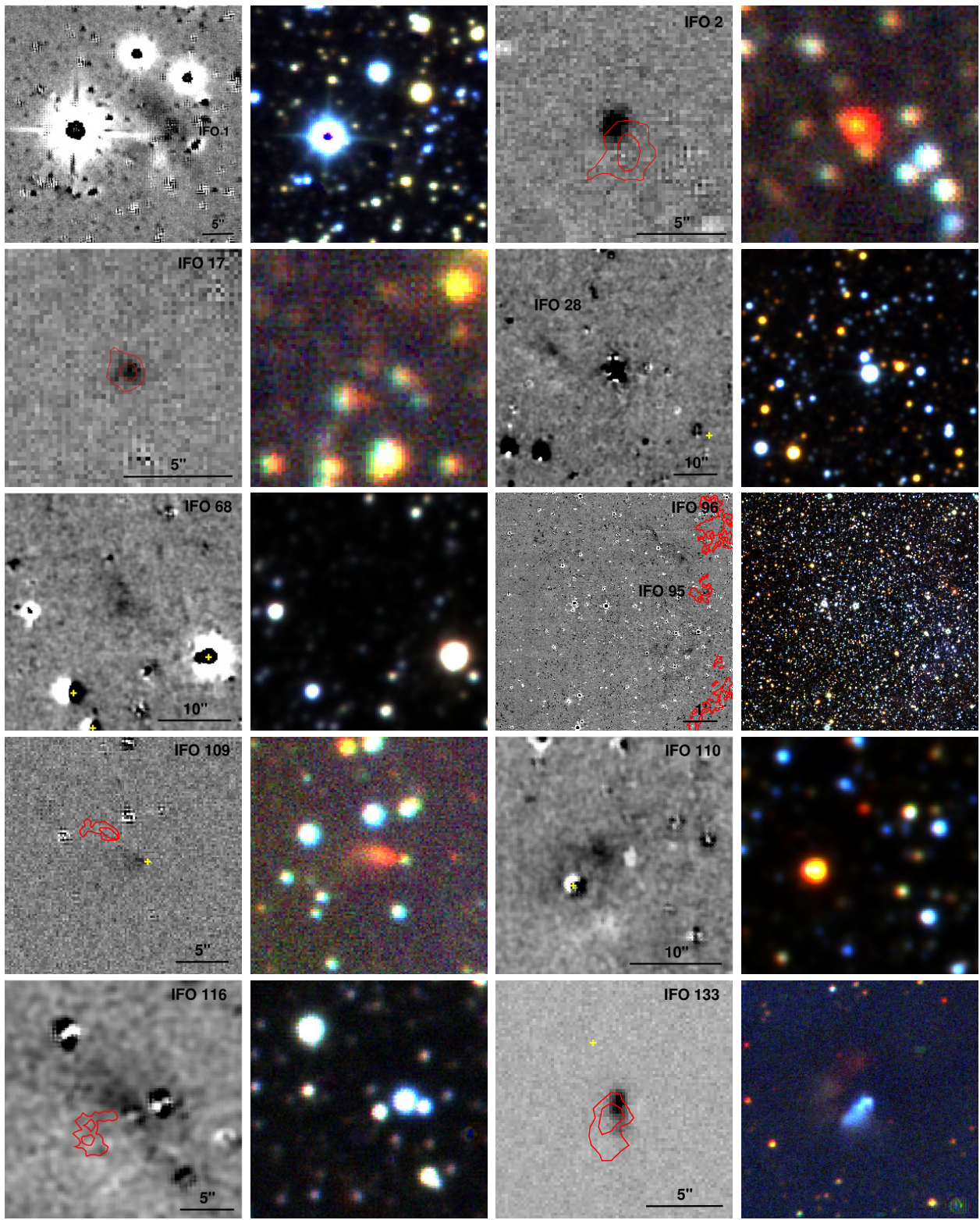

**Figure B.3.** IFOs with counterpart candidates unknown in continuum-subtracted [Fe II] images as in Fig. 1. IFO 95 is a PN-IFO inside the field of view. The format for these images is the same as that of Fig. 1.

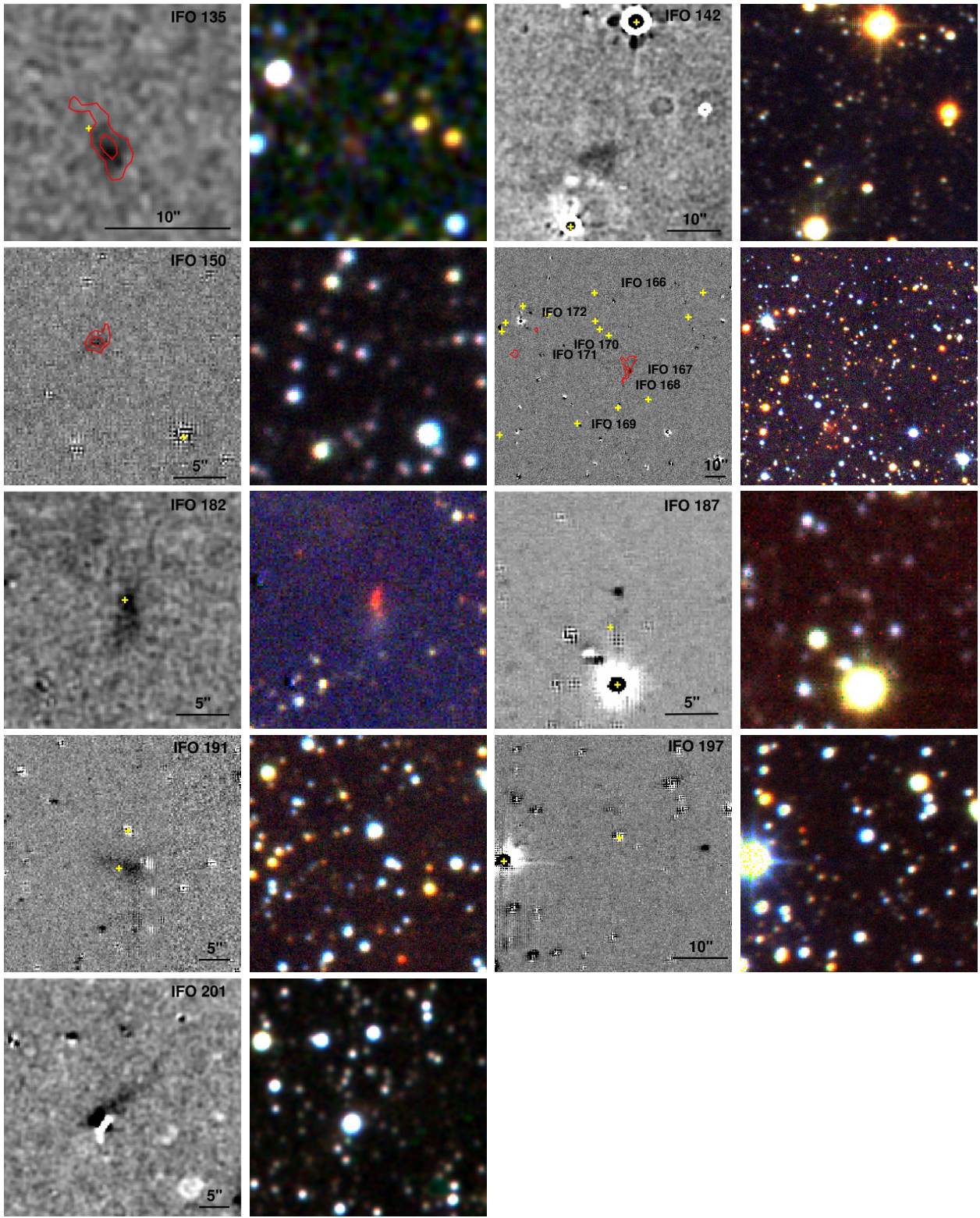

**Figure B.3.** – *Continued*